# Tailoring the Morphology of Cellulose Nanocrystals via Controlled Aggregation

*Kévin Ballu[a], Jia-Hui Lim[b], Thomas G. Parton[c], Richard M. Parker[a], Bruno Frka-Petesic[a,d], Alexei A. Lapkin[e,f], Yu Ogawa[b]*, Silvia Vignolini[a,c]**

[a] Yusuf Hamied Department of Chemistry, University of Cambridge, Cambridge CB2 1EW, UK

[b] University of Grenoble Alpes, CNRS, CERMAV, Grenoble 38000, France

[c] Department of Sustainable and Bio-inspired Materials, Max Planck Institute of Colloids and Interfaces, Potsdam 14476, Germany

[d] International Institute for Sustainability with Knotted Chiral Meta Matter (WPI-SKCM$^2$), Hiroshima University, Hiroshima 739-8526, Japan

[e] Department of Chemical Engineering and Biotechnology, University of Cambridge, Cambridge CB3 0AS, UK

[f] Innovative Center in Digital Molecular Technologies, Yusuf Hamied Department of Chemistry, University of Cambridge, Cambridge CB2 1EW, UK

* Yu Ogawa: yu.ogawa@cermav.cnrs.fr

* Silvia Vignolini: sv@mpikg.mpg.de

ABSTRACT

Cellulose nanocrystals (CNCs) are elongated nanoparticles derived from natural cellulose, with potential applications ranging from rheological modifiers and emulsion stabilizers to photonic pigments and sensors. For most applications, precise control over CNC morphology and surface chemistry is essential, but the relationship between process parameters, CNC characteristics, and their resulting behavior is poorly understood. Here, we investigate the impact of centrifugation and ionic strength on CNC morphology after dialysis using transmission electron microscopy, small-angle X-ray scattering and scanning electron diffraction. We find that the centrifugation step commonly applied during CNC purification promotes the formation of compact composite nanoparticles made of aligned crystallites, referred to as "bundles", that are associated preferentially along their hydrophobic faces. In stark contrast, transient exposure to high ionic strength leads to fractal-like, irregular composite nanoparticles. We then examine the consequence of these morphological differences on the cholesteric self-organization of the CNCs: aligned bundles reduce the cholesteric pitch in suspension, causing a blue-shift in the color of dish-cast photonic films, while misaligned particles promote gelation, producing colorless films. This study reveals the importance of sample history, in particular, the often-disregarded purification steps, on CNC characteristics and their ensemble behavior, thereby unlocking new routes for tailoring this promising nanomaterial.

# INTRODUCTION

Cellulose nanocrystals (CNCs) are elongated crystalline nanoparticles derived from native cellulose.[1,2] CNCs have recently attracted significant interest as a sustainably-sourced nanomaterial for numerous potential applications.[3] At present, the most industrially relevant method used for CNC extraction is the hydrolysis of wood cellulose with concentrated sulfuric acid (~ 10 M). During this process, the cellulose fibers are degraded and disassembled into nanoscale fragments while sulfate half-ester groups ($–OSO_3H$) are grafted onto the exposed surfaces.[4,5] Then, the reaction is quenched via dilution with water and the nanoparticles are purified. At the laboratory scale, this purification process usually consists of several rounds of centrifugation and redispersion in deionized water, which allows for the separation of the acidic supernatant from the CNC pellet, followed by dialysis against deionized water to remove excess ions. The resulting CNC suspension is a colloidally stable mixture of isolated cellulose crystallites and composite multi-crystallite particles.[6–8]

Among the composite particles, the presence of raft-like particles made of aligned crystallites—often referred to as bundles—influences key characteristics of the final suspension. For example, an increasing proportion of CNC bundles has been correlated with a decrease in the cholesteric pitch and a narrowing of the concentration range of the biphasic regime.[8] Moreover, the presence of bundles could also modify the overall behavior of CNCs as Pickering agents, due to their distinctive morphology and amphiphilicity.[9–11] On this basis, it has been proposed that the crystallites that make up the composite particles are preferentially associated along a specific crystal plane.[9–11] Yet, the substructure of these objects has never been observed, leaving it unclear which specific crystal face, if any, might be involved in a potential preferential orientation.

Whether these laterally associated composite particles predate the hydrolysis (and are therefore related to the source),[6] or are mostly formed through aggregation during the production process remains an open question. Conventional DLVO theory suggests that the high ionic strength conditions during the hydrolysis and the initial rounds of centrifugation (*i.e.* $[H_2SO_4] \approx 4\text{-}10$ M) should cause irreversible aggregation of the CNCs. Nevertheless, CNC pellets can be readily redispersed upon lowering the ionic strength and are assumed to do so spontaneously during dialysis, with final suspensions exhibiting very little sedimentation. This discrepancy with theory is rarely mentioned in the literature, with few studies investigating the impact of hydrolysis conditions (*e.g.* acid concentration, temperature) on the CNC morphology and properties,[2] and fewer studies considering the influence of the suspension history after hydrolysis.[12–15] This likely results in the oversight of experimental factors that are crucial in determining the CNCs characteristics. For example, little attention has been given to the centrifugation parameters used during purification (i.e., relative centrifugal force, duration), with reported conditions ranging from intense centrifugation (e.g. 20,000g for 20 min) to none at all,[16–18] and with studies often not even specifying the parameters used.

In this work, we investigate the impact of the post-hydrolysis centrifugation step on the morphology of CNCs and compare it with the irreversible aggregation induced by divalent cations. Using dynamic light scattering (DLS), transmission electron microscopy (TEM), small-angle X-ray scattering (SAXS), and viscometry, we show that CNC purification by centrifugation favors the formation of bundles, *i.e.* compact laterally aligned composite particles (**Figure 1a**). We further show by scanning nanobeam electron diffraction (SNBED) that these bundles are preferentially associated through their hydrophobic faces. Conversely, destabilization by exposure to calcium chloride ($CaCl_2$) leads to an irreversible increase in the CNC size due to the formation of randomly associated fractal-like composite particles held by $Ca^{2+}$ crosslinking (**Figure 1b**). We then investigated the consequences of aggregation on the cholesteric self-assembly behavior. We found that increasing the presence of bundles reduces the pitch, resulting in a blue-shift of the films obtained by dish-casting the suspension, while the presence of irregular particles made using $CaCl_2$ instead promote gelation at a lower concentration, leading to colorless films (**Figure 1c**).

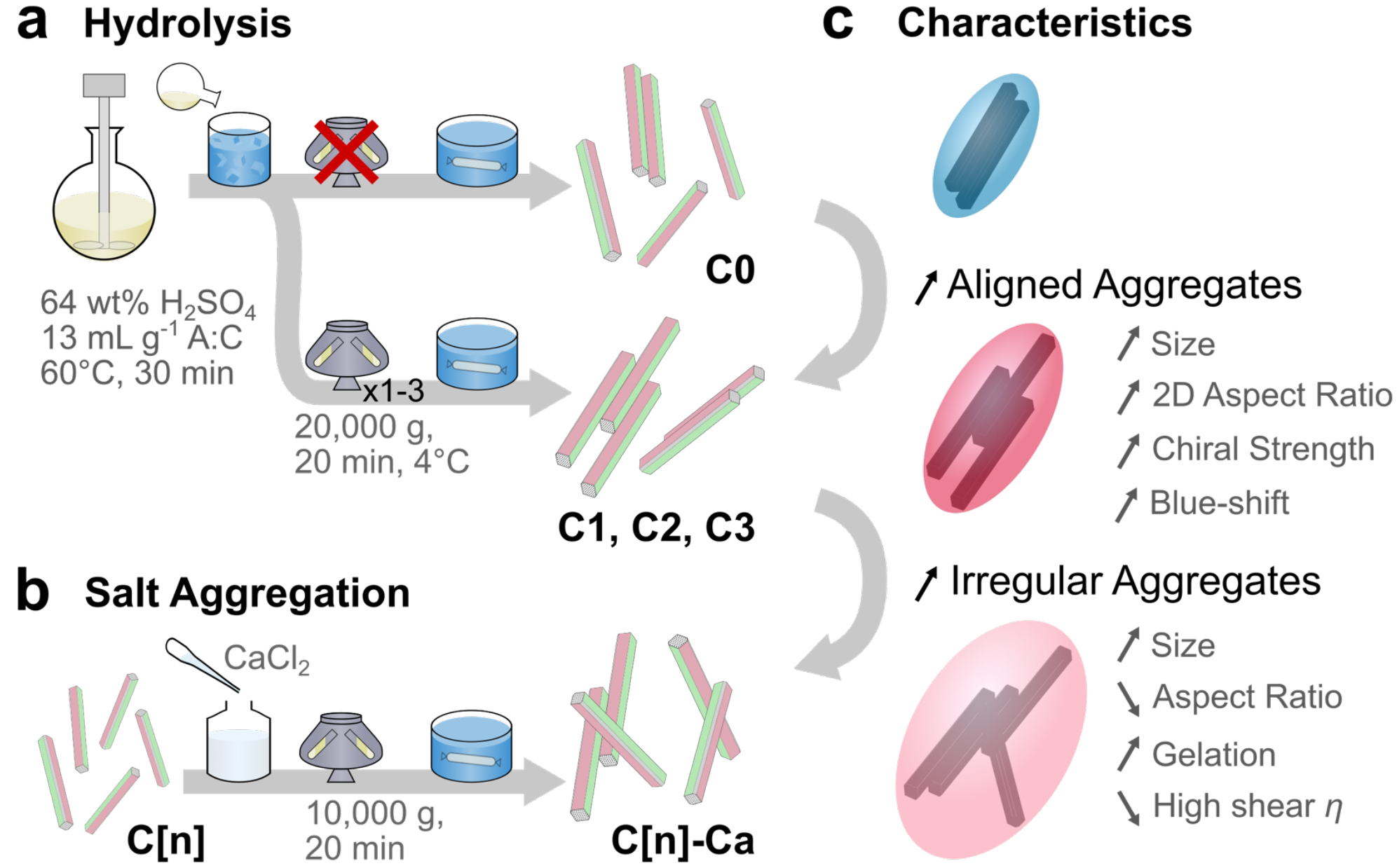

**Figure 1**. Scheme summarizing the preparation of **(a)** never centrifuged (**C0**) and centrifuged (**C1, C2, C3**) CNCs, and **(b)** salt aggregated CNCs subsequently produced by $CaCl_2$ addition followed by centrifugation (**C0-Ca** and **C3-Ca**), and **(c)** corresponding key differences in CNC characteristics relative to **C0**.

## RESULTS AND DISCUSSION

### Self-limiting CNC Aggregation by Centrifugation

To explore the influence of post-hydrolysis centrifugation on CNC morphology, cotton-derived cellulose was hydrolyzed into CNCs according to the conditions summarized in **Figure 1a** and detailed in the Experimental Section. After quenching the hydrolysis reaction, part of the mixture was isolated and never centrifuged (**C0**), while the remaining mixture was subjected to multiple cycles of centrifugation and re-dispersal of the pellet in ultrapure water. Aliquots of the mixture were isolated after one, two, or three cycles of centrifugation and redispersion, resulting in samples **C1**, **C2** and **C3** respectively.

The effect of centrifugation on the CNC size after dialysis was investigated by measuring the Z-average hydrodynamic diameter ($D_H$) of each suspension by DLS. As shown in **Figure 2**, the first round of centrifugation (from **C0** to **C1**) led to a significant increase in $D_H$, while subsequent rounds (from **C1** to **C2** or from **C2** to **C3**) had no significant effect. This indicates that the first centrifugation step applied during CNC purification leads to composite particle formation.

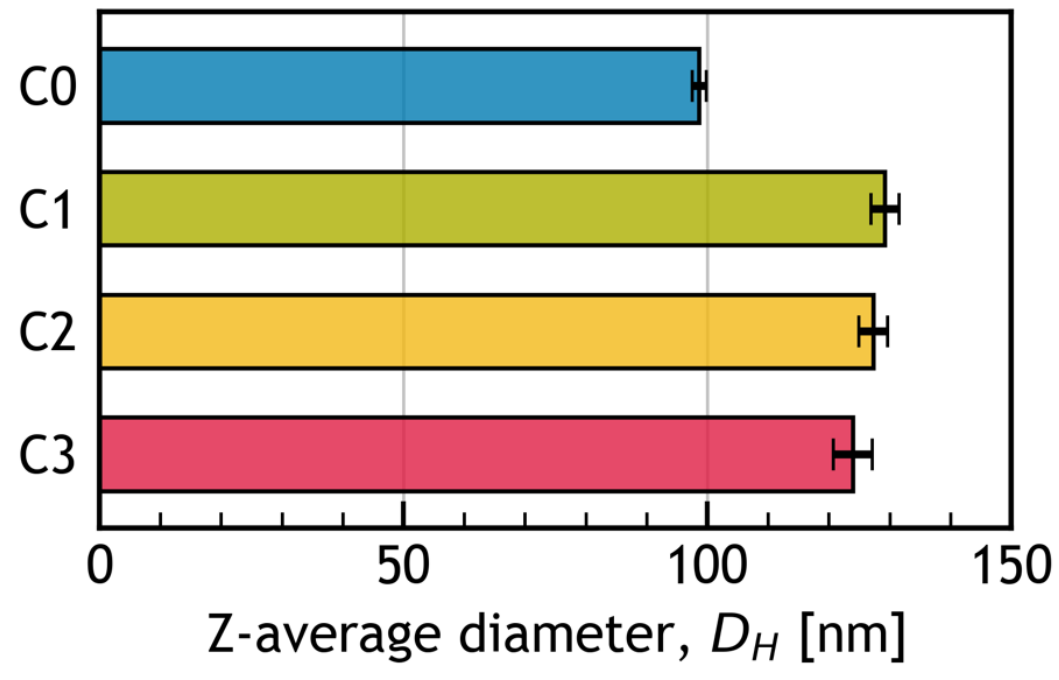

**Figure 2**. Z-average diameter ($D_H$) of the never centrifuged CNCs (**C0**) and CNCs centrifuged once (**C1**), twice (**C2**) or three times (**C3**), as determined by dynamic light scattering (DLS).

To evaluate the colloidal stability of the CNCs, the surface charge for each suspension was determined by conductometric titration against sodium hydroxide. The centrifuged (**C1**, **C2**, **C3**) and never-centrifuged CNCs (**C0**) all exhibited a similar amount of sulfate half-ester groups on their surface ($264 \pm 5$ mmol kg$^{-1}$). The similar surface charge exhibited by all the samples suggests that the formation of composite particles did not cause the –$OSO_3H$ groups to become buried between internal crystallite interfaces. These typical surface charge values, along with the high ζ-potential ($\zeta \approx -49$ mV, see Supporting Information (SI) **Table S1**), are indicative of good colloidal stability at low ionic strength.

However, the ionic strength ($I$) of the medium during the first and second centrifugation steps (with [$H_2SO_4$] > 0.1 M) is significantly greater than the thresholds sufficient to induce aggregation of sulfated CNCs (typically reported to be < 0.03 M).[19–21] Consequently, the CNCs are expected to be colloidally unstable during both these steps. Yet, no further size increase between C1 and C2 was observed after dialysis. This suggests that centrifugation under high ionic strength triggers irreversible CNC aggregation, but only up to a certain size limit, likely reached during the first centrifugation. Further aggregation beyond this point appears reversible, as it can be undone by lowering the ionic strength through extensive dialysis against water. Moreover, all the suspensions previously experienced a stronger ionic strength environment during the hydrolysis, indicating that the irreversible formation of composite particles during centrifugation cannot be attributed solely to electrostatic destabilization.

## Irreversible CNC Aggregation by Salt Addition

The role of electrostatic destabilization on the irreversible formation of composite CNC particles was investigated by increasing the ionic strength of CNC suspensions, followed by redispersion at low ionic strength to retain only irreversible aggregates. For these experiments, the **C0** and **C3** suspensions were pH-neutralized with sodium hydroxide and concentrated to yield **C0-Na** and **C3-Na** respectively.

The protocol for transiently increasing the ionic strength consisted of four steps, as summarized in **Figure 3a**: (i) preparation of a CNC suspension at fixed ionic strength ($I_{agg}$) and CNC weight fraction ($w_{agg}$ = 6.5 wt%), (ii) centrifugation (10,000 g for 20 min), (iii) redispersion by mixing, (iv) and dilution to 0.1 wt% CNC and with an ionic strength as close as possible to 1 mM for DLS measurement. These parameters were selected following a series of experiments to optimize the protocol (see Section **S2.1** of the SI, **Table S2**, **Figure S2**, and **Figure S3**). Interestingly, the size of the irreversible aggregates increased during the first hour of exposure to elevated ionic strength, after which the size plateaued (see **Table S2** and **Figure S3**).

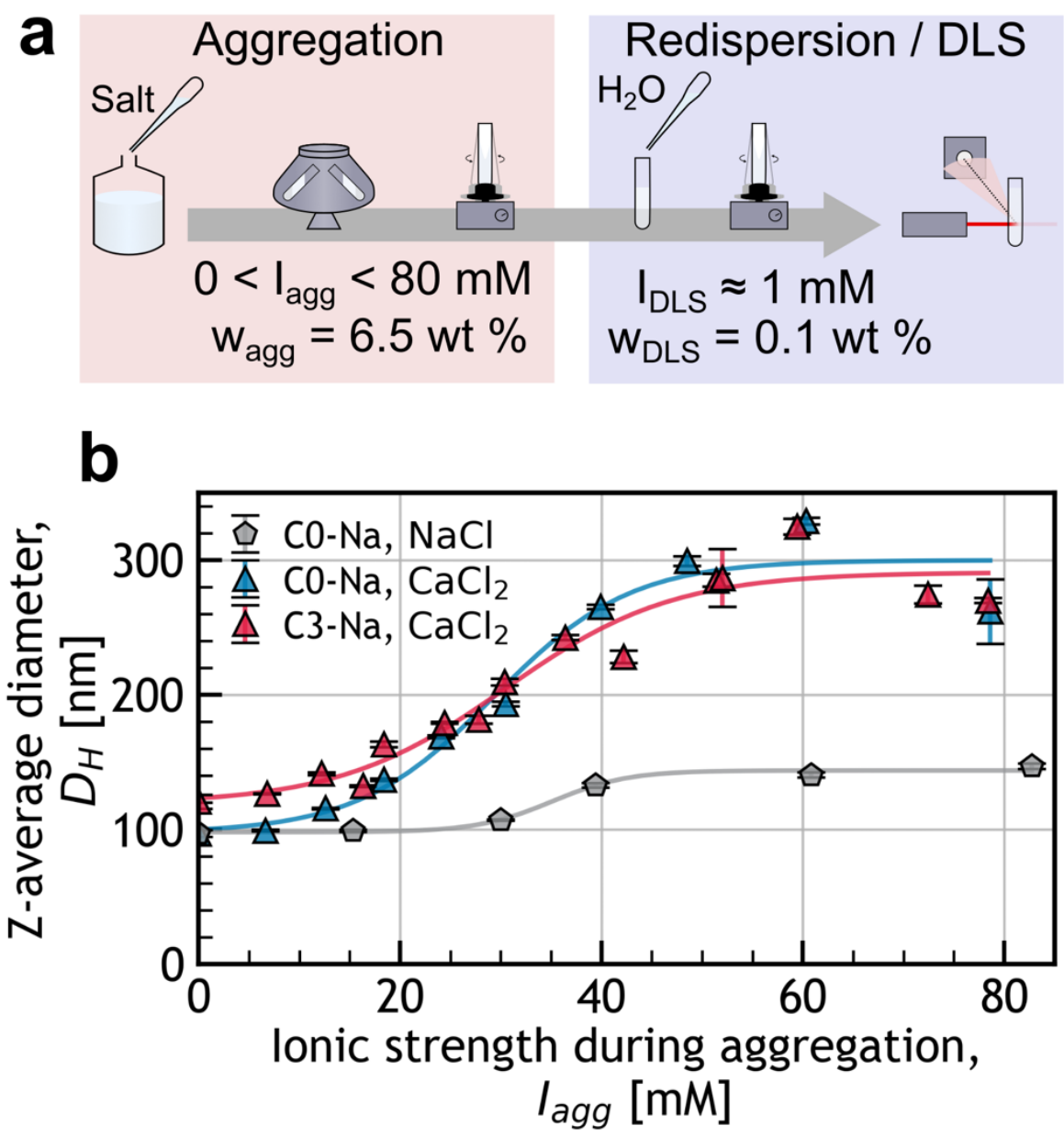

**Figure 3**. **(a)** Scheme summarizing the calcium aggregation process and **(b)** the impact of salt type (NaCl vs $CaCl_2$) and transiently raised ionic strength ($I_{agg}$) at fixed weight fraction ($w_{agg}$ = 6.5 wt%) during the aggregation step on the resulting Z-average diameter ($D_H$), as measured after dilution to 0.1 wt% in an ionic strength as close as possible to 1 mM. The lines of best fit are presented as visual guides and were obtained from a logistic function.

In an initial experiment, the never-centrifuged CNCs were mixed with NaCl at varying transient ionic strengths ($0 \leq I_{agg} \leq 80$ mM). As shown in **Figure 3b**, no change in $D_H$ was observed after the centrifugation of **C0-Na** without added salt ($I_{agg} = 0$). This indicates that the size increase observed between **C0** and **C1** was only made possible by the high ionic strength of the medium. Then, a moderate increase of particle size was first observed around $I_{agg}$ = 30 mM before quickly reaching a plateau of ~ 145 nm for $I_{agg} \geq 50$ mM (see also **Figure S4** for corresponding intensity-based DLS size histograms). This suggests that the size of the irreversible aggregates formed by exposure to high ionic strength is self-limited, in agreement with the previous DLS measurements (**Figure 2**). Our findings contrast with a previous study reporting that NaCl-induced aggregation of CNCs was reversible upon dialysis against deionized water.[20] However, since the CNCs in that study were already centrifuged during purification, we suspect that they had already reached their maximum self-limiting size prior to salt-induced aggregation.

To determine whether this size limit could be overcome, we investigated the effect of a divalent cation by using calcium chloride ($CaCl_2$) instead of NaCl. In this case, the CNCs exhibited a much greater increase in $D_H$ than that observed with NaCl at equivalent $I_{agg}$ (**Figure 3b**). The $D_H$ values followed a sigmoid-like curve with $I_{agg}$, starting to increase from approx. 100-120 nm at $I_{agg}$ = 12 mM before reaching a plateau around 290 nm above $I_{agg}$ = 50 mM. Above $I_{agg}$ = 40 mM, the variability of the measurements seemingly increased significantly, and macroscopic gel-like objects were visible in the sample cuvette for $I_{agg} \geq 50$ mM. The destabilization with $CaCl_2$ was repeated using centrifuged CNCs (**C3-Na**), leading to similar results (**Figure 3b**). Although the particle sizes were initially slightly larger than the never-centrifuged sample, both samples exhibited comparable sizes for $I_{agg} \geq 20$ mM.

These results indicate that the size of the irreversible aggregates is dependent on the cation used, in agreement with previous studies comparing CNC aggregation induced by monovalent or polyvalent cations.[4,20–23] In the present case, the difference could be attributed to the irreversible formation of calcium cross-links between CNCs. Furthermore, the similar trends observed for **C0-Na** versus **C3-Na**, suggest that the mechanism of calcium-induced aggregation is different from centrifugation-induced aggregation. The two methods thus provide independent pathways to produce composite CNCs.

## Dialyzed Calcium Aggregated CNC Suspensions

To explore the differences in particle characteristics between the two aggregation pathways, **C0-Na** and **C3-Na** were centrifuged in the presence of calcium chloride ($I_{agg} = 40$ mM) followed by dialysis against water, to respectively produce **C0-Ca** and **C3-Ca**, collectively referred to as Ca-CNCs (see protocol illustrated in **Figure 1b**). Elemental analysis of **C3-Ca** yielded a calcium content of 133 mmol $kg^{-1}$, corresponding to exactly half of the sulfate half-ester groups measured for its parent sample (**C3**) (264 ± 5 mmol $kg^{-1}$, see **Table S1**). Given the divalent character of $Ca^{2+}$ ions, this corresponds to 266 meq $kg^{-1}$ of $Ca^{2+}$, indicating that all the charged groups on the surface of **C3-Ca** were in the form of calcium salts and that the charged groups were not hidden by the calcium-induced destabilization.

The hydrodynamic diameter of these Ca-CNCs was confirmed to be substantially greater than for **C0** and **C3** (**Table S1**), and a comparison of the DLS-derived count rate versus Z-average diameter suggests that the Ca-CNCs are considerably less compact than the **C0, C1, C2** and **C3** samples (**Figure S1** and Section **S1.1** of the SI). However, as a single, ensemble-averaged quantity, the hydrodynamic size provided by DLS measurements cannot distinguish more subtle morphological differences between samples.

## CNC Particles Morphology (TEM)

TEM was used to compare the morphological properties of the individual CNC particles, as exemplified in **Figure 4a**. Visual inspection of the images revealed a variety of particle types in each sample, including composite particles constituted of aligned subunits (referred to as bundles) and irregular composite particles, which are both commonly reported for wood and cotton-derived CNCs.[6,8] Notably, **C0-Ca** and **C3-Ca** appeared to contain a higher proportion of larger irregular composite particles compared to **C0** and **C3**.

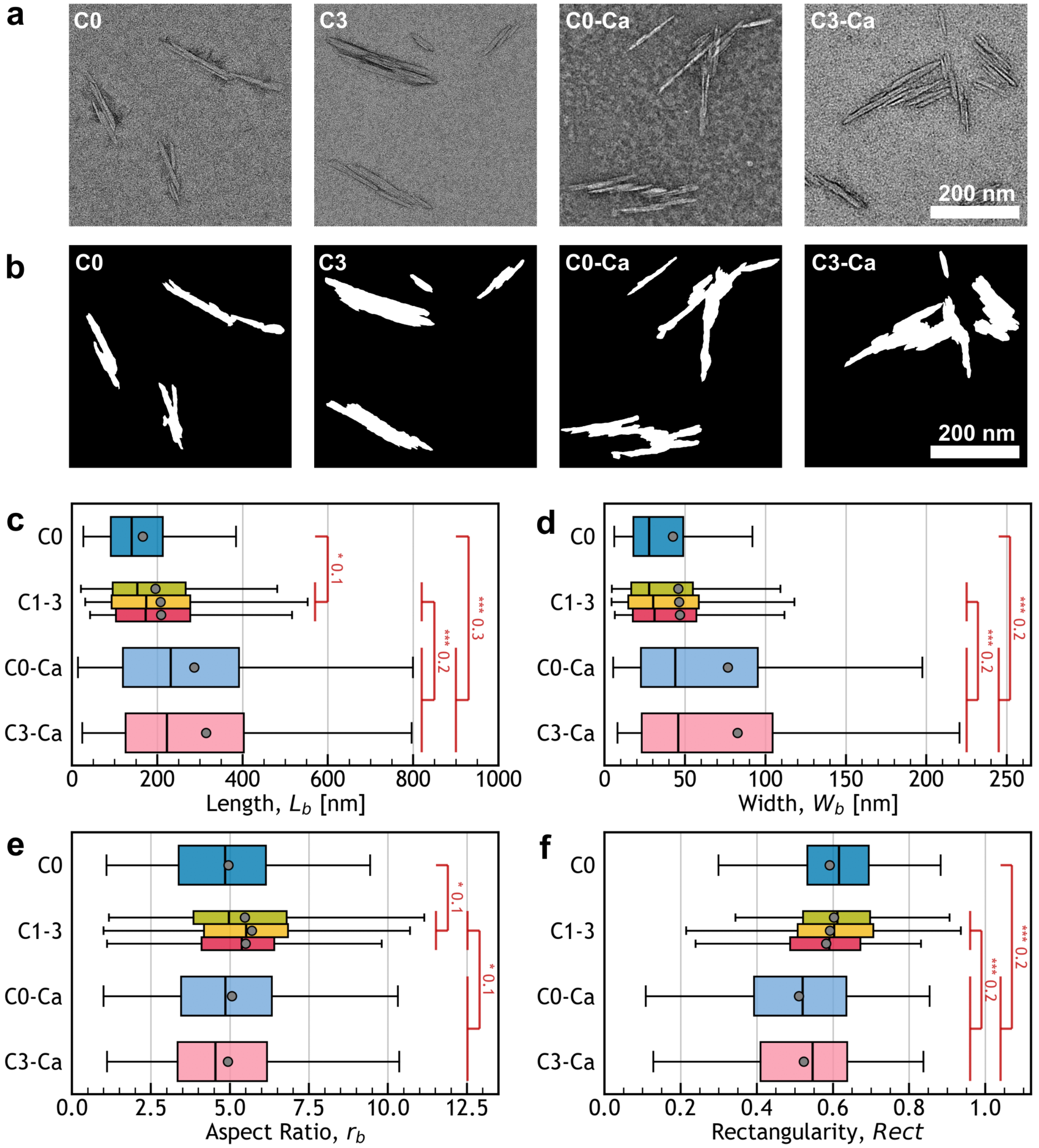

**Figure 4.** Analysis of the CNC morphology from TEM images. **(a)** Typical objects observed by TEM and **(b)** examples of the corresponding contoured shapes used for dimension extraction. **(c-f)** Boxplots of the morphological parameters extracted: **(c)** bounding box lengths ($L_b$), **(d)** bounding box widths ($W_b$), **(e)** bounding box aspect ratios ($r_b$), and **(f)** rectangularity (*Rect*). Significance of the pairwise Mann–Whitney *U* test is indicated by the following p-value thresholds: * ($0.05 > p > 0.01$), ** ($0.01 > p > 0.001$), and *** ($p < 0.001$). The importance of the effect is indicated by the absolute point biserial coefficient (number in red), with values of 0.5, 0.3, and 0.1 often considered as obvious (large), subtle (medium), and merely statistical (small) effects, respectively (see Section **S4.2** of the SI for more details). Grey-filled circles indicate the average values and outliers are not displayed; full data are presented in **Figure S5** and **Figure S6** (SI).

To quantitatively compare the particles, their morphological properties were extracted. In the literature, the presence of composite particles leads to inconsistencies regarding what is considered an individual CNC during particle analysis, leading to significant operator bias during image analysis.[2] Considering each discrete object on the TEM grid as a single CNC, whether it appears as a single crystallite or as a composite object formed by multiple overlapping crystallites, minimizes user bias from particle selection and produces morphological distributions consistent with ensemble-averaged size values.[8] Using this approach, the outline of more than 225 particles per sample were traced, as exemplified in **Figure 4b**.

The distributions for the bounding box length ($L_b$) and width ($W_b$) of the CNC outlines are shown as box plots in **Figure 4c-d**. As typically observed for CNCs, both parameters exhibit high variance, making it challenging to determine whether the observed differences in mean values are statistically significant. Moreover, while the histograms for $L_b$ and $W_b$ are often modeled as log-normal distributions,[6,24] a thorough investigation revealed that most samples in this study exhibit signs of deviation from log-normality (see **Figure S5**, **Table S3** and Section **S3.2** of the SI). Therefore, the series were compared pairwise using the non-parametric Mann–Whitney *U* test (see **Figure 4c-d** and Section **S4.2** of the SI) which determines the statistical significance of a difference between the means of two distributions of arbitrary type.[25] The magnitude of the difference in mean values was also quantified using the point biserial correlation coefficient *r* (detailed calculation in Section **S4.2** of the SI).[26]

As indicated in **Figure 4c-d**, the never centrifuged sample (**C0**) exhibited significantly shorter particles than the centrifuged CNCs (**C1**, **C2**, **C3**) while the centrifuged CNCs exhibited similar lengths and widths. Similarly, the Ca-CNCs (**C0-Ca** and **C3-Ca**) exhibited similar dimensions that were significantly longer and wider than all the other CNCs. These trends are in agreement with the previous DLS measurements.

The morphology of the particles was further examined by calculating the bounding box aspect ratio ($r_b$) and rectangularity (*Rect*) defined respectively as:

$$r_b = \frac{L_b}{W_b} \tag{1}$$

and

$$Rect = \frac{A}{L_b W_b} \tag{2}$$

where *A* is the outline area.[8] The distributions for $r_b$ and *Rect* for each sample were investigated graphically and statistically. As presented in **Figure 4e-f**, $r_b$ and *Rect* followed skewed normal distributions for all samples (see **Figure S6** for histograms and **Table S3**). Compared to **C1**, **C2** and **C3**, **C0** exhibits a lower aspect ratio but a similar rectangularity, while Ca-CNCs display both a lower aspect ratio and a lower rectangularity.

Cryo-TEM of the CNC suspensions was used to estimate whether the observed morphologies originate from drying artifacts arising from standard TEM grid preparation.[27] In the cryo-TEM images, **C0** and **C3** mainly appeared laterally associated with rare occurrences of irregular composite particles, whereas **C3-Ca** exhibited significantly more irregular objects (**Figure S9**). As such, these observations qualitatively confirm the divergence of morphology between the different samples determined from the standard TEM analysis.

In summary, the morphological parameters extracted from TEM images for **C0**, **C1**, **C2** and **C3** indicate that centrifugation during purification leads to the irreversible formation of composite particles while retaining a similar shape. For elongated flat particles, this scaling infers the formation of compact composite particles constituted of aligned subunits (*i.e.* bundles). Moreover, the comparable morphological parameters of centrifuged CNCs (**C1**, **C2**, **C3**) further suggest that the formation of bundles occurred only during the first round of centrifugation (**C0** to **C1**) and was likely due to the presence of a saturation point beyond which particle association is negligible.

Concerning the CNCs aggregated with $CaCl_2$ (**C0-Ca** and **C3-Ca**), we hypothesize that they underwent random and irreversible association of crystallites into fractal-like irregular composite particles cross-linked by calcium ions. For charged elongated rods, extended DLVO theory predicts that crossed association offers a lower energetic barrier to overcome compared to parallel association.[28,29] Therefore, despite the parallel association being more stable, crossed association is favored upon rapid aggregation, leading to bigger and less elongated particles. Moreover, this crossed-association mechanism is favored at high ionic strength,[30] and was previously observed for charged CNCs.[21,31,32]

## CNC Ensemble Morphology (SAXS)

To validate the trends in individual particle morphology observed from TEM images, we characterized the same samples in suspension using SAXS. The SAXS spectra were collected over $0.008 \leq q \leq 0.247$ nm$^{-1}$, corresponding to distances in real space between 750 and 30 nm ($\sim 2\pi/q$). The corresponding concentration normalized intensities, $I_c(q)$, are presented in **Figure 5a**. At very low $q$ ($q < 0.013$ nm$^{-1}$), **C3**, **C0-Ca** and **C3-Ca** all exhibited a lack of linearity of ln [$I(q)$] as a function of $q^2$. This was previously attributed to CNC size polydispersity,[6] and prevents reliable extraction of radii of gyration ($R_g$) from a Guinier analysis. In the Porod regime ($0.06 \leq q \leq 0.2$ nm$^{-1}$), the samples exhibited power law exponents of -1.9, -2.0, -2.1 and -2.1 for **C0**, **C3**, **C0-Ca** and **C3-Ca** respectively. This suggests a rather elongated but flattened shape, with increasing degree of branching in this sequence. A similar trend was previously observed in the same $q$ region upon increasing CNC aggregation through salt addition.[31]

At low concentration, the SAXS intensity profile of a sample is mainly determined by the shape of the particles in suspension. To extract this morphological information, a SAXS intensity profile was modeled for each sample by generating populations of polydisperse rectangular prisms, as illustrated in **Figure 5b**. The lengths and widths of the objects were randomly generated from the sample size distributions obtained by TEM, while their thickness ($T$) was set to a single, arbitrary value (see Section **S4.1** of the SI for more details).[6] For each sample, this process was repeated over a range of $T$ values to identify the thickness $T_b$ that best matched the considered experimental SAXS curve (see **Figure S10** and Section S4.2 of the SI). The resulting $T_b$ for **C0**, **C3**, **C0-Ca** and **C3-Ca** were 3.7, 8.5, 6.0 and 6.3 nm respectively. The corresponding modeled intensity profiles are shown in **Figure 5c**. The excellent agreement between the modeled and the experimental SAXS profiles indicates that the dimensions extracted from TEM are representative of the particles in suspension.

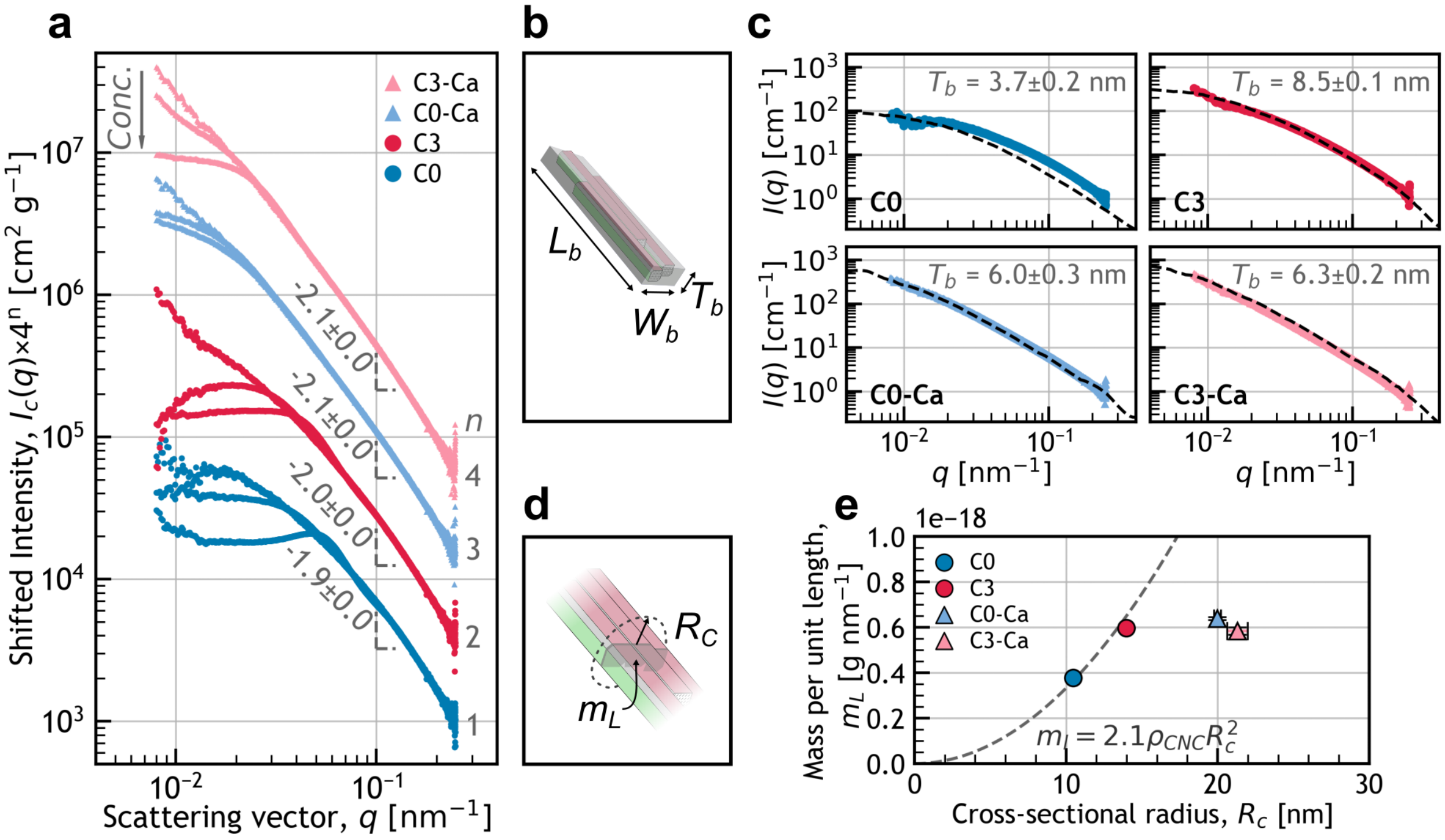

**Figure 5**. SAXS analysis of the CNCs. **(a)** Concentration normalized intensity, $I_c(q)$ at different concentrations (*Conc.*, starting from the bottom **C0**: 1.0, 0.5 and 0.1 wt%; **C3**: 1.0, 0.5 and 0.1 wt%; **C0-Ca**: 0.77, 0.5 and 0.1 wt% and **C3-Ca**: 1.0, 0.5 and 0.1 wt%, all rescaled vertically for clarity) and corresponding Porod exponents (measured at $0.06 \leq q \leq 0.2$ nm$^{-1}$). **(b)** Rectangular prism model used to compute SAXS intensity profiles from the TEM-extracted length ($L_b$) and width ($W_b$), and the corresponding best fitting thickness ($T_b$). (**c**) Scattering intensity profiles *I(q)* at 0.1 wt% for each sample reported with best thickness and corresponding best fitting model (black dashed line). (**d**) Schematic illustrating the particle cross-sectional mass ($m_L$) and cross-sectional radius of gyration ($R_c$) extracted from the cross-sectional Guinier analysis. **(e)** Corresponding evolution of $m_L$ as a function of $R_c$, and best fit of $m_L = a\ \rho_{CNC}\ R_c^2$ for **C0** and **C3**. See Section **S4** of the SI for more details.

The extracted $T_b$ value for **C0** ($T_b$ = 3.7 nm) appears unexpectedly small compared to previous studies on cotton CNCs and crystallites.[6,31] This likely reflects a bias introduced by the modeling approach. Indeed, applying the bounding box dimensions to a rectangular prisms model for samples containing a significant proportion of irregular composite particles (*i.e.* a low rectangularity) leads to an overestimation of the particle volume (see **Figure 4b**, **f**). Since the SAXS intensity scales with the actual volume (or mass) of cellulose contained in each particle (at a given $\Phi$), a lower apparent thickness $T_b$ compensates for such volume overestimation.

Interestingly, although **C3** exhibited larger bounding box dimensions than **C0**, it also had a greater thickness ($T_b$ = 8.5 *vs.* 3.7 nm). This suggests that **C3** contains larger particles with a comparable rectangularity, consistent with a higher proportion of bundles in this sample. In contrast, **C0-Ca** and **C3-Ca** exhibited intermediate thickness values ($T_b$ = 6.0 and 6.3 nm respectively), despite displaying even larger bounding box dimensions. This discrepancy likely results from a greater proportion of irregular composite particles in these samples, as indicated by their lower rectangularity (see **Figure 4f**).

To compare the cross-sectional density of the CNCs while minimizing assumptions, a cross-sectional Guinier analysis at intermediate $q$ was conducted by assuming a cylindrical particle shape (see **Figure S11** and Section **S4.3** of the SI for more details). The corresponding cross-sectional radius of gyration ($R_c$) and the cross-sectional particle mass ($m_L$, g nm$^{-1}$) are shown in **Figure 5d, e.** The $R_c$ increased from **C0** to **C3** and then further to Ca-CNCs, indicating an increase in the cross-section of the CNCs in this order. For a compact particle, an increase in cross-section is expected to cause a proportional increase of $m_L$ with $R_c^2$. Fitting of $m_L$ against $\rho_{cell}\, R_c^2$ for **C0** and **C3** leads to a proportionality factor of 2.1 ± 0.1, supporting that these samples have a similar compactness, in agreement with bundle formation upon centrifugation.

Concerning the Ca-CNCs samples, they exhibited $m_L$ values that were similar to each other and comparable to that of **C3**. However, their $R_c$ was significantly larger, revealing a distinctive decrease in cross-sectional density upon calcium-induced aggregation, in accordance with the formation of randomly associated particles. Interestingly, **C0-Ca** and **C3-Ca** presented a similar cross-sectional density, despite their respective parent samples having different morphologies. Yet, if calcium-induced aggregation only caused random association of the particles, then **C0-Ca** should contain fewer bundles than **C3-Ca** and therefore exhibits a lower cross-sectional mass and apparent $T_b$, which is not observed here. Instead, it seems that calcium-induced aggregation also triggers some lateral association between crystallites, within the irregular particles, to the same extent as the centrifugation during purification, rendering **C0-Ca** and **C3-Ca** structurally comparable.

Overall, the morphological parameters extracted from the SAXS analysis are consistent with the TEM analysis. This further supports the formation of compact laterally associated particles upon centrifugation in sulfuric acid, and the formation of randomly associated fractal-like particles upon calcium-induced aggregation. Moreover, this ensemble measurement demonstrates that the reported morphological differences between these samples are significant in suspension, despite the CNC polydispersity.

## Relative Orientation of Crystallites within CNCs (SNBED)

The morphological characterization techniques discussed above only consider the overall particle morphology and the alignment of the long axes of the crystallites within the composite particles. However, the relative orientation of the crystallite cross-section with respect to adjoining crystallites can also have an important impact on the CNC properties. In cotton fibers, native crystallites are mainly found in the cellulose Iβ crystalline allomorph and are believed to possess approximately hexagonal cross-sections, as illustrated in **Figure 6a**.[33] Consequently, the vast majority of the exposed crystal surfaces corresponds to the (110) and (1-10) planes, with the remaining surfaces corresponding to the (100) plane. The greater density of hydroxy groups on the (110) and (1-10) planes is expected to make these faces more hydrophilic than the (100) plane, which is therefore described as "hydrophobic" in comparison.

The relative orientation of the cross-section of the crystallites within composite CNCs cannot be determined by conventional morphological characterization such as TEM or AFM. Therefore, we characterized our samples using scanning nanobeam electron diffraction (SNBED), a method of four-dimensional scanning transmission electron microscopy (4D-STEM). This technique utilizes a focused electron beam to obtain the local 2D electron diffraction (ED) patterns for each position scanned across a TEM grid, as illustrated in **Figure 6b**.[34] In this work, each ED pattern was collected by an electron beam probe with a diameter of 25 nm, with all crystallites within the probe diameter contributing to the obtained diffraction information. Each scan position containing an ED pattern (constituting a "pixel") was indexed according to the cellulose Iβ unit cell.[35] This information was used to determine the local crystallographic orientation with respect to the incident electron beam, as illustrated in **Figure 6b**. Example SNBED data are provided in **Figure S12**.

The CNC diffraction patterns for **C0**, **C3**, **C0-Ca** and **C3-Ca** could be divided into three distinct categories, corresponding to either the [010], [110] or [1-10] zone axes pointing normal to the TEM grid surface. None of the diffraction patterns corresponded to the [100] zone axis, indicating that crystallites were never observed with their hydrophobic faces oriented against the grid. This is likely due to either poor adhesion between the hydrophobic (100) plane and the hydrophilic glow-discharged carbon film used as the TEM grid, or cross-sectional anisotropy of the particles. The lack of observation of other crystallographic planes suggests the CNCs are relatively well-faceted in the planes corresponding to the observed zone axes.

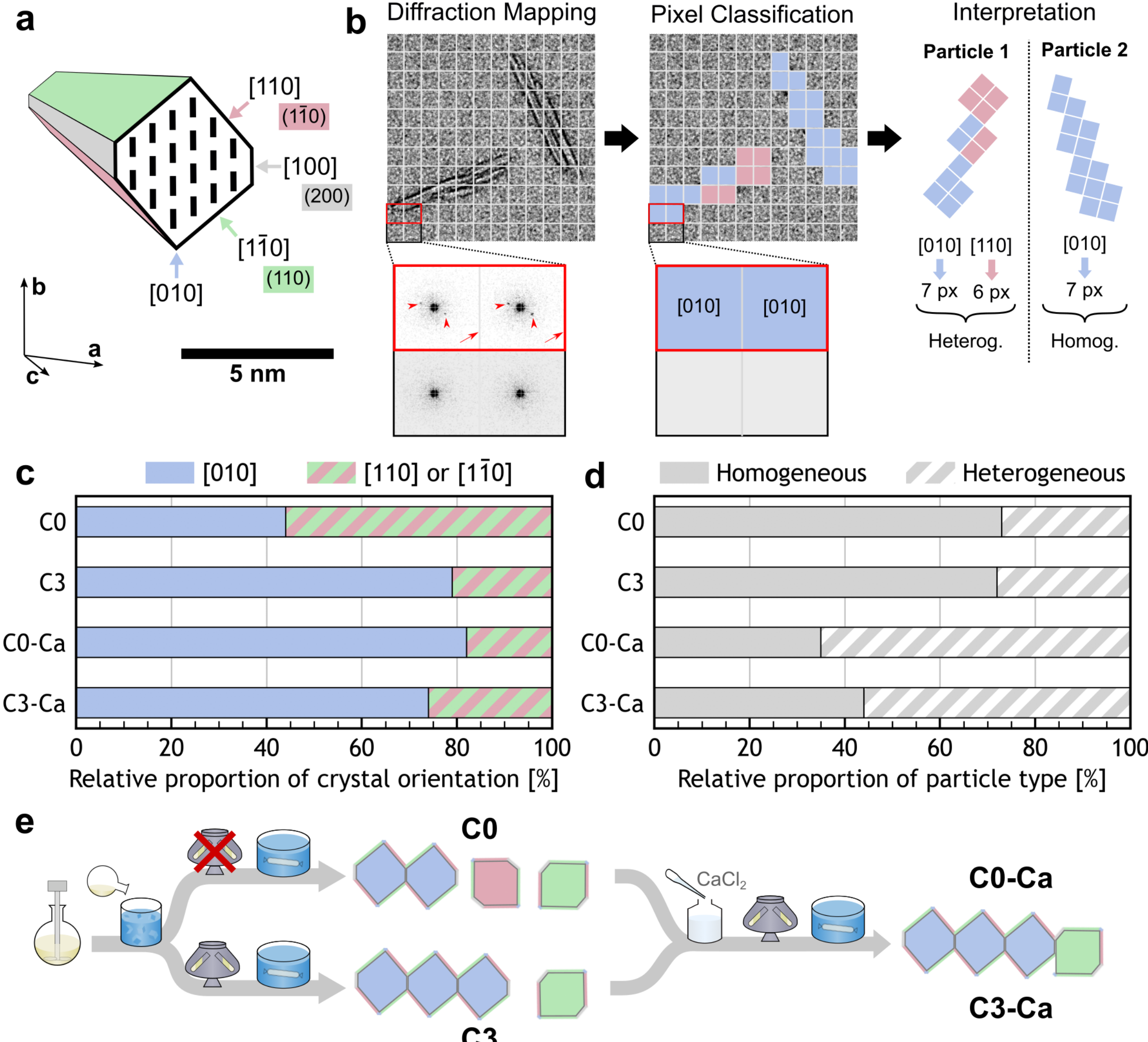

**Figure 6**. Local orientation of the crystallites constituting the cellulose nanocrystals probed by scanning nanobeam electron diffraction (SNBED). **(a)** Cross-section of a native cellulose Iβ crystal from cotton, with the crystal zone-axes directions and their corresponding crystal planes labelled. **(b)** Illustration of the diffraction pattern mapping process, with pixel classification (illustrative stained TEM image) according to crystal orientation ([010] in blue, [110] in pink or [1-10] in green), and particle classification with particles displaying a mix of crystal orientations classified as Heterogeneous (**Particle 1**) and pixels from particles displaying only one orientation classified as Homogeneous (**Particle 2**). The pixels have a dimension of 25x25 $nm^2$. **(c)** Relative proportion of cellulose crystal plane orientations (based on pixel number) and **(d)** relative proportion of particle type for the different CNC samples. **(e)** Schematic summarizing the preparation of the different CNC samples and the resulting impact on the crystallite orientations and particle types observed by SNBED (dialysis steps were hidden for clarity).

**Figure 6c** shows the relative amount of each crystallite orientation for each sample as a proportion of the SNBED pixels for which diffraction patterns were observed. For the never-centrifuged CNCs (**C0**), the proportion of pixels corresponding to the [010] zone axis (pointing normal to the TEM grid) was $P_{[010]}$ = 44%, with the remaining 56% divided between the [110] and [1-10] zone axes. In contrast, for the centrifuged CNCs (**C3**) and the Ca-CNCs (**C0-Ca** and **C3-Ca**), a strong majority of the measured pixels corresponded to the [010] zone axis (respectively $P_{[010]}$ = 79, 82, and 74%). This observation was corroborated by selected area electron diffraction (SAED) conducted on the same suspensions on more densely deposited grids and on much larger areas (~ 0.8 $\mu m^2$) containing a large ensemble of particles, and averaged over multiple locations across the TEM grid (see **Figure S13**). These results indicate that for all the samples, the crystallites are more likely to be facing the TEM grid on the (010) crystal plane than on any other plane.

The greater $P_{[010]}$ for **C3** compared to **C0** suggests that the formation of bundles increases the number of crystallites with their (010) plane facing the grid (**Figure 6c**). This observation cannot be explained simply by different relative interactions between specific crystal faces and the TEM grid (see discussion in Section **S5.2** of the SI). Instead, we propose that the greater $P_{[010]}$ for **C3** relative to **C0** arises from preferential association of the crystallites along their hydrophobic (100) plane when they join to form bundles, as illustrated in **Figure 6e**. Such a raft-like structure is more likely to settle on the grid along the (010) plane and would therefore exhibit a single homogeneous [010] zone axis diffraction pattern, despite being composed of multiple distinct crystallites.

Previous studies have postulated that crystallites in bundle-like CNCs are preferentially associated along these hydrophobic faces.[7,10] Such preferential orientation is plausible, as it would lead to a minimization of the free energy through a lowering of the proportion of hydrophobic faces exposed with water. Indeed, simulations showed that the work of adhesion between two (100) surfaces in water is greater than between two (110) surfaces or between a (100) and a (110) surface.[36] Moreover, this association mechanism is consistent with the preservation of the number of charged groups per mass upon bundle formation from **C0** to **C3** (see **Table S1**), as an association along the hydrophobic (expected to be non-charge-bearing) faces would maintain the number of exposed charged groups. The impact of this preferential association on the tensioactive ability of the CNCs was investigated through pendant drop experiments of aqueous CNC suspensions in oil (see **Figure S15** and Section **S6** of the SI). Compared to **C0-Na**, **C3-Na** displayed a statistically significant but small decrease in interfacial tension (~ 2 $mN\ m^{-1}$). This difference likely arises from the balance of competing effects including amphiphilicity and cross-sectional morphology of the particles, but also surface coverage and interparticle interactions.

The SNBED dataset was further analyzed to assess the heterogeneity of the crystalline orientations within individual CNCs. For this, the data were classified based on the number of extracted crystal orientations within an individual particle: particles composed of pixels displaying two or more distinctive crystallographic orientations were classified as "heterogeneous" (illustrated by particle 1 in **Figure 6b**), while particles containing pixels all sharing the same crystallographic orientation were classified as "homogeneous" (illustrated by particle 2 in **Figure 6b**).

The relative occurrence of each particle type is shown in **Figure 6d**. For the never-centrifuged (**C0**) and the centrifuged (**C3**) samples, nearly three-quarters of the particles were homogeneous (73 and 72% respectively). This further suggests that bundle-like CNCs are made of preferentially oriented crystallites. In contrast, much fewer homogeneous particles were observed for **C0-Ca** and **C3-Ca** (35 and 44% respectively), which is consistent with these samples containing a large proportion of randomly associated crystallites.

The SNBED data for **C0-Ca** and **C3-Ca** were then compared to their parent samples **C0** and **C3**. Compared to **C3**, **C3-Ca** exhibited a similar $P_{[010]}$ with a higher proportion of heterogeneous crystal orientations (**Figure 6c**, **d**). These observations could be explained by the random association of new crystallites to pre-existing raft-like particles, leading to an increase of heterogeneous particles with a similar $P_{[010]}$ (**Figure 6e**). However, the $P_{[010]}$ value for **C0-Ca** was substantially higher than for **C0**, but comparable to **C3** and **C3-Ca** (**Figure 6c** and **Figure 6d**). This suggests that calcium-induced aggregation of never-centrifuged CNCs has the combined effect of reducing the proportion of hydrophobic planes (to the same extent as centrifugation-induced bundling in **C3**) while also forming heterogeneous particles (**Figure 6e**). This conclusion agrees with the similar cross-sectional density of **C0-Ca** and **C3-Ca** in the analysis of the SAXS data. This lateral association is also suggested by the size increase exhibited by CNC suspensions concentrated in the presence of NaCl (see **Figure S17** and discussion in Section **S8** of the SI), and has been hypothesized to explain the formation of a stiffer network upon gelation of CNCs in the presence of additional cations.[23]

## Liquid Crystalline Ordering of the CNC Suspensions and Films

To determine whether the bundles formed by centrifugation display any enhanced chiral strength,[8] the liquid crystalline properties of the different CNC suspensions were characterized. In suspension, CNCs spontaneously phase-separate into a denser anisotropic liquid crystalline phase and a lighter isotropic phase. With increasing CNC concentration, the proportion of anisotropic phase increases until the entire suspension becomes anisotropic. The volume fraction of the anisotropic phase ($\varphi_{ani}$) as a function of the CNC volume fraction ($\Phi$) for **C0-Na**, **C3-Na** and **C3-Ca** is presented in **Figure 7a**. Compared to the never-centrifuged CNCs (**C0-Na**), the biphasic regime of **C3-Na** was narrower and shifted to lower concentrations. According to the Onsager model for achiral hard rods, this is indicative of rods having a larger aspect ratio,[37] which is in accordance with our morphological analyses indicating that centrifuged CNCs have an aspect ratio greater than or equal to the never-centrifuged CNCs.

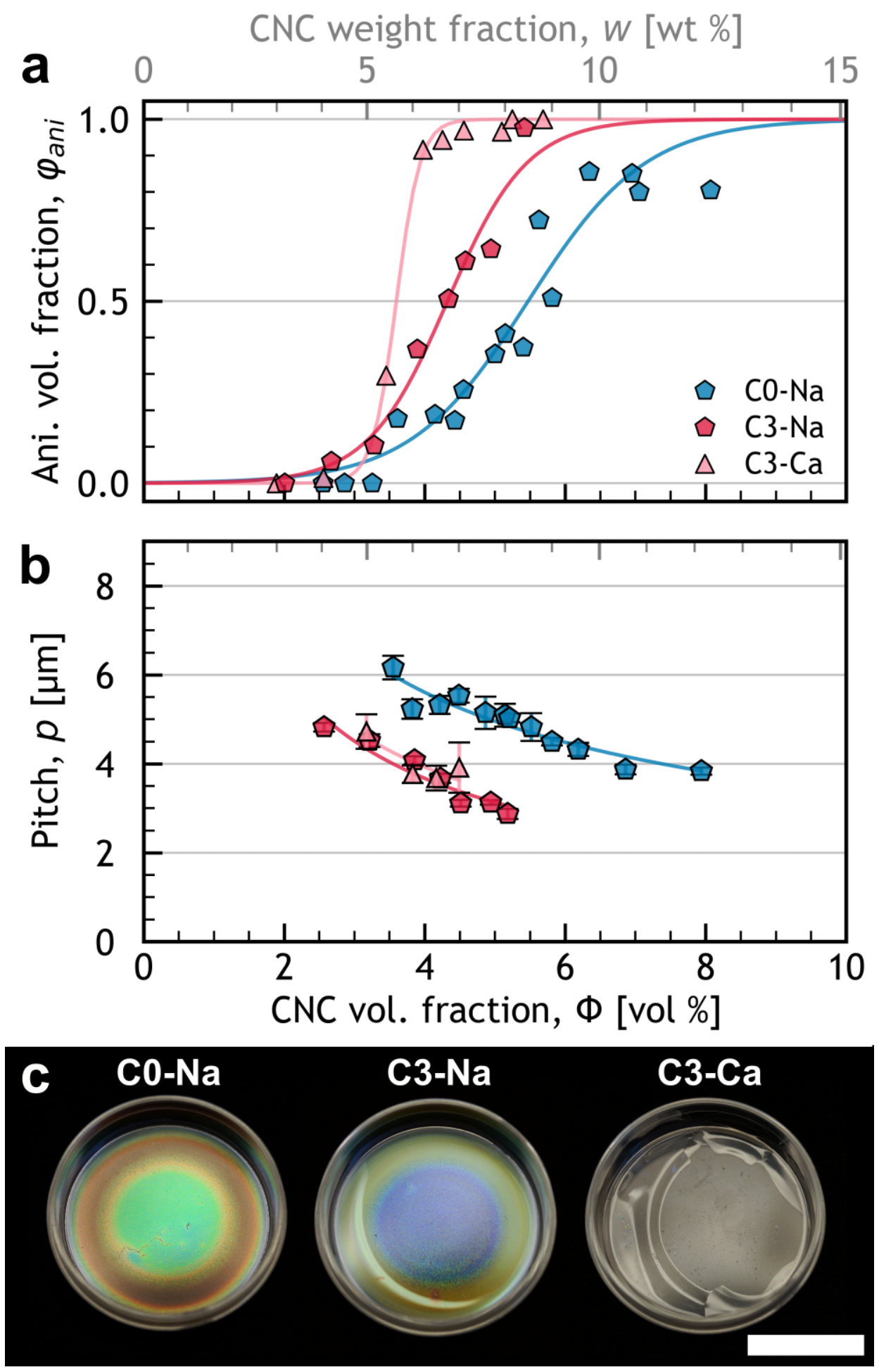

**Figure 7**. Liquid crystalline properties of different CNC suspensions: **(a)** volume fraction of the anisotropic phase ($\varphi_{ani}$) and **(b)** pitch ($p$) as a function of total CNC volume fraction ($\Phi$) and weight fraction ($w$) for **C0-Na**, **C3-Na** and **C3-Ca** (fitting lines are shown as a guide). **(c)** Photographs of corresponding dish-cast films (cast suspension: 2.7 mL with [CNC] = 1.5 wt% and [NaCl] = 1.5 mM) taken on a dark background. Scale bar 2.0 cm.

The anisotropic phase formed by the CNCs displays a left-handed 'cholesteric' structure, also commonly called 'chiral nematic', whereby the elongated constituents point locally parallel to one another in a direction that twists helicoidally into a left-handed periodic structure. The distance over which the local orientation of the CNCs completes a full rotation about the helical axis is called the pitch ($p$, in µm). The pitch, is related to the chiral interactions between the CNCs, is known to decrease as $\Phi$ increases.[38,39] For cotton CNCs, $p$ is found to be inversely proportional to $\Phi$, which implies that the strength of the chiral interactions between neighboring CNCs can be measured through a parameter defined as the chiral strength ($\kappa$, in µm$^{-1}$, and defined as positive for simplicity) expressed as:[40]

$$\frac{2\pi}{p} = \kappa\,\Phi \tag{3}$$

The evolution of the pitch as a function of the CNC volume fraction in **Figure 7b** reveals that **C3-Na** had a smaller pitch than **C0-Na**. The corresponding chiral strengths obtained by fitting **Equation** 3 are 26 ± 1 $\mu m^{-1}$ for **C0-Na** and 44 ± 2 $\mu m^{-1}$ for **C3-Na**, suggesting a stronger chiral interaction for the centrifuged CNCs. These observations are in agreement with the previously reported positive correlation between chiral strength and bundle proportion in the CNC population.[8] These results suggest that the centrifugation step leading to **C3**, and commonly applied in most laboratory-made CNCs, is responsible for the formation of additional bundles that increase the effective left-handed chiral strength of the suspension and decrease the cholesteric pitch.

The ability of CNCs to self-organize into cholesteric structures can be harnessed to make structurally colored films. Such a film is produced by dish-casting a CNC suspension, leading to the formation of cholesteric structure of decreasing pitch upon slow drying into a solid film. When the pitch in the resulting film is comparable to the wavelength ($\lambda$) of visible light, a selective reflection occurs for $\lambda = n\,p$, where $n$ is the average refractive index of cellulose. The reflected wavelength $\lambda$ is thus proportional to the pitch $p$ in the film, and from **Equation** 3, it is also proportional to $1/\kappa$.[8,41,42] Consequently, an increase of the chiral strength of the CNCs is expected to cause a blue-shift of the film color. To illustrate the practical significance of this behavior, the suspensions were used to produce such CNC films, as presented in **Figure 7c**. The samples **C0-Na** and **C3-Na** both formed structurally colored films, with the **C0-Na** film exhibiting a green center with a red edge while the **C3-Na** exhibited a blue center with a greenish edge. This shows that using centrifugation as a purification step causes a significant blue-shift of the subsequent CNC film and is of significant practical importance for applications that exploit the chiral self-assembly properties of CNCs.

The interpretation of the evolution of the measured anisotropic volume fraction and pitch for **C3-Ca** is less straightforward. This sample exhibited a steep increase of $\varphi_{ani}$ with increasing CNC volume fraction ($\Phi$) (**Figure 7a**), which is usually indicative of early gelation.[43] Between crossed polarizers, this sample displayed distinctive shear-alignment above 6 wt% (**Figure S16**), indicating that the sample was kinetically arrested and was not able to relax over time, leading to an apparent $\varphi_{ani}$ close to 100%. Such a gelation at a low CNC volume fraction can be caused by the greater hydrodynamic volume induced by the irregular composite particles. Despite these clear signs of gelation, **C3-Ca** exhibited some fingerprint patterns at intermediate $\Phi$ arising from isolated tactoids, suggesting that local cholesteric ordering occurred in the transient stage between sample preparation and gelation.[2] In these regions, the evolution of $p$ as a function of $\Phi$ followed the same trend as for **C3-Na**. This suggests that calcium-induced aggregation did not increase the chiral strength of the CNCs, but that a subpopulation of chiral bundles is still present. The coexistence of subpopulations of irregular composite particles and of chiral bundles is consistent with the broadening of the size distribution toward larger sizes while maintaining a significant proportion of small sizes (see **Figure 4c**, **d**). Nevertheless, the film made from **C3-Ca** appeared colorless and slightly hazy (**Figure 4e**), which we can ascribe to early gelation promoted by the irregular particles. The behavior of this sample illustrates the critical difference between the two types of composite particles.

More generally, these results highlight the practical consequences of CNC suspension history for their self-organization behavior. To illustrate this point, the self-assembly of CNC samples with different bundle content and ionic strength history was investigated. The results, presented in **Figure S17**, demonstrate that transient exposure to an increased ionic strength (NaCl) by concentration followed by dilution during capillary preparation leads to an increase of particle size and a convergence of their self-assembly behavior (see discussion in Section **S8** of the SI for more details).

## High-shear Viscosity of the CNC Suspensions

The characterization of the morphological properties and the self-organization behavior of the different samples suggests that the different types of composite particles exhibit different hydrodynamic behaviors, with potential importance for their use as rheological modifiers. Moreover, the rheological properties of diluted suspensions can provide indirect information about the particle morphology and are also exempt from drying artifact and statistical noise. For these reasons, the evolution of the relative viscosity ($\eta_r$) of the samples was investigated as a function of the particle concentration $c$ (g mL$^{-1}$), as reported in **Figure 8.** Both **C0-Na** and **C3-Na** displayed a similar evolution of the viscosity with the concentration, in accordance with previous studies.[44] However, despite exhibiting earlier gelation, the relative viscosity of **C3-Ca** was always lower than for **C0-Na** and **C3-Na**. This result indicates that **C3-Ca** exhibits a greater dependence of the viscosity with the CNC concentration and/or the shear rate than **C0-Na** and **C3-Na**.[45,46]

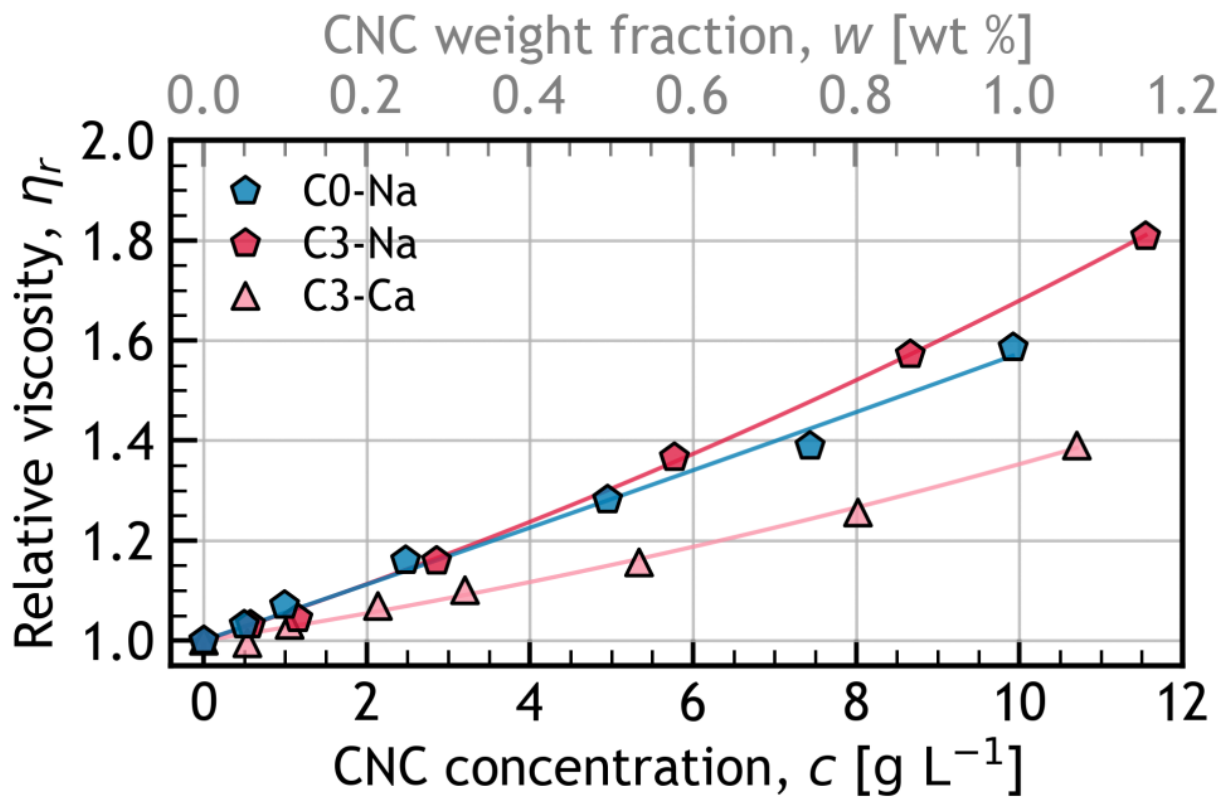

**Figure 8**. Plot of the relative viscosity ($\eta_r$) as a function of the CNC concentration ($c$) with best fit using **Equation** 4.

The viscometry measurements were used to extract the intrinsic viscosity $[\eta]$ (mL $g^{-1}$) and the dimensionless Huggins coefficient ($k_H$), using the Huggins equation (see **Table S15**):[47]

$$\eta_r \approx 1 + [\eta]c + k_H[\eta]^2c^2 \quad (4)$$

This analysis revealed that **C3-Na** and **C0-Na** exhibited similar intrinsic viscosities (54 ± 2 mL $g^{-1}$ and 56 ± 6 mL $g^{-1}$ respectively) which were greater than that of **C3-Ca** (25 ± 3 mL $g^{-1}$). This trend is agrees with a decrease in the aspect ratio of the particles.[48] Assuming the CNCs can be modeled as prolate ellipsoids, the intrinsic viscosity can be used to extract their shape factor *r*, defined as their 3D aspect ratio, according to:[49,50]

$$[\eta] = \frac{8}{15} \frac{(r^4 - 1)}{\rho_{cell}\, r^2 \left[\frac{(2r^2 - 1)\cosh^{-1}(r)}{r\sqrt{r^2 - 1}} - 1\right]} \quad (5)$$

where $\rho_{cell}$ is the CNC volumetric mass density (taken to be 1.6 g $mL^{-1}$)[51]. Both **C3-Na** and **C0-Na** had the same shape factor ($r$ = 35), which was greater than the shape factor exhibited by **C3-Ca** ($r$ = 22) (see **Table S15**). These results indicate a similarity in 3D aspect ratio between never-centrifuged and centrifuged CNCs, while highlighting a lower aspect ratio for the Ca-CNCs, consistent with the other morphological analyses. The values of shape factors obtained by viscometry are significantly larger than the aspect ratios obtained from the analysis of TEM images. This discrepancy can be explained by the presence of electroviscous effects that cause a significant increase in apparent aspect ratio when the measurements are performed without salt addition,[52,53] which was the case for this study.

The Huggins coefficients ($k_H$), which appear as a second order additive correction factors in **Equation** 4 and characterize particle interactions**,** were extracted and reported in **Table S15**. Both **C3-Na** and **C0-Na** exhibited similarly low $k_H$ values (0.5 ± 0.1 and 0.1 ± 0.2, respectively), which are both distinctly smaller than for **C3-Ca** (1.6 ± 0.8). All were in the range previously reported for CNCs,[54] and other comparable charged elongated nanoparticles.[48] The significantly greater $k_H$ displayed by **C3-Ca** indicates that Ca-CNCs displayed much stronger mutual interactions (either repulsive or attractive). This is consistent with the lower absolute value of the ζ-potential of Ca-CNCs (ζ ≈ -35 mV) compared to that of pH neutralized CNCs (ζ ≈ -43 mV) (see **Table S1**). Together, the greater $k_H$ and lower ζ-potential of Ca-CNCs are indicative of a lower colloidal stability of the Ca-CNCs, in agreement with the formation of bigger particles and the adsorption of tightly bound calcium cations on the particle surface.[55]

In conclusion, calcium-induced aggregation of CNCs can be of interest for rheological applications where a high viscosity or gel-like behavior is desired at rest. However, this increase in viscosity comes at the cost of reduced colloidal stability. Overall, the rheological properties of Ca-CNCs could be investigated further to assess their dynamic and concentration-dependent behavior.

## Fragmentation of CNC Aggregates

Mild ultrasonication (*e.g.* using a bath sonicator) is often used to redisperse loose CNC aggregates, with a minor impact on the morphology of the crystallites.[2,56] However, significantly increasing the ultrasonication dose eventually leads to near-complete fragmentation of the composite particles back into individual crystallites.[8,57] Consequently, ultrasonication is accompanied by a drastic reduction of the Z-average diameter of the CNCs.[8,24,58,59]

According to DLVO theory, irregular composite CNCs are less stable than laterally aligned CNCs.[28] Therefore, randomly associated composite particles should be easier to fragment than the laterally associated bundles. To explore this, the different CNC suspensions were exposed to increasing ultrasonication dose ($u$, J mL$^{-1}$) and the Z-average diameter ($D_H$) was monitored by DLS (**Figure 9**). For all samples, $D_H$ decreased sharply with increasing $u$, as previously observed for CNCs.[8,24,58,59] Eventually, all samples reached a similar plateau of $D_H^{\infty} = 59 \pm 2$ nm for $u > 445$ J mL$^{-1}$, despite having different initial sizes. At low dose ($u < 30$ J mL$^{-1}$) the samples showed two different behaviors with **C3** and **C0** displaying a moderate size change (~ 5 nm mL J$^{-1}$) while **C3-Ca** and **C0-Ca** exhibit a greater size drop (~ 18 nm mL J$^{-1}$).

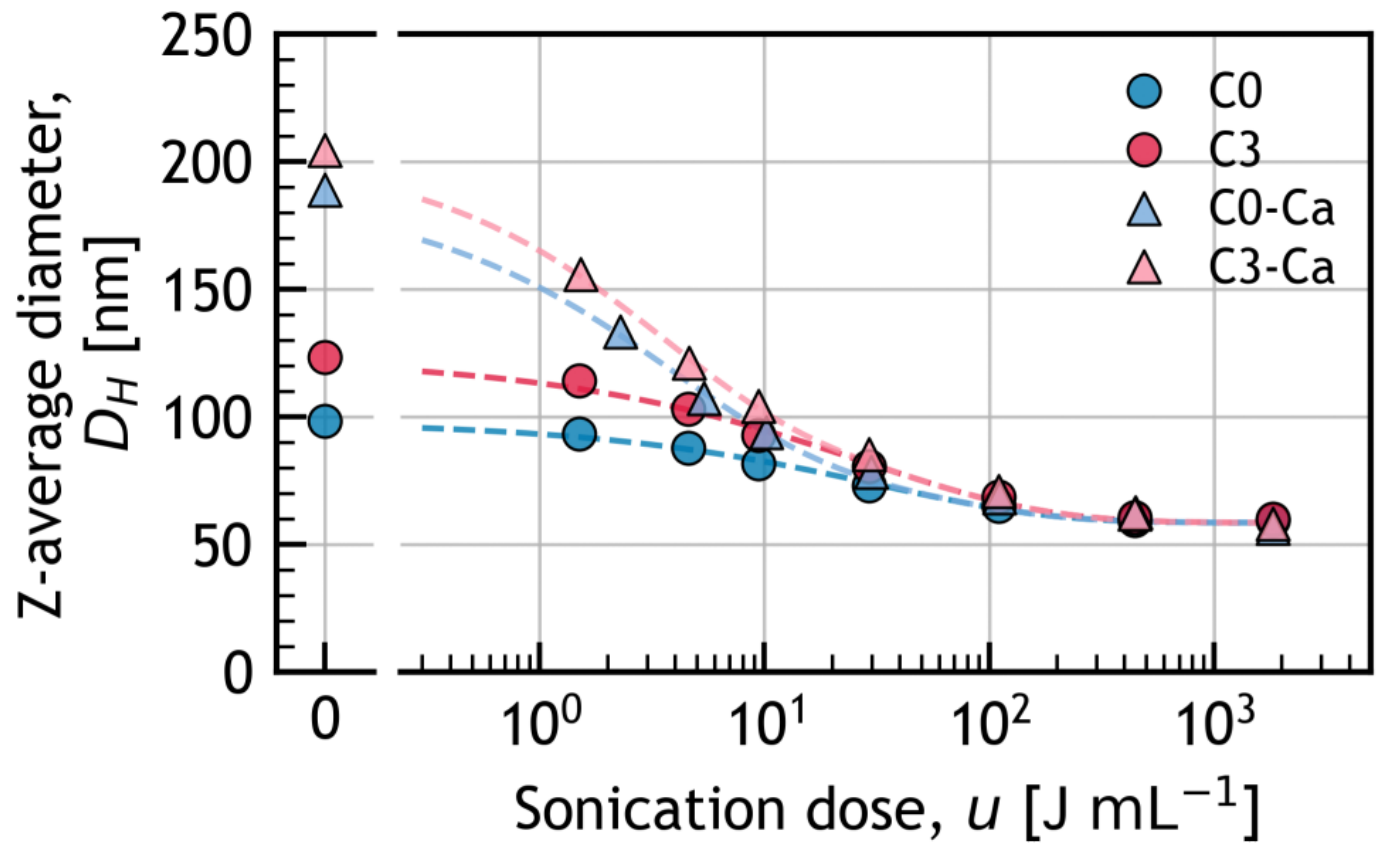

**Figure 9**. Impact of ultrasonication dose ($u$) on the Z-average diameter ($D_H$) fitted using **Equations** S24 and S25 (dashed line, see Section S10 of the SI for more details).

The evolution of the particle size as a function of particle type and sonication dose was also modeled by using a modified dissociation expression (**Figure 9** see Section S10 of the SI for more details). This analysis further shows that the ultrasonication dose has a greater impact on the size change of calcium induced composite particles than it has on bundles (see **Table S16**).

This result suggests that the calcium-induced irregular composite particles are less tightly bound than the bundle-like particles. This is in accordance with the higher energetic configuration and weaker energetic barrier predicted by DLVO theory for cross-associated CNCs compared to laterally associated CNCs.[28] This analysis shows that both irregular composite particles and bundle objects are broken down by prolonged ultrasonication and thus should be avoided when the presence of composite particles is desired.

## CONCLUSIONS

In this study, we explored the impact of centrifugation and transient exposure to high ionic strength on the properties of CNCs, through a thorough characterization of their individual morphologies, internal structures, and collective chiral properties in suspensions. The impact of these treatments was quantified after extensive dialysis against ultrapure water, and compared to the control suspension using DLS, zetametry, TEM, SAXS, SNBED, viscometry, and other experimental comparative techniques.

We showed that the first centrifugation step applied in most CNC extraction processes at the laboratory scale is responsible for the formation of additional CNC "bundles" with raft-like morphology. These raft-like particles stemmed from the preferential lateral association of crystallites through their hydrophobic faces, which occurred without loss of the number of exposed sulfate half-ester groups. Despite their lower proportion of hydrophobic surfaces, the bundles did not significantly modify the surfactant ability of the suspension. However, the bundles enhanced the chiral strength of the suspension, leading to a clear blue-shift of the corresponding, structurally colored, dish-cast film. This step could be further explored and adjusted via the centrifugation force, duration, and ionic environment to provide further control on the CNC chirality.

An alternative aggregation pathway was explored through the destabilization of purified CNC induced by $CaCl_2$ addition followed by dialysis against ultrapure water. The corresponding CNCs were in the form of calcium salts and exhibited a clear increase of the proportion of irregular composite particles as indicated by their increased average size, irregular morphology, lower 3D aspect ratio, and irregular internal structure. Consequently, these particles exhibited a smaller $\zeta$-potential and promoted gelation at lower volume fraction while exhibiting a lower viscosity at high shear. As demonstrated in this study, this aggregation pathway can be adjusted with salt concentration and type, with potential applications as rheological modifiers.

This work raises broader questions about both the origin of bundles and the chiral character of CNCs. Cellulose fibers often display a chiral arrangement in the plant cell wall,[60] and drying treatments of the cellulose source prior to hydrolysis can cause irreversible fiber aggregation (*i.e.* hornification).[61] As a result, it has often been suggested that chiral CNC bundles are inherited from these pre-existing structures in the plant tissue. By showing that exposure to high ionic strength during centrifugation favors the formation of bundles that enhance chirality, this work implies that the bundles initially present in never-centrifuged suspensions could also originate from the same mechanism during the hydrolysis and quenching steps, since these conditions can also trigger irreversible aggregation of the cellulose crystallites (when they assemble through their hydrophobic faces). Consequently, centrifugation at high ionic strength may not create a new type of chiral enhancing bundles but rather amplifies their formation. Quantitative investigations using alternative cellulosic sources (e.g. never-dried starting material), pre-treatment (*e.g.* hornification) or hydrolysis conditions (e.g. temperature, acid) could be of interest to assess the origin and chiral properties of the bundles present in never-centrifuged suspensions.

To conclude, our results highlight how overlooked variations in the CNC extraction protocol can be critical in terms of suspension behavior. Importantly, while these findings are based on the analysis of CNCs obtained from sulfuric acid hydrolysis of cotton, they are likely applicable to different sources and methods. This encourages further investigation of CNCs extracted from wood or obtained by other production methods, which also typically display a dominant proportion of bundled particles.[6,62–64] These results are especially important to consider when transferring findings from laboratory-made to industrial CNCs, as the latter are typically purified by ultrafiltration instead of centrifugation,[65] which is expected to impact their particle morphology and colloidal properties. As such, this work pinpoints previously unexplored possibilities with immediate practical considerations for commercial applications of CNCs, where their chiral liquid crystalline properties or their amphiphilicity are directly exploited, but also in any other situation where the CNC morphology and surface chemistry are key.

## METHODS

**Materials**: Sulfuric acid ($H_2SO_4$, ≥ 95%, analytical grade), sodium hydroxide (NaOH, 99%, pellets), sodium chloride (NaCl, ≥ 99.5%, laboratory grade), calcium chloride ($CaCl_2$, fused granular) and hydrogen peroxide ($H_2O_2$, > 30 w/v%, laboratory reagent grade) were provided by Fisher Scientific. Hexadecane (Sigma-Aldrich reagent plus 99%) was passed through basic silica before use. All water used in this work was type 1 ultrapure water (Milli-Q, Millipore, Synergy UV system).

**Data processing** was performed with custom-made Python scripts. Statistical analyses and data fitting were performed with the Scipy and LMFIT libraries, respectively.

**CNCs suspensions** (**C0**, **C1**, **C2**, **C3**) were produced by sulfuric acid hydrolysis of cotton derived filter paper (Whatman No. 1). Shredded filter paper (15 g, CookWork coffee grinder) was introduced in a 63.9± 0.1 wt% sulfuric acid solution (300 g, $\rho$ = 1.543 ± 0.001 g mL$^{-1}$, hydrometer ISO 650, 1.500 – 1.600, Scientific Laboratory Supplies) preheated to 60 °C. After 30 min of hydrolysis under vigorous mechanical stirring, the medium was quenched with ice-cold water (300 mL). Part of the mixture was set aside and the rest centrifuged (20,000 g, 20 min, 4 °C, Lynx 6000 Thermo Scientific, T29-8x50 rotor), with the resulting pellet redispersed in water. This process was repeated to produce aliquots of CNCs that had never been centrifuged (**C0**), centrifuged once (**C1**), twice (**C2**), and three times (**C3**). All samples were then dialyzed against deionized water (MWCO 12-14 kDa, Medicell membrane), with the water changed at least once a day until the conductivity was stable (around 2 weeks). The suspensions were passed through cellulose nitrate filters (8 then 0.8 μm, Sartorius)

**CNC mass fraction** was obtained from gravimetric analysis by drying the suspensions in an oven (65 °C, > 40 h). The mass of dry CNC was at least 15 mg, measurements were made in triplicate.

**CNC surface sulfate half-ester groups** were quantified by conductometric titration of CNC suspensions diluted in water (approx. 180 mL) with the addition of NaCl solution (0.1 M, 2 mL). An automatic titrator (Metrohm, 800 Dosino) was used to inject NaOH solution (10 mM Titripur®, 50 μL $min^{-1}$) while monitoring the conductivity (856 conductivity module). The number of CNC surface sulfate half-ester groups was deduced from the first equivalence point.

**Concentrated CNC suspensions** (**C0-Na**, **C3-Na**) were prepared by neutralizing the suspensions with 1 molar equiv. of NaOH per CNC surface sulfate half-ester group followed by concentrating with a rotavapor (35 °C, 20 mbar).

**Salt-induced irreversible aggregates** were prepared by diluting concentrated CNC suspension in water before pipetting $CaCl_2$ or NaCl aqueous solution to form a 2 – 6.5 wt% CNC suspension in an ionic strength of 0 to 80 mM. The mixture was centrifuged (10,000 g for 20 min, Minispin Eppendorf), redispersed, then diluted for Zetasizer Measurements measurement as described below. The ionic strength of the CNCs due to their surface ions was neglected.

**Dialyzed calcium aggregated CNC suspensions** (**C0-Ca**, **C3-Ca**) were obtained by preparing 6.5 wt% CNC suspensions of **C0-Na** or **C3-Na** containing $CaCl_2$ (13.2 ± 0.2 mM), followed by centrifugation (10,000 g for 20 min), redispersion (to approx. 1 wt% CNC), and dialysis against water (≥ 2 weeks). For self-assembly measurements, **C3-Ca** was first concentrated with a rotavapor (35 °C, 20 mbar), then further concentrated by evaporation under ambient conditions.

**Calcium content** was measured by inductively coupled plasma-optical emission spectrometry (ICP-OES, Thermo Fisher Scientific iCAP 7400 Duo ICP Spectrometer). Freeze-dried CNCs (~ 20 mg) were further dried overnight in an oven (60 °C). The precisely weighed CNCs were digested for 1 h in freshly prepared piranha solution (3:1 *v/v* $H_2SO_4$:$H_2O_2$). A known amount of the mixture was diluted in water to obtain a solution of (approx. 3.6 wt% acid, approx. 2 g $L^{-1}$ CNCs) that was used for the measurement. ICP Standard (Sigma-Aldrich) were diluted with approx. 2% nitric acid (TraceMetal™ Grade, Fisher) in water (TraceSelect™ for Trace Analysis, Honeywell Riedel-de Haen™) to make the standard curve. Analysis performed on Qtegra software.

**Zetasizer Measurements** were performed on dilute CNC suspensions (0.1 wt%), NaCl was used to set the ionic strength as close as possible to 1 mM. The Z-average diameters were estimated in backscattering geometry (173°, 633 nm, Malvern Zetasizer Nano ZS) from three runs of ten measurements after an initial waiting time of 5 min for the temperature to equilibrate (20 – 22 °C). The zeta potential was acquired after the Z-average measurement, through three runs of ten measurements and analyzed using the Smoluchowski equation. Data are presented as mean ± standard deviation.

**TEM** was performed with a Talos F200X G2 microscope (FEI, 200 kV, CCD camera). A drop of CNC suspension (0.002 wt%, in 1 mM NaCl) was deposited on a glow-discharged carbon-coated copper grid. After 2 min, the excess solution was blotted with filter paper. Then, a drop of uranyl acetate aqueous solution (2 wt%) was deposited and let to sit for 1.5 min before blotting again. Particles ($N \geq 225$) were manually outlined using Fiji (ImageJ), with all touching objects considered as a discrete CNC particle. Outlines were processed using the Shape Filter plugin to extract the bounding box length ($L_b$), width ($W_b$), and the particle area ($A$).[66] The bounding box aspect ratio ($r_b$) and rectangularity ($Rect$) were calculated from **Equation** 1 and **Equation** 2 respectively. Values were compared using a Mann–Whitney $U$ test and the magnitude of differences between samples was quantified using the point biserial correlation coefficient, for more details see Section **S4.2** of the SI.

**Cryo-TEM** was performed using a JEOL JEM 2100Plus (Jeol, Japan), operated at 200 kV, equipped with a Gatan RIO 16 camera (Gatan Inc., U.S.A.). Cryo-frozen samples were prepared using an EM GP2 Automatic Plunge Freezer (Leica microsystem, Germany) to vitrify the sample by rapid immersion in liquid ethane. The images presented were contrast enhanced to highlight the CNCs.

**SAXS** measurements were performed at the ID02 beamline of European Synchrotron Radiation Facility (ESRF) using X-rays of 12.23 keV. Two-dimensional X-ray scattering patterns were recorded on an Eiger2 4M pixel detector (Dectris) at a detector distance of *ca.* 10 m. The CNC suspensions were sealed in glass capillaries of an outer diameter of 1 mm and wall thickness of 0.13 mm. The obtained 2D data were then azimuthally averaged using SAXSutilities.[67]

**SNBED** data were acquired in a low-dose condition optimized for cellulose crystals as described previously, using a JEOL 2100F operating at 200 kV equipped with a NanoMEGAS ASTAR system. The nanobeam configuration consisted of a converged electron probe of 25 nm. An ED pattern was recorded at every probe position using a Cheetah Medipix3 direct electron detector (manufactured by Amsterdam Scientific Instruments) with a 0.5 ms exposure time per probe position. The diffraction datasets were analyzed using a dedicated ASTAR® software to perform: (i) crystal orientation identification through correlation with templates (*i.e.* pre-computed theoretical patterns) and (ii) Virtual–Bright (VBF) and Virtual–Dark field (VDF) image reconstruction that consists of plotting the intensity fluctuations of the transmitted beam (VBF) and user-selected diffraction positions (VDF) over the scanned area.

**Selected area ED** experiments were carried out with a JEM-2100Plus TEM operated at 200 kV using a selected area aperture of a diameter of 1 μm. The CNC suspensions were deposited on a glow-discharged carbon-coated copper grid. The excess liquid was removed by filter-paper blotting, then the grids were dried in air. Two-dimensional ED patterns were recorded from areas of CNCs with local orientation along the fiber axis on a MerlinEM hybrid pixel detector (Quantum Detectors) with an exposure time of 10 fps in a continuous acquisition mode. Equatorial ED profiles were obtained using the first 10 consecutive frames of the recorded datasets using an in-house program.

**Interfacial tension** between CNC suspensions (0.9 wt%) in 20 mM NaCl and hexadecane was measured through the pendant drop method with a FTA1000 Analyzer System (with a 22 gauge needle). For each sample, four photographs per drop (approx. 35 µL) were taken every minute starting 1 min after its formation, for at least six drops ($N$ = 6x4 pictures). The surface tension ($\gamma$) was calculated for each picture through the method of a selected plane. The in-house software developed for this purpose is available on Github: DropPyTension (see Section **S6** of the SI for more details).[68,69] The corresponding Bond numbers were in [0.16, 0.22] and the apparent Worthington numbers in [0.66, 0.93], indicating that the measurement conditions were satisfactory to extract $\gamma$ with accuracy.[70]

**Liquid crystalline properties** were investigated by observation of CNC suspensions in glass capillaries. The flat capillaries (CM Scientific, ID = 0.3 x 6.0 $mm^2$) were filled with a series of suspension dilutions prior to sealing with nail varnish and marking the initial meniscus position (so that any evaporation could be accounted for). The anisotropic volume fraction and corrected concentration were measured from the analysis of images taken after two weeks and again after at least a further week to confirm that no further evolution occurred. The pitch was measured after two weeks using polarized optical microscopy. Images were recorded in brightfield transmission configuration using a Zeiss Axio microscope in Koehler illumination equipped with a 50x objective (Nikon T Plan SLWD, NA 0.4) and a CMOS camera (UI- 3580LE-C-HQ, IDS). At least four images were recorded at different locations in the anisotropic phase, with three pitch measurements performed per image ($N \geq 12$).

**Structurally colored films** were made by slowly drying 2.7 mL of 1.5 wt% CNC suspension with 1.5 mM of NaCl in polystyrene petri dishes (35 mm diameter). The pictures were taken against a black background using a camera oriented at approximately 30° from the normal of the film that was under diffuse illumination.

**Relative viscosity** was calculated from flow time measurements at low particle concentration by using:

$$\eta_r = \frac{t\rho}{t_0\rho_0} \approx \frac{t}{t_0} \qquad (6)$$

where $t$ and $t_0$ are the flow times of the suspension and of the solvent respectively and $\rho$ and $\rho_0$ are the densities of the suspension and the solvent respectively. Flow times were measured in triplicate with an Ubbelohde viscometer (Technico, 0.05 cSt $s^{-1}$) in air at 20 °C. The flow time of water was estimated to be 18.65 ± 0.09 s (measured five times in triplicate, $N$ = 15). The intrinsic viscosity, and the Huggins coefficient, were extracted by fitting the relative viscosity as a function of the CNC concentration according to **Equation** 4 and the 3D aspect ratio of the samples was estimated from the intrinsic viscosity by using **Equation** 5. The choice of approach and equation derivation are described in more detail in Section **S9.1** of the SI.

**Ultrasonication** of dilute CNC solutions (30 – 40 mL, 0.1 wt% CNCs, 1 mM NaCl) was performed in an ice bath using a Fisherbrand Ultrasonic disintegrator (20 kHz, Ø = 12.7 mm, pulses 2:1 sec On:Off, 40% amplitude). Samples were ultrasonicated for regular intervals of increasing time in between which aliquots were removed for analysis (1 mL). The dose received by each sample (J $mL^{-1}$) was calculated for each step from ultrasonication time divided by the volume of the sample and multiplied by the true power delivered by the probe to the sample (20 W, determined by calorimetry by Parton *et al.*[8,71]).

ASSOCIATED CONTENT

**Supporting Information**

The supporting information is available free of charge at the end of this document.

DLS derived count rate, Additional aggregation experiments, TEM distributions and statistics, Cryo-TEM images, SAXS analysis details, SNBED illustration and discussion, SAED, Interfacial tension experiments, Capillary images, Additional self-assembly experiments, Viscometry analysis, Detail on the fit of Z-average diameter as a function of sonication

**Data Availability Statement**

The data that support the findings of this study are openly available in the University of Cambridge data repository at https://doi.org/10.17863/CAM.110504.

AUTHOR INFORMATION

**Corresponding Authors**

**Yu Ogawa** − Email: yu.ogawa@cermav.cnrs.fr
**Silvia Vignolini** − Email: sv@mpikg.mpg.de

**Conflict of interest**

The authors declare no financial conflict of interest.

**Author Contributions**

The presented work was conceived by T.P., B.F.P., R.P. and K.B. and supervised by S.V., Y.O., B.F.P., R.P. and A.L., the investigations were performed by K.B. and J.H.L. and the original draft was written by K.B. before being reviewed and edited by all the authors. All the authors have given approval to the final version of the manuscript.

## Funding Sources

This work was funded by: EPSRC CDT, Automated Chemical Synthesis Enabled by Digital Molecular Technologies EP/S024220/1 (K.B., A.L.); EPSRC Bio-derived and Bio-inspired Advanced Materials for Sustainable Industries EP/W031019/1 (K.B., R.P., B.F.P., S.V.); EPSRC EP/T517847/1 (T.P.); ERC Horizon 2023 Marie Skłodowska-Curie grant 101154876 CINCOS (T.P.); ERC Horizon 2022 Proof of Concept Grants (ID: 101082172) (R.P., S.V.); Hiroshima University WPI-SKCM$^2$ (B.F.P.); EP/P030467/1 (for multi-user TEM equipment call). This project was cofounded by the European Regional Development Fund via the project "Innovation Centre in Digital Molecular Technologies" (A.L.). J.H.L. and Y.O. acknowledge Agence Nationale de la Recherche (ANR grant number: ANR-21-CE29-0016-1) and Glyco@Alps (ANR-15-IDEX-02) for their financial support and the NanoBio-ICMG platform (FR 2607) for granting access to the electron microscopy facility.

## ACKNOWLEDGMENT

The authors would like to thank Y. Nishiyama (University of Grenoble Alpes) for his critical feedback on the SAXS analysis, H. Greer (University of Cambridge) for her assistance in acquiring the TEM images, and N. Howard (University of Cambridge) for performing the ICP-OES measurements. An AI-based tool was used to edit parts of this manuscript.

# SUPPORTING INFORMATION

# Tailoring the Morphology of Cellulose Nanocrystals via Controlled Aggregation

*Kévin Ballu[a], Jia-Hui Lim[b], Thomas G. Parton[c], Richard M. Parker[a], Bruno Frka-Petesic[a,d], Alexei A. Lapkin[e,f], Yu Ogawa[b]*, Silvia Vignolini[a,c]**

[a]Yusuf Hamied Department of Chemistry, University of Cambridge, Cambridge CB2 1EW, United Kingdom

[b]University of Grenoble Alpes, CNRS, CERMAV, 38000 Grenoble, France

[c]Department of Sustainable and Bio-inspired Materials, Max Planck Institute of Colloids and Interfaces, 14476 Potsdam, Germany

[d]International Institute for Sustainability with Knotted Chiral Meta Matter (WPI-SKCM$^2$), Hiroshima University, Hiroshima 739-8526, Japan

[e]Department of Chemical Engineering and Biotechnology, University of Cambridge, Cambridge CB3 0AS, United Kingdom

[f]Innovative Center in Digital Molecular Technologies, Yusuf Hamied Department of Chemistry, University of Cambridge, Cambridge CB2 1EW, United Kingdom

## S1. CNC Hydrodynamic and Surface Properties

**Table S1**. Summary of the CNC characteristics: Z-average diameter (*$D_H$*), ζ-potential, surface sulfate half-ester groups ($-OSO_3^-$) per CNC mass from titration measurements, corresponding surface charge per surface area, and calcium content from elemental analysis. The surface charge per surface area was calculated by considering spherical CNCs with a diameter equal to their measured Z-average diameter.

| Sample | $D_H$ [nm] | ζ-potential [mV] | Sulfate half-ester [mmol kg$^{-1}$] | Surface charge [e nm$^{-2}$] | Ca content [mmol kg$^{-1}$] |
|---|---|---|---|---|---|
| **C0** | 99 ± 1 | -48 ± 5 | 264 ± 5 | 4.2 ± 0.1 | N/A |
| **C1** | 129 ± 1 | -50 ± 8 | 269 ± 5 | 5.6 ± 0.1 | N/A |
| **C2** | 127 ± 1 | -49 ± 5 | 267 ± 5 | 5.5 ± 0.1 | N/A |
| **C3** | 124 ± 2 | -50 ± 8 | 264 ± 5 | 5.3 ± 0.1 | N/A |
| **C0-Na** | 94 ± 1 | -41 ± 4 | N/A | N/A | N/A |
| **C3-Na** | 114 ± 1 | -44 ± 1 | N/A | N/A | 6 |
| **C0-Ca** | 189 ± 1 | -35 ± 1 | N/A | N/A | N/A |
| **C3-Ca** | 216 ± 9 | -35 ± 2 | N/A | N/A | 133 |

## S1.1 CNC Dynamic Light Scattering (DLS) Derived Count Rate

For a diluted suspension of scattering particles (in the single-scattering regime), the scattered light intensity measured in a fixed solid angle segment $\mathrm{d}\Omega = \sin(\theta)\,\mathrm{d}\theta\,\mathrm{d}\varphi$ is proportional to the product of the differential scattering cross-section of the particles ($\mathrm{d}\sigma_s/\mathrm{d}\Omega$) and their number density ($n_s$):

$$I(\theta,\varphi) \propto \frac{\mathrm{d}\sigma_s}{\mathrm{d}\Omega}\, n_s \tag{S1}$$

For comparing CNC samples at equal mass concentration $\gamma_s$ (g L$^{-1}$) the scattering intensity is proportional to the differential cross-section per scatterer mass:

$$I(\theta,\varphi) \propto \frac{\mathrm{d}\sigma_s}{\mathrm{d}\Omega}\,\frac{\gamma_s}{m_s} \tag{S2}$$

When performing a dynamic light scattering (DLS) measurement with a Malvern Zetasizer, it is possible to access the derived count rate (DCR) in kilocounts per second (kcps) which is directly proportional to the scattering intensity (multiplied by a device-dependent constant $c$),[1] and thus to the differential scattering cross section:

$$DCR = c\,I \propto \frac{\mathrm{d}\sigma_s}{\mathrm{d}\Omega}\, n_s \tag{S3}$$

The scattering cross-section of a particle depends on its size, shape and optical contrast with the surrounding medium. Consequently, comparison of the DCR with the Z-average hydrodynamic diameter of different samples, measured at similar concentration, can give an indication of the relative compactness of the particles.

**Figure S1** shows values for the DCR of the samples versus an effective "hydrodynamic volume" given by the Z-average diameter cubed ($D_H^3$). All samples were measured at similar CNC mass concentration (0.1 wt%) and should therefore have identical volume fraction but varying number density. All the non-Ca samples could be placed on a single line DCR $= m\,D_H^3$, whereas for the Ca-CNCs samples the increase of DCR per $D_H^3$ is comparatively smaller. This result suggests that the effective density of the **C0**, **C1**, **C2** and **C3** samples (*i.e.* mass enclosed within its effective hydrodynamic volume) is similar, while the Ca-CNCs have a significantly lower effective density. This finding is consistent with the hypothesis that centrifugation-induced aggregation leads to the formation of compact aligned particles while the calcium-induced aggregation yields randomly associated particles.

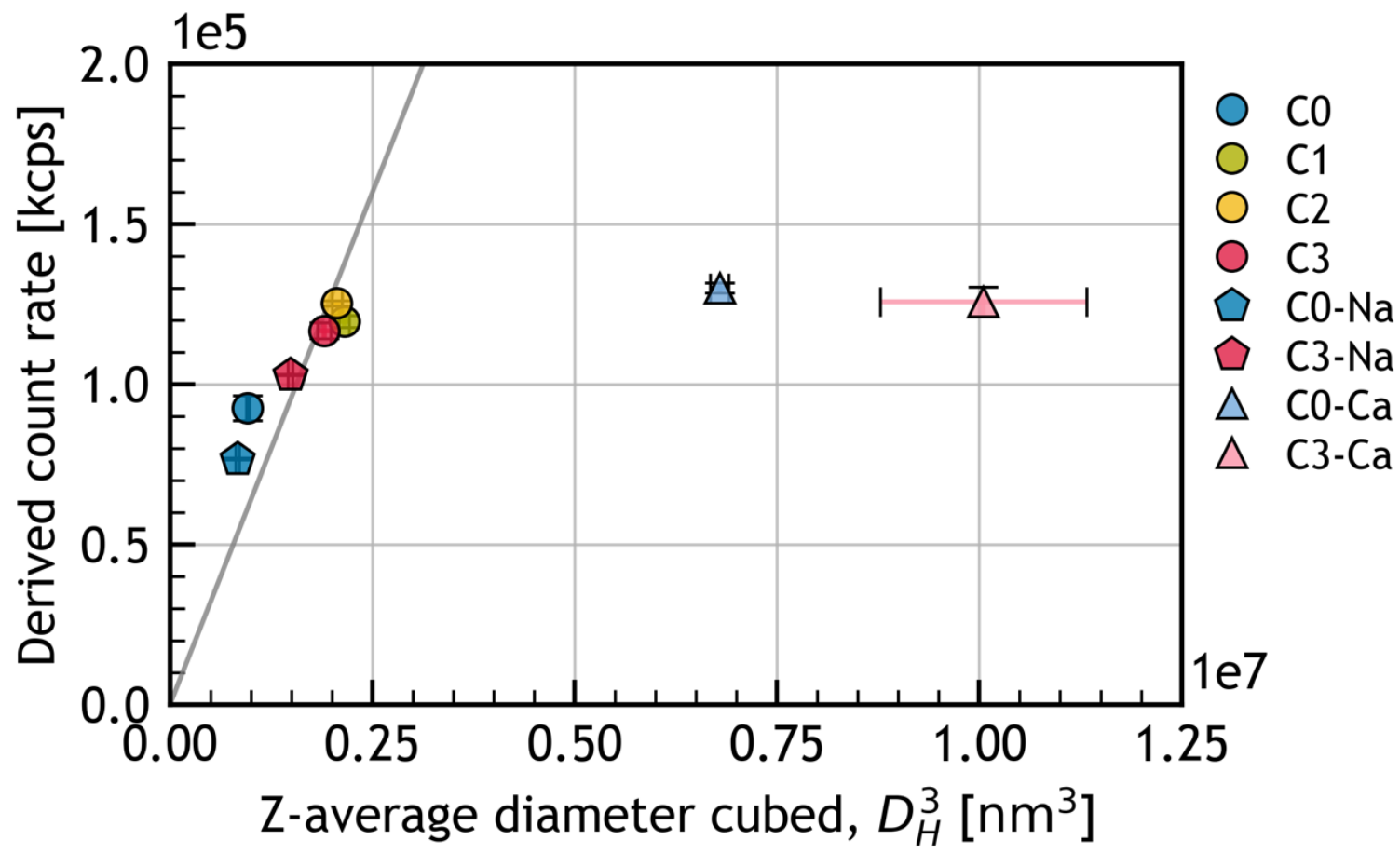

**Figure S1.** Evolution of the dynamic light scattering (DLS) derived count rate with the Z-average diameter cubed ($D_H^3$) for different CNC samples measured at similar concentration (0.1 wt% CNCs). The linear fitting line is presented as a guide to the eye.

## S2. Salt-induced CNC aggregation

### S2.1 Influence of process parameters on salt-induced CNC aggregation

The impact of process parameters during the salt-induced aggregation of CNCs was investigated by DLS measurements. A CNC suspension of **C3-Na** was prepared with added $CaCl_2$ to reach a transiently raised ionic strength $I_{agg}$ = 51 mM and a CNC weight fraction of $w_{agg}$ = 6.5 wt%. After a given time (0 h to 5 days) or centrifugation treatment (one or two 20 min cycles at 10,000 g), the suspension was redispersed and diluted in deionized water to a stable CNC suspension with $w_{DLS}$ = 0.1 wt% and at an ionic strength $I_{DLS}$ = 1 mM. The Z-average diameter ($D_H$) of the resulting particles measured by DLS are presented in **Table S2**. In these conditions, the waiting time before redispersion and dilution seemed to have a negligible impact beyond 1 h. Similarly, applying any number of centrifugation cycles led to similar particle sizes, indicating that this does not influence the size of the particles after dilution. Consequently, the aggregation step of the salt-induced aggregation experiments was carried out using one centrifugation cycle.

**Table S2**. Impact of the aggregation time ($t_{agg}$) and number of centrifugation cycles ($N_{centri}$) on the Z-average diameter ($D_H$) of diluted CNCs after salt-induced aggregation of **C3-Na** at $w_{agg}$ = 6.5 wt% CNC and $I_{agg}$ = 51 mM of ionic strength induced by $CaCl_2$ addition.

| $t_{agg}$ | $N_{centri}$ | $D_H$ [nm] |
|---|---|---|
| 0 h | | 209 ± 1 |
| 1 h | | 293 ± 2 |
| 3 h | 0 | 299 ± 4 |
| 24 h | | 270 ± 2 |
| 196 h | | 275 ± 3 |
| N/A | 1 | 285 ± 5 |
| | 2 | 274 ± 2 |
| 50 days | 1 | 274 ± 6 |

The impact of the CNC weight fraction during aggregation ($w_{agg}$) at fixed transiently raised ionic strength ($I_{agg}$) was also investigated. **C3-Na** suspensions over a range of CNC weight fraction (2.0 ≤ $w_{agg}$ ≤ 6.5 wt%) were centrifuged at $I_{agg}$ = 36 mM, followed by redispersion and dilution ($w_{DLS}$ = 0.1 wt%, $I_{DLS}$ = 1 mM) before measuring their $D_H$. As shown in **Figure S2**, lowering the CNC weight fraction led to an increase of $D_H$ after aggregation from 243 to 311 nm. This effect could be due to the decrease of viscosity, facilitating the formation of contact points between the particles. Alternatively, it is possible that at higher weight fraction, CNC are better packed, leading to denser aggregates. Therefore, the next aggregation experiments, presented in **Figure 3b**, were all carried out with a fixed CNC weight fraction during aggregation of $w_{agg}$ = 6.5 wt%.

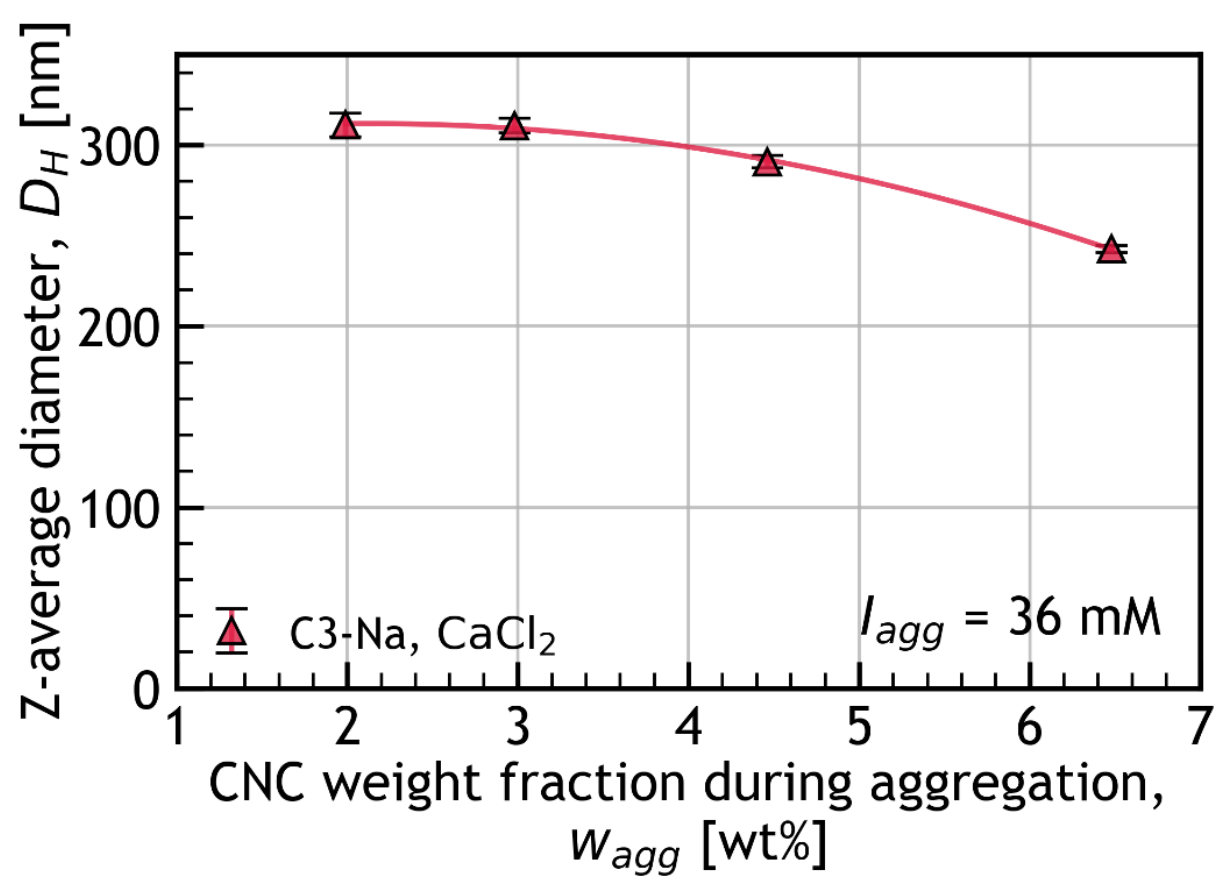

**Figure S2**. Evolution of the Z-average diameter ($D_H$) as a function of the CNC (**C3-Na**) mass fraction ($w_{agg}$) during calcium aggregation at fixed ionic strength ($I_{agg}$ = 36 mM). The line of best fit, presented as a visual guide, was obtained from a second order polynomial.

## S2.2 Influence of salt type and ionic strength on salt-induced CNC aggregation

The aggregation protocol presented in **Figure 3a** was followed to investigate the impact of the salt-type and ionic strength on the properties of the CNC suspensions after centrifugation and on their $D_H$ after dilution. We compared the impact of NaCl and $CaCl_2$ at various transiently raised ionic strengths ($0 \leq I_{agg} \leq 80$ mM) for both **C0-Na** and **C3-Na** ($w_{agg} = 6.5$ wt%) on the properties of the suspension in two steps. First, after the aggregation step (i.e. after salt addition and centrifugation), leading to biphasic systems made up of a bottom birefringent arrested phase, and an upper clear liquid layer. Then, after redispersion and dilution to a stable CNC suspension with $w_{DLS} = 0.1$ wt% and at an ionic strength ($I_{DLS}$) as close as possible to 1 mM.

After the aggregation step, the bottom phase of the CNC suspension was qualitatively inspected. For both salts, increasing $I_{agg}$ from 0 to 80 mM led to gelation of the bottom layer, that required an increasing effort to be redispersed. The volume of the gel layer decreased with increasing $I_{agg}$, indicating an increase of the density of the arrested phase. Overall, these qualitative inspections indicate that for all salt types and CNC samples, increasing the ionic strength led to the formation of a stronger gel.

In the second step, the obtained CNC suspensions were redispersed, diluted and their $D_H$ was measured. The time between sample centrifugation and their dilution for measurement varied from 0 to 63 days without significant impact on the $D_H$ trend, as presented in **Figure S3**. The evolution of $D_H$ was highly dependent on the salt type and transiently raised ionic strength, as illustrated in **Figure 3b**.

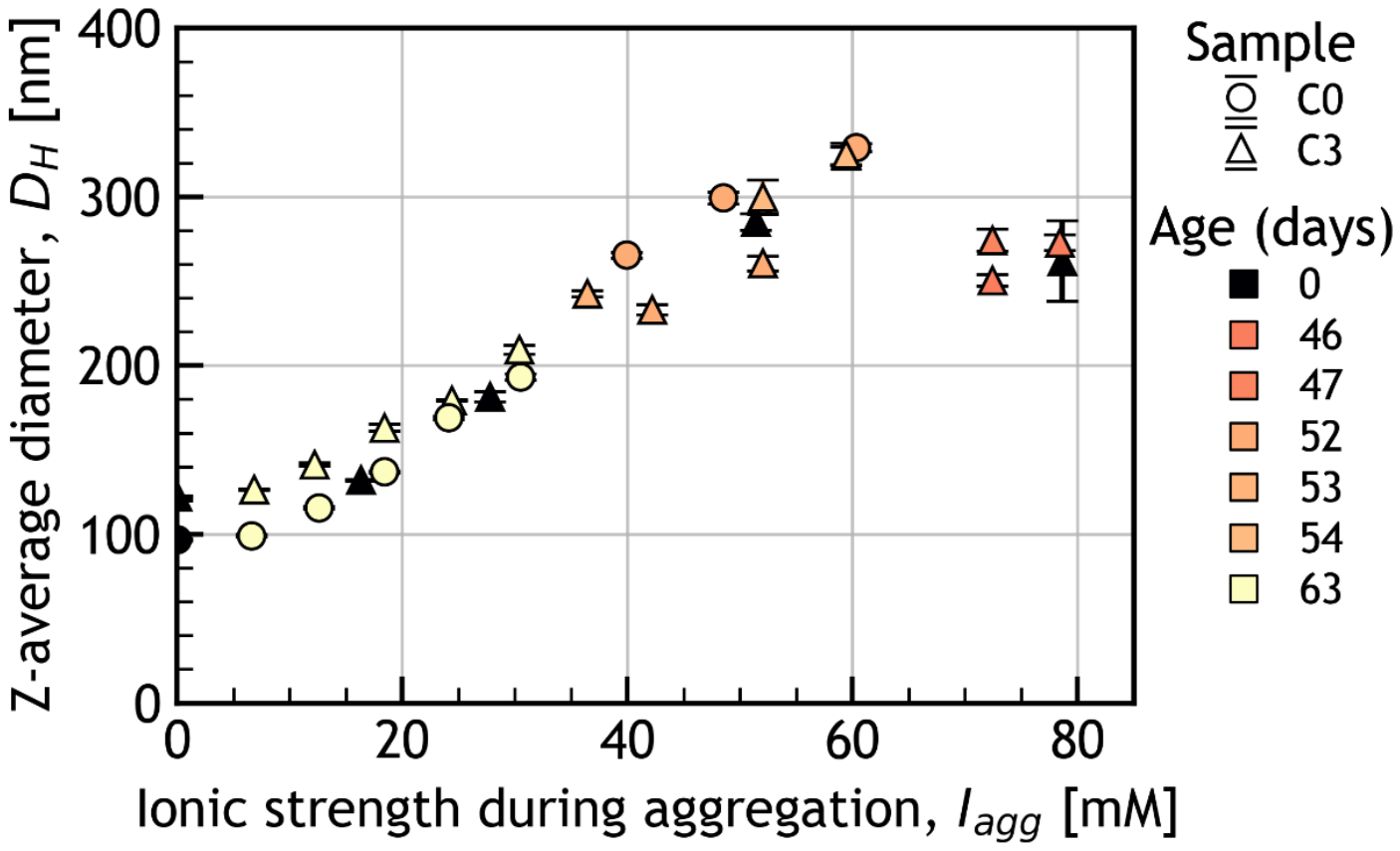

**Figure S3**. Evolution of the CNC Z-average size ($D_H$) after dilution following calcium-induced aggregation as a function of ionic strength ($I_{agg}$), CNC sample, and waiting time before dilution.

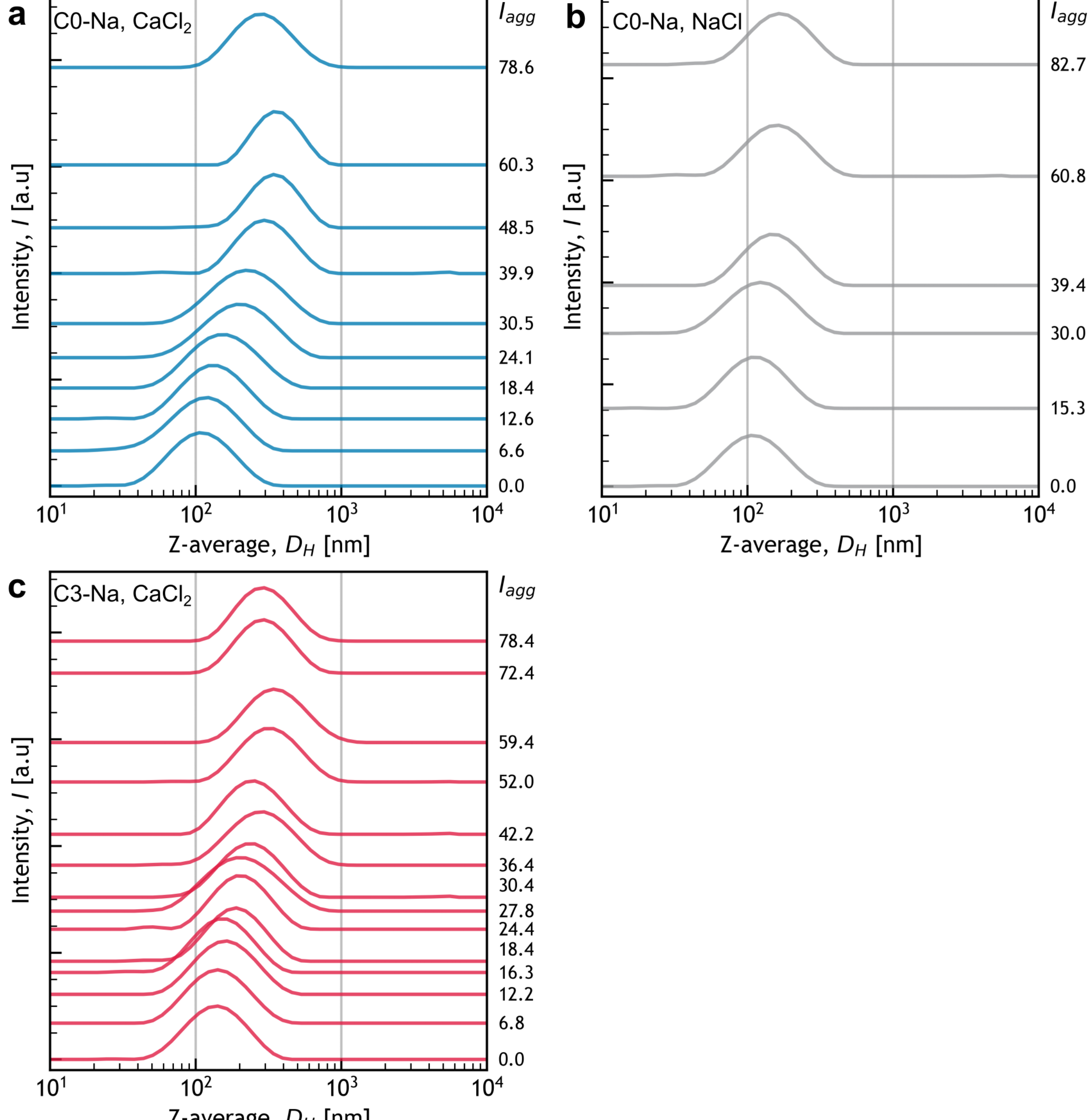

**Figure S4**. Evolution of the DLS scattered intensity ($I_{abs}$) as a function of the Z-average diameter ($D_H$) after dilution following salt-induced aggregation as a function of the transiently raised ionic strength ($I_{agg}$) for **(a) C0-Na** with NaCl, **(b) C0-Na** with $CaCl_2$ and **(c) C3-Na** with $CaCl_2$. Measurements were performed after redispersion and dilution to $w_{DLS}$ = 0.1 wt% and an ionic strength ($I_{DLS}$) as close as possible to 1 mM.

## S3. TEM and Cryo-TEM analyses

### S3.1 Histograms of TEM Values

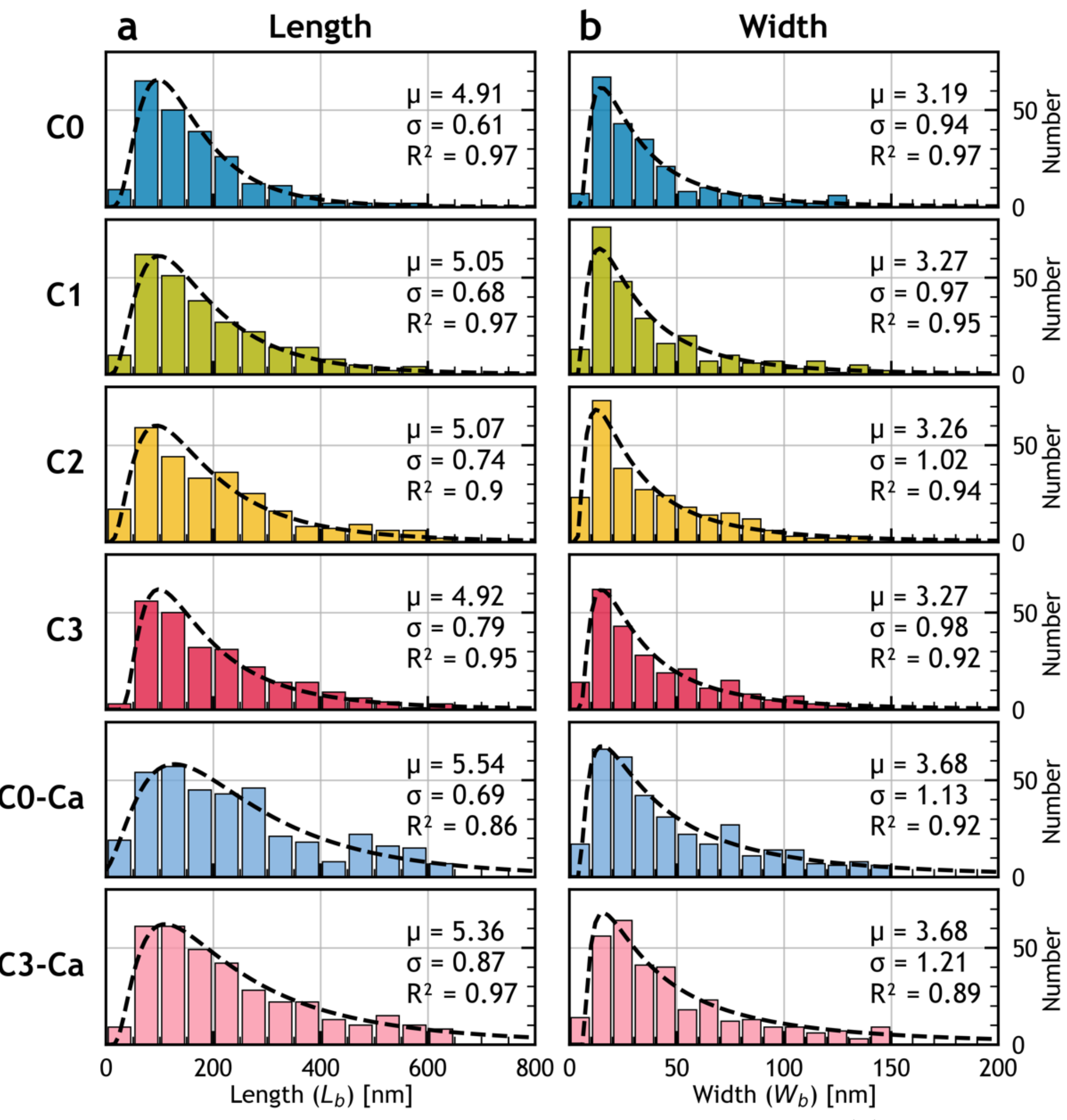

**Figure S5**. Histograms of CNC dimensions from analysis of TEM images (**a**) bounding box length ($L_b$) and (**b**) bounding box width ($W_b$). The data were fitted with a log-normal distribution and the corresponding fitting parameters are presented for each plot.

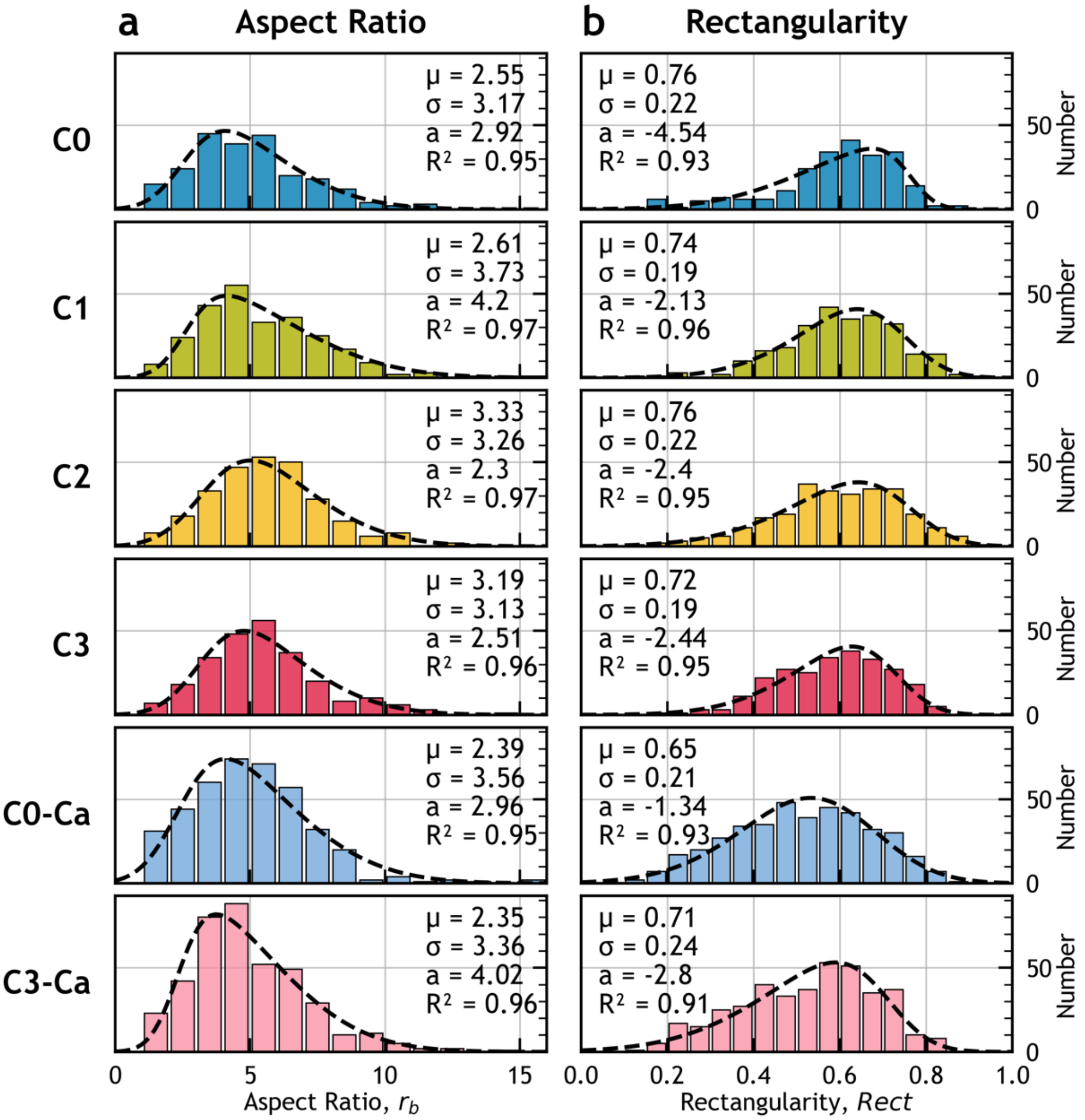

**Figure S6**. Histograms of CNC shape parameters from analysis of TEM images (**a**) bounding box aspect ratio ($r_b$) and (**b**) rectangularity (*Rect*). The data were fitted by skewed-normal distributions, the corresponding fitting parameters are presented for each plot.

## S3.2 Distribution of the TEM Values

The values obtained by TEM for the bounding box lengths ($L_b$) and bounding box widths ($W_b$) seem to follow a log-normal distribution (see **Figure S5**). If that is the case, then ln ($L_b$) and ln ($W_b$) should follow a normal distribution. The normality of a distribution can be investigated graphically through a normal quantile-quantile (Q-Q) plot, showing the quantiles of the investigated distribution against the quantiles of a normal distribution. If the points follow a perfectly linear evolution, then the investigated distribution follow a normal distribution. **Figure S7** shows the normal Q-Q plot for ln ($L_b$) and ln ($W_b$), for all the samples most of the data followed a linear evolution with small deviations at low and high quantiles indicating that the underlying distributions are close to normal.

Graphical investigations of normality are often used in conjunction with analytical tests such as the Shapiro–Wilk test. This test compares the distribution of a dataset to a normal distribution through calculation of the statistic W, $\in [0;1]$, for which a value of one indicates normality of the data, and a corresponding *p*-value below the chosen confidence level indicates that there is evidence that the dataset is not normally distributed.[2] As presented in **Table S3**, for ln ($L_b$) and ln ($W_b$) most samples display some deviation from normality when using a confidence threshold of 0.05.

The same investigations were applied to the corresponding bounding box aspect ratios ($r_b$) and rectangularities (*Rect*) that seemed to follow a skewed normal distribution (see **Figure S6**). The normal Q-Q plots for $r_b$ and *Rect* presented in **Figure S8** exhibit some constant deviation from normality and all the corresponding Shapiro–Wilk test indicate non-normality (**Table S3**). More precisely $r_b$ and *Rect* values displayed a right and left skew respectively.

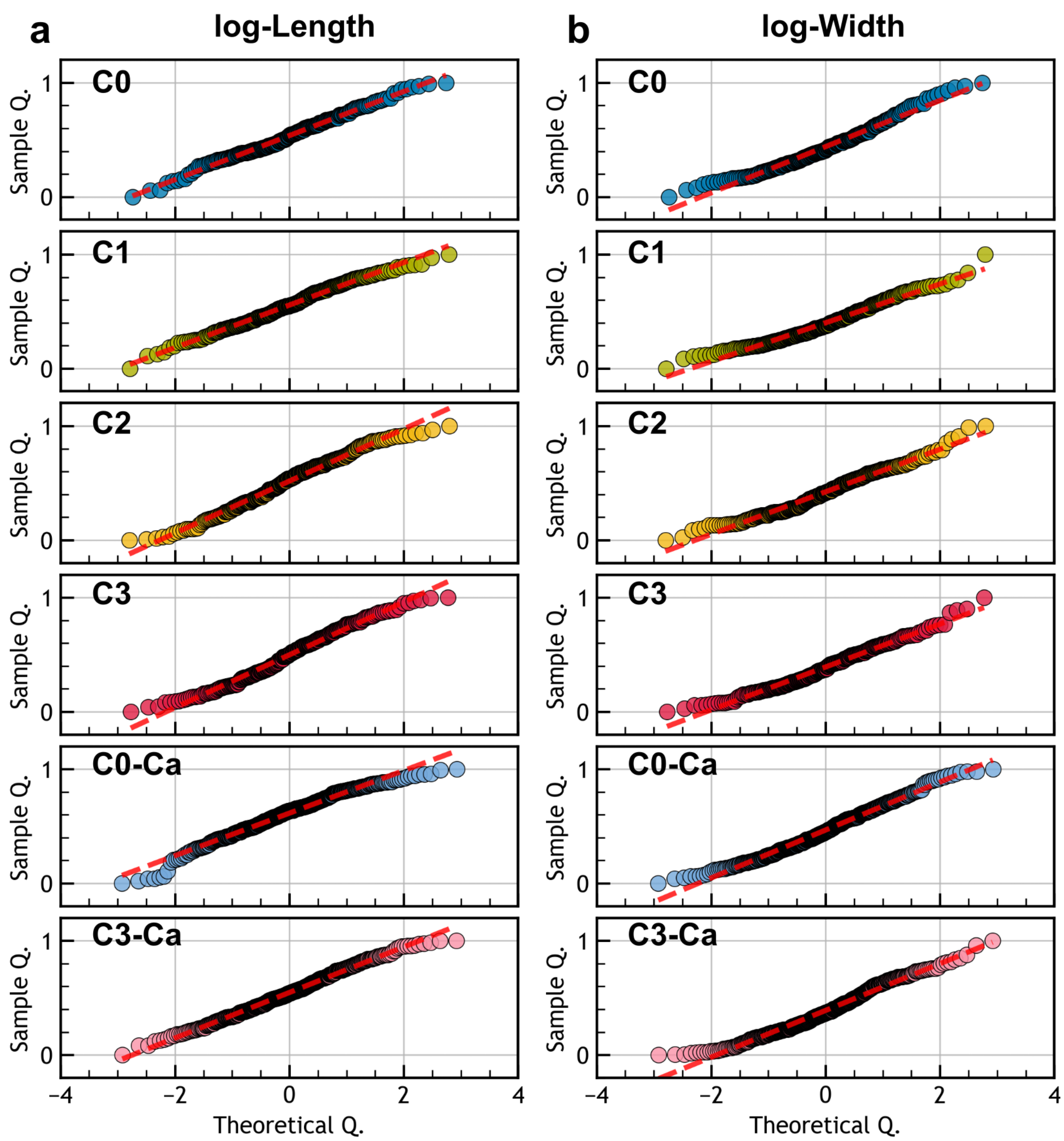

**Figure S7**. Normal quantile-quantile (Q-Q) plots of the CNC shape parameters from analysis of TEM images (**a**) logarithm of bounding box length, ln ($L_b$), and (**b**) logarithm of bounding box width, ln ($W_b$).

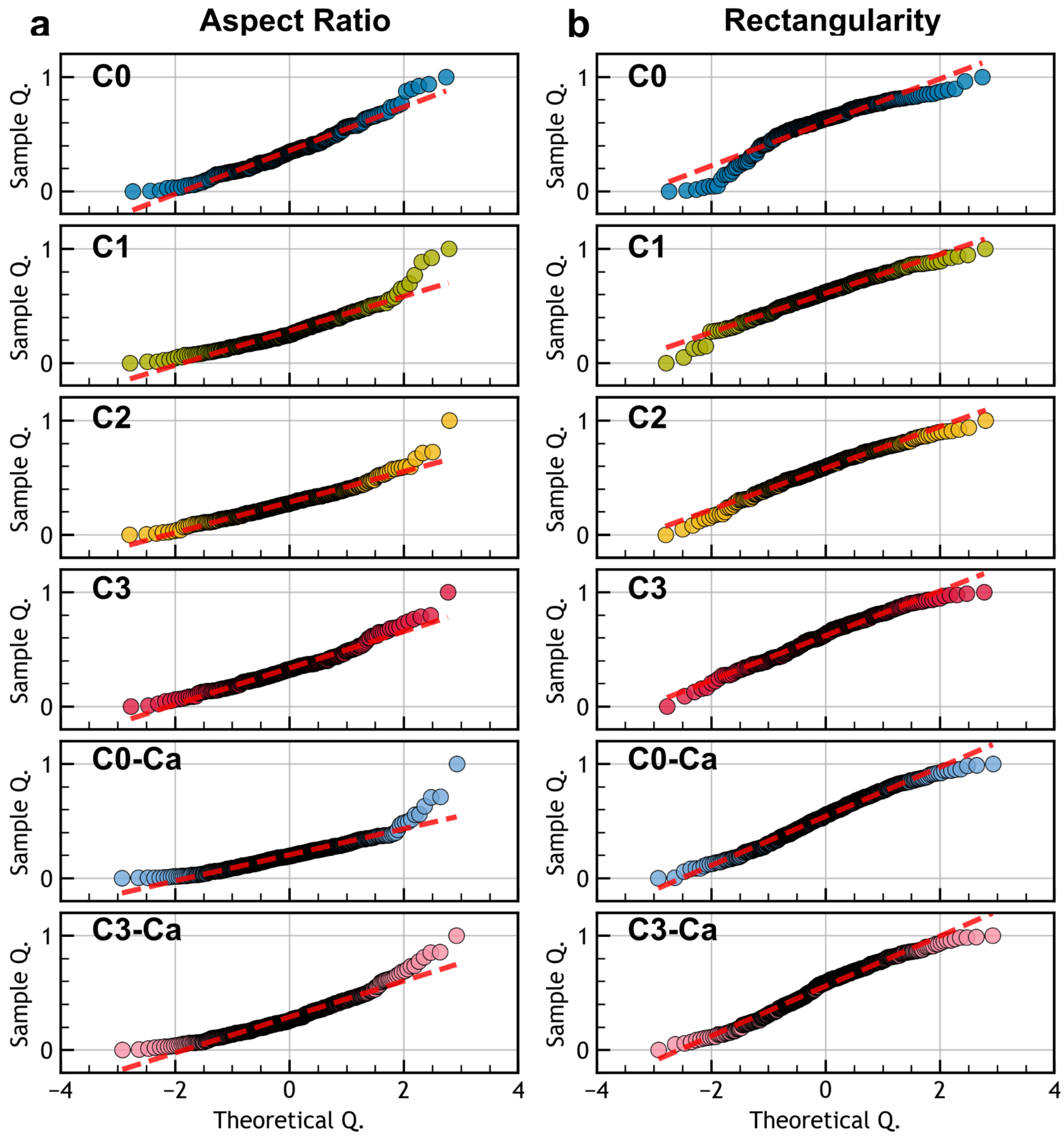

**Figure S8**. Normal quantile-quantile (Q-Q) plots of the CNC shape parameters from analysis of TEM images (**a**) bounding box aspect ratio ($r_b$) and (**b**) rectangularity (*Rect*).

**Table S3**. Shapiro–Wilk test of normality $W$ statistic and corresponding $p$-values for the logarithm of the bounding box lengths ($L_b$) and widths ($W_b$) and for the aspect ratio ($r_b$) and rectangularity (*Rect*) for each sample (*i.e.* the $p$-value can be viewed as the probability to obtain these specific sample distributions if the corresponding population was normally distributed). The distribution of the data highlighted in red does not differ significantly from a normal distribution with a confidence threshold of 0.05.

| Sample | $\ln(L_b)$ | $\ln(W_b)$ | $r_b$ | *Rect* |
|---|---|---|---|---|
| **C0** | 0.993<br>(0.346) | 0.976<br>(0.001) | 0.971<br>(0.000) | 0.922<br>(0.000) |
| **C1** | 0.992<br>(0.155) | 0.977<br>(0.000) | 0.933<br>(0.000) | 0.983<br>(0.004) |
| **C2** | 0.984<br>(0.005) | 0.982<br>(0.002) | 0.958<br>(0.000) | 0.983<br>(0.003) |
| **C3** | 0.982<br>(0.003) | 0.984<br>(0.007) | 0.967<br>(0.000) | 0.984<br>(0.008) |
| **C0-Ca** | 0.978<br>(0.000) | 0.987<br>(0.001) | 0.921<br>(0.000) | 0.987<br>(0.001) |
| **C3-Ca** | 0.993<br>(0.087) | 0.980<br>(0.000) | 0.952<br>(0.000) | 0.979<br>(0.000) |

## S4.2 Statistical Analysis of TEM Values

The statistical significance of the differences of average values between samples can be estimated through different tests. The most common test for such a comparison is the Student t-test. This test requires the underlying populations to be normally distributed, which in our case is uncertain. Consequently, pairwise comparison between samples was performed using a Mann–Whitney $U$ test, as it does not assume normality, does not necessitate any transformation of the data and is generally more powerful.[3] Note that the corresponding Student t-tests were also performed for comparison, leading to equivalent results (see **Table S5**, **Table S7**, **Table S9**, and **Table S11**).

The size of the effects was also estimated through calculation of the point biserial correlation coefficient, $r$, according to:[4]

$$r = \frac{z}{\sqrt{N}} \tag{S4}$$

where $N$ is the pooled sample size and $z$ is the normalized $U$ statistics calculated from:

$$z = \frac{U - 0.5 n_1 n_2}{\sqrt{\frac{n_1 n_2 (n_1 + n_2 + 1)}{12}}} \tag{S5}$$

with $n_1$ and $n_2$ the sample sizes. The results of the Mann–Whitney $U$ test and the corresponding r for $L_b$, $W_b$, $r_b$ and *Rect* are presented respectively in **Table S4**, **Table S6**, **Table S8**, and **Table S10**, and summarized in **Figure 4**.

**Table S4**. Mann–Whitney *U* statistics, corresponding *p*-value and point biserial correlation coefficient resulting from the pairwise comparison of $L_b$ for all samples. Data highlighted in red indicates that the corresponding samples do not differ significantly with a confidence threshold of 0.05.

| | **C0** | **C1** | **C2** | **C3** | **C0-Ca** | **C3-Ca** |
|---|---|---|---|---|---|---|
| **C0** | / | 32384 (0.043)<br>**-0.09** | 34667 (0.009)<br>**-0.12** | 32849 (0.001)<br>**-0.15** | 60965 (0.000)<br>**-0.28** | 60493 (0.000)<br>**-0.30** |
| **C1** | 32384 (0.043)<br>**0.09** | / | 36022 (0.548)<br>-0.03 | 34029 (0.246)<br>-0.05 | 64522 (0.000)<br>**-0.20** | 64121 (0.000)<br>**-0.22** |
| **C2** | 34667 (0.009)<br>**0.12** | 36022 (0.548)<br>0.03 | / | 32521 (0.573)<br>-0.02 | 43106 (0.000)<br>**-0.17** | 41304 (0.000)<br>**-0.19** |
| **C3** | 32849 (0.001)<br>**0.15** | 34029 (0.246)<br>0.05 | 32521 (0.573)<br>0.02 | / | 59359 (0.000)<br>**-0.16** | 38783 (0.000)<br>**-0.17** |
| **C0-Ca** | 60965 (0.000)<br>**0.28** | 64522 (0.000)<br>**0.20** | 43106 (0.000)<br>**0.17** | 59359 (0.000)<br>**0.16** | / | 77495 (0.601)<br>-0.02 |
| **C3-Ca** | 60493 (0.000)<br>**0.30** | 64121 (0.000)<br>**0.22** | 41304 (0.000)<br>**0.19** | 38783 (0.000)<br>**0.17** | 77495 (0.601)<br>0.02 | / |

**Table S5**. Absolute Student-t statistics and corresponding *p*-value resulting from the pairwise comparison of ln ($L_b$) for all samples. Data highlighted in red indicates that the corresponding samples do not differ significantly with a confidence threshold of 0.05.

| | **C0** | **C1** | **C2** | **C3** | **C0-Ca** | **C3-Ca** |
|---|---|---|---|---|---|---|
| **C0** | / | 2.05 (0.041) | 2.53 (0.012) | 3.51 (0.000) | 7.28 (0.000) | 8.56 (0.000) |
| **C1** | 2.05 (0.041) | / | 0.52 (0.603) | 1.38 (0.167) | 5.05 (0.000) | 6.27 (0.000) |
| **C2** | 2.53 (0.012) | 0.52 (0.603) | / | 0.82 (0.411) | 4.39 (0.000) | 5.57 (0.000) |
| **C3** | 3.51 (0.000) | 1.38 (0.167) | 0.82 (0.411) | / | 3.71 (0.000) | 4.94 (0.000) |
| **C0-Ca** | 7.28 (0.000) | 5.05 (0.000) | 4.39 (0.000) | 3.71 (0.000) | / | 1.19 (0.234) |
| **C3-Ca** | 8.56 (0.000) | 6.27 (0.000) | 5.57 (0.000) | 4.94 (0.000) | 1.19 (0.234) | / |

**Table S6**. Mann–Whitney *U* statistics, corresponding *p*-value and point biserial correlation coefficient resulting from the pairwise comparison of $W_b$ for all samples. Data highlighted in red indicates that the corresponding samples do not differ significantly with a confidence threshold of 0.05.

| | **C0** | **C1** | **C2** | **C3** | **C0-Ca** | **C3-Ca** |
|---|---|---|---|---|---|---|
| **C0** | / | 29398 (0.932)<br>-0.00 | 30499 (0.995)<br>0.00 | 29804 (0.232)<br>-0.05 | 56773 (0.000)<br>**-0.21** | 57477 (0.000)<br>**-0.24** |
| **C1** | 29398 (0.932)<br>0.00 | / | 34637 (0.852)<br>0.01 | 33844 (0.295)<br>-0.05 | 64491 (0.000)<br>**-0.20** | 65297 (0.000)<br>**-0.24** |
| **C2** | 30499 (0.995)<br>-0.00 | 34637 (0.852)<br>-0.01 | / | 31518 (0.249)<br>-0.05 | 41250 (0.000)<br>**-0.20** | 38551 (0.000)<br>**-0.23** |
| **C3** | 29804 (0.232)<br>0.05 | 33844 (0.295)<br>0.05 | 31518 (0.249)<br>0.05 | / | 59734 (0.000)<br>**-0.17** | 37217 (0.000)<br>**-0.20** |
| **C0-Ca** | 56773 (0.000)<br>**0.21** | 64491 (0.000)<br>**0.20** | 41250 (0.000)<br>**0.20** | 59734 (0.000)<br>**0.17** | / | 76159 (0.349)<br>-0.03 |
| **C3-Ca** | 57477 (0.000)<br>**0.24** | 65297 (0.000)<br>**0.24** | 38551 (0.000)<br>**0.23** | 37217 (0.000)<br>**0.2** | 76159 (0.349)<br>0.03 | / |

**Table S7**. Absolute Student-t statistics and corresponding *p*-value resulting from the pairwise comparison of ln ($W_b$) for all samples. Data highlighted in red indicates that the corresponding samples do not differ significantly with a confidence threshold of 0.05.

| | **C0** | **C1** | **C2** | **C3** | **C0-Ca** | **C3-Ca** |
|---|---|---|---|---|---|---|
| **C0** | / | 0.21 (0.835) | 0.01 (0.989) | 1.09 (0.277) | 5.90 (0.000) | 6.94 (0.000) |
| **C1** | 0.21 (0.835) | / | 0.22 (0.829) | 0.87 (0.385) | 5.63 (0.000) | 6.66 (0.000) |
| **C2** | 0.01 (0.989) | 0.22 (0.829) | / | 1.08 (0.283) | 5.76 (0.000) | 6.78 (0.000) |
| **C3** | 1.09 (0.277) | 0.87 (0.385) | 1.08 (0.283) | / | 4.80 (0.000) | 5.85 (0.000) |
| **C0-Ca** | 5.90 (0.000) | 5.63 (0.000) | 5.76 (0.000) | 4.80 (0.000) | / | 1.09 (0.278) |
| **C3-Ca** | 6.94 (0.000) | 6.66 (0.000) | 6.78 (0.000) | 5.85 (0.000) | 1.09 (0.278) | / |

**Table S8**. Mann–Whitney *U* statistics, corresponding *p*-value and point biserial correlation coefficient resulting from the pairwise comparison of $r_b$ for all samples. Data highlighted in red indicates that the corresponding samples do not differ significantly with a confidence threshold of 0.05.

| | **C0** | **C1** | **C2** | **C3** | **C0-Ca** | **C3-Ca** |
|---|---|---|---|---|---|---|
| **C0** | / | 32692 (0.026)<br>**-0.10** | 36685 (0.000)<br>**-0.17** | 32436 (0.003)<br>**-0.14** | 46531 (0.613)<br>-0.02 | 43841 (0.751)<br>0.01 |
| **C1** | 32692 (0.026)<br>**0.10** | / | 37759 (0.112)<br>-0.07 | 33140 (0.535)<br>-0.03 | 47209 (0.043)<br>**0.08** | 44161 (0.004)<br>**0.11** |
| **C2** | 36685 (0.000)<br>**0.17** | 37759 (0.112)<br>0.07 | / | 35444 (0.249)<br>0.05 | 63658 (0.000)<br>**0.15** | 64770 (0.000)<br>**0.19** |
| **C3** | 32436 (0.003)<br>**0.14** | 33140 (0.535)<br>0.03 | 35444 (0.249)<br>-0.05 | / | 43810 (0.009)<br>**0.10** | 57202 (0.000)<br>**0.14** |
| **C0-Ca** | 46531 (0.613)<br>0.02 | 47209 (0.043)<br>**-0.08** | 63658 (0.000)<br>**-0.15** | 43810 (0.009)<br>**-0.10** | / | 82397 (0.323)<br>0.04 |
| **C3-Ca** | 43841 (0.751)<br>-0.01 | 44161 (0.004)<br>**-0.11** | 64770 (0.000)<br>**-0.19** | 57202 (0.000)<br>**-0.14** | 82397 (0.323)<br>-0.04 | / |

**Table S9**. Absolute Student-t statistics and corresponding *p*-value resulting from the pairwise comparison of $r_b$ for all samples. Data highlighted in red indicates that the corresponding samples do not differ significantly with a confidence threshold of 0.05.

| | **C0** | **C1** | **C2** | **C3** | **C0-Ca** | **C3-Ca** |
|---|---|---|---|---|---|---|
| **C0** | / | 2.59 (0.010) | 3.8 (0.000) | 2.84 (0.005) | 0.6 (0.547) | 0.08 (0.936) |
| **C1** | 2.59 (0.010) | / | 1.06 (0.289) | 0.12 (0.906) | 2.19 (0.029) | 2.94 (0.003) |
| **C2** | 3.8 (0.000) | 1.06 (0.289) | / | 0.99 (0.321) | 3.49 (0.001) | 4.33 (0.000) |
| **C3** | 2.84 (0.005) | 0.12 (0.906) | 0.99 (0.321) | / | 2.45 (0.014) | 3.26 (0.001) |
| **C0-Ca** | 0.6 (0.547) | 2.19 (0.029) | 3.49 (0.001) | 2.45 (0.014) | / | 0.77 (0.440) |
| **C3-Ca** | 0.08 (0.936) | 2.94 (0.003) | 4.33 (0.000) | 3.26 (0.001) | 0.77 (0.440) | / |

**Table S10**. Mann–Whitney *U* statistics, corresponding *p*-value and point biserial correlation coefficient resulting from the pairwise comparison of *Rect* for all samples. Data highlighted in red indicates that the corresponding samples do not differ significantly with a confidence threshold of 0.05.

| | **C0** | **C1** | **C2** | **C3** | **C0-Ca** | **C3-Ca** |
|---|---|---|---|---|---|---|
| **C0** | / | 29523 (0.868)<br>-0.01 | 29961 (0.730)<br>0.02 | 25770 (0.130)<br>0.07 | 31253 (0.000)<br>**0.26** | 32523 (0.000)<br>**0.22** |
| **C1** | 29523 (0.868)<br>0.01 | / | 33933 (0.557)<br>0.03 | 29224 (0.080)<br>0.08 | 34806 (0.000)<br>**0.28** | 36449 (0.000)<br>**0.24** |
| **C2** | 29961 (0.730)<br>-0.02 | 33933 (0.557)<br>-0.03 | / | 35385 (0.263)<br>0.05 | 70043 (0.000)<br>**0.25** | 66313 (0.000)<br>**0.21** |
| **C3** | 25770 (0.130)<br>-0.07 | 29224 (0.080)<br>-0.08 | 35385 (0.263)<br>-0.05 | / | 36859 (0.000)<br>**0.22** | 59164 (0.000)<br>**0.18** |
| **C0-Ca** | 31253 (0.000)<br>**-0.26** | 34806 (0.000)<br>**-0.28** | 70043 (0.000)<br>**-0.25** | 36859 (0.000)<br>**-0.22** | / | 75276 (0.227)<br>-0.04 |
| **C3-Ca** | 32523 (0.000)<br>**-0.22** | 36449 (0.000)<br>**-0.24** | 66313 (0.000)<br>**-0.21** | 59164 (0.000)<br>**-0.18** | 75276 (0.227)<br>0.04 | / |

**Table S11**. Absolute Student-t statistics and corresponding *p*-value resulting from the pairwise comparison of *Rect* for all samples. Data highlighted in red indicates that the corresponding samples do not differ significantly with a confidence threshold of 0.05.

| | **C0** | **C1** | **C2** | **C3** | **C0-Ca** | **C3-Ca** |
|---|---|---|---|---|---|---|
| **C0** | / | 0.89 (0.374) | 0.08 (0.938) | 0.71 (0.475) | 6.45 (0.000) | 5.47 (0.000) |
| **C1** | 0.89 (0.374) | / | 0.83 (0.405) | 1.75 (0.081) | 7.98 (0.000) | 6.94 (0.000) |
| **C2** | 0.08 (0.938) | 0.83 (0.405) | / | 0.82 (0.413) | 6.73 (0.000) | 5.72 (0.000) |
| **C3** | 0.71 (0.475) | 1.75 (0.081) | 0.82 (0.413) | / | 6.27 (0.000) | 5.20 (0.000) |
| **C0-Ca** | 6.45 (0.000) | 7.98 (0.000) | 6.73 (0.000) | 6.27 (0.000) | / | 1.14 (0.256) |
| **C3-Ca** | 5.47 (0.000) | 6.94 (0.000) | 5.72 (0.000) | 5.20 (0.000) | 1.14 (0.256) | / |

## S3.3 Cryo-TEM images

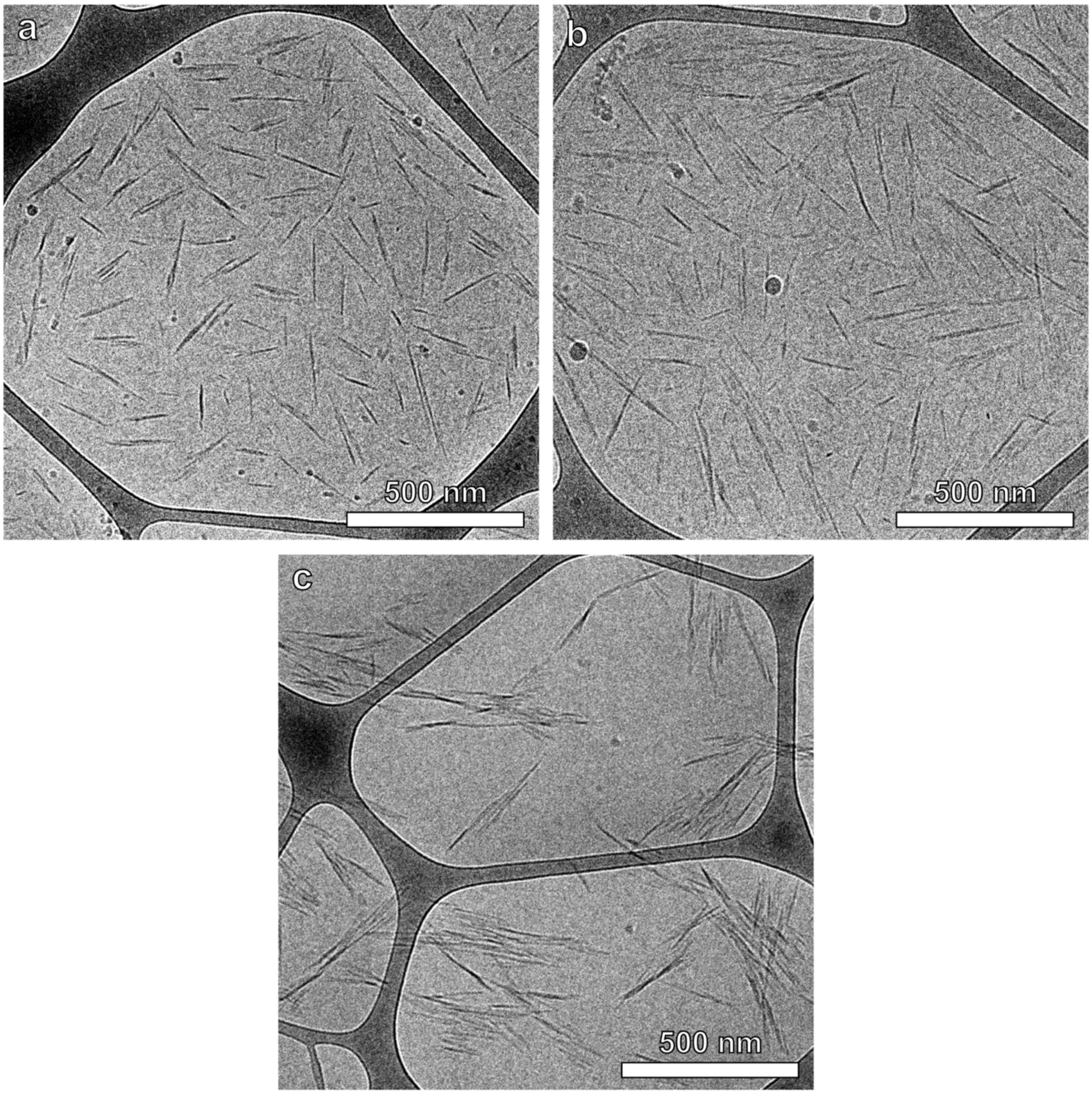

**Figure S9**. Typical images from Cryo-TEM of (**a**) never centrifuged CNCs (**C0**), (**b**) triply-centrifuged CNCs (**C3**) and (**c**) Ca-aggregated CNCs (**C3-Ca**).

## S4. SAXS

### S4.1 SAXS Profile Modeling Expression

At low particle concentration, the SAXS intensity profile of a monodisperse sample, $I(q)$ ($cm^{-1}$), can be expressed in terms of the orientationally averaged form factor $P(q)$ as:[5]

$$I(q) = \Phi\ \Delta\rho^2 P(q) V_p^2 \tag{S6}$$

where $\Phi$ (v/v%) is the particle volume fraction, $V_p$ ($cm^3$) is the volume the particles and $\Delta\rho$ ($cm^{-2}$) is the scattering contrast, given by:[6]

$$\Delta\rho = \left(\frac{n_{e,0}\rho_{cell}}{M_0} N_A - \tilde{\rho}_{solv}\right) r_0 \tag{S7}$$

with $n_{e,0}$ and $M_0$ are respectively the number of electron and the molar mass of a repeating unit (86 e and 162.1 g $mol^{-1}$ respectively), $\rho_{cell}$ is the cellulose mass density (taken as 1.6 g $cm^{-3}$ for a Iβ cellulose crystal),[7] $N_A$ is the Avogadro constant (6.02 $10^{23}$ $mol^{-1}$), $\tilde{\boldsymbol{\rho}}_{\boldsymbol{solv}}$ is the electronic number density of water (3.34 $10^{23}$ $e$ $cm^{-3}$),[6] and $r_0$ is the scattering length of an electron (2.82 $10^{-13}$ cm).[6]

For polydisperse systems, the orientationally averaged form factor contribution from each particle $P_i(q)$ must be accounted for. The total intensity then becomes:

$$I(q) = \Phi\ \Delta\rho^2 \frac{1}{N\langle V\rangle} \sum_{i=1}^{N} P_i(q) V_i^2 \tag{S8}$$

where $N$ is the number of particles, $V_i$ the volume of particle $i$., and $\langle V\rangle$ is the average particle volume. The orientationally averaged form factor for each particle $P_i(q)$ can be estimated by averaging the form factor $F_i$ of the particle over several orientations ($M$), as expressed by:

$$P_i(q) = \frac{1}{M} \sum_{j=1}^{M} \left|F_i(q, \theta_j, \varphi_j)\right|^2 \tag{S9}$$

where the form factor of a rectangular prism of length $L_i$, width $W_i$ and thickness $T_i$ is given by:[5,6]

$$F_i(q,\theta,\varphi) = \operatorname{sinc}\left(\frac{L_i\ q\cos(\theta)}{2}\right) \operatorname{sinc}\left(\frac{W_i\ q\sin(\theta)\sin(\varphi)}{2}\right) \operatorname{sinc}\left(\frac{T_i\ q\sin(\theta)\cos(\varphi)}{2}\right) \tag{S10}$$

with $\operatorname{sinc}(x) = \sin(x)/x$.

For each sample, the SAXS profile at $c$ = 0.1 wt % was resampled to 100 $q$-values evenly spaced in logarithmic scale. For each $q$, 800 particles were generated with their length and width randomly drawn from the distribution of $L_b$ and $W_b$ (extracted from TEM), and with a fixed thickness ($T$). The orientationally averaged form factor $P_i(q)$ for each particle was computed over 200 random orientations using **Equation** S9. The total SAXS intensity at each $q$ was then obtained by averaging over the 800 particles according to **Equation** S8.

## S4.2 Thickness Extraction

Using the method above, we modeled for each sample their corresponding SAXS profile over a range of thickness values $T$ to extract the best fitting thickness. In practice, we first evaluate for each sample the coefficient of determination $R^2$ between the experimental and modeled profiles for each $T$ value (**Figure S10**). Then, we fit the dependence of $R^2$ with $T$ using a second order polynomial function. This allows for the extraction of the best thickness ($T_b$) corresponding to the optimum value of $T$ that maximizes $R^2$. The resulting SAXS profiles, modeled for each sample at their respective $T_b$, are presented in **Figure 5c**.

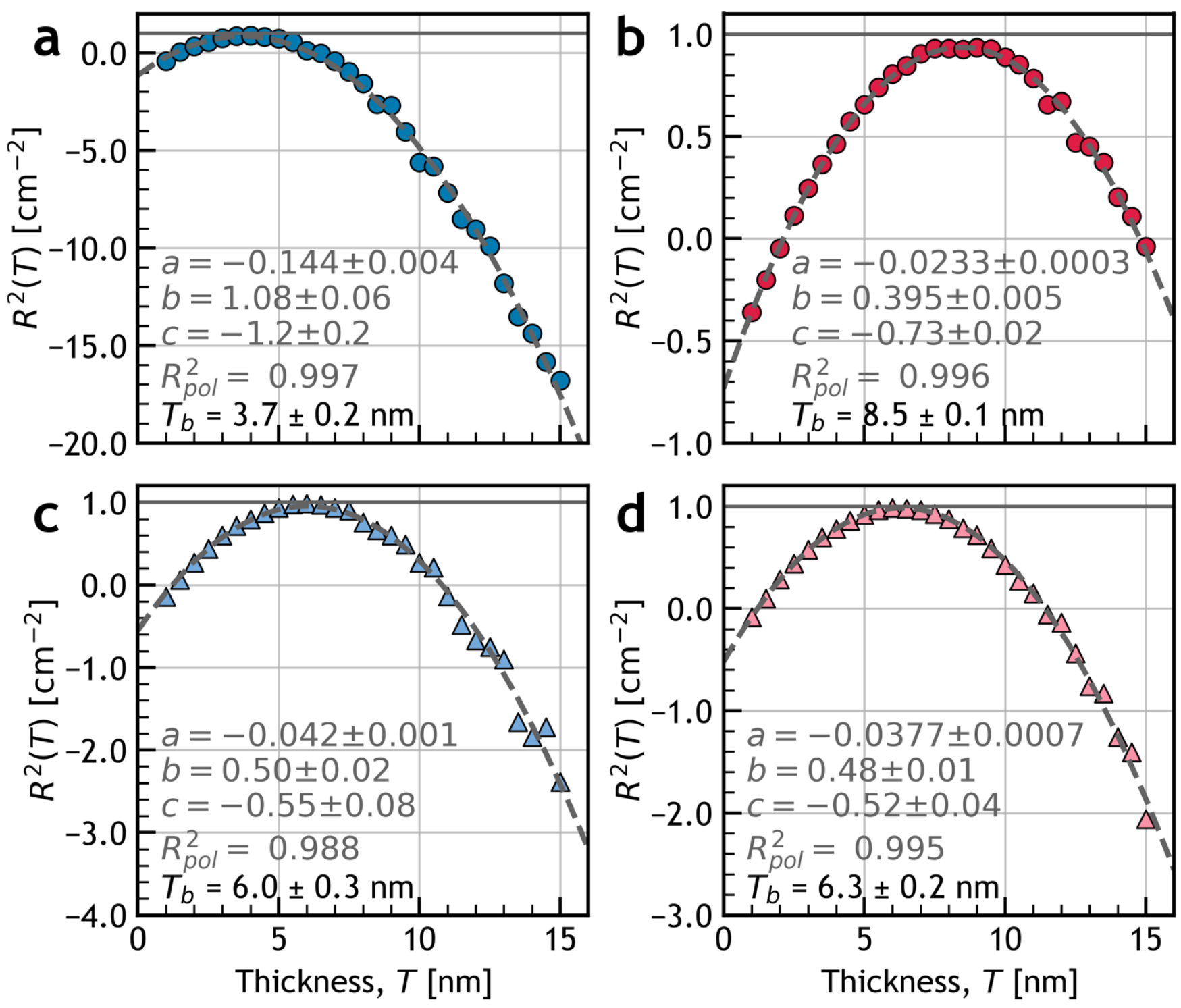

**Figure S10**: Evolution of the coefficient of determination ($R^2$) between the experimental SAXS profiles and the unoptimized modeled profiles over a range of thickness values ($T$). Best fit obtained from a second order polynomial function $R^2 = aT^2+bT+c$ (dashed black line), corresponding parameters, coefficient of determination ($R_{pol}^2$), and best-fit thickness ($T_b$) indicated. The horizontal grey line indicates $R^2 = 1$.

### S4.3 Cross-sectional Guinier Analysis

Similarly, for elongated rods at intermediate $q$, the scattered intensity can be expressed as:[6,8]

$$I(q) = \frac{1}{q}\pi c \frac{\Delta\rho^2}{\rho_{cell}^2} m_L \exp\left(-\frac{q^2 R_C^2}{2}\right), \qquad \frac{2\pi}{L} < q < \frac{1}{R_c} \tag{S11}$$

where $m_L$ (g cm$^{-1}$) is the particle mass per unit length and $R_c$ (cm) is the cross-sectional gyration radius. This analysis was performed for all the samples (see **Figure S11**) and the corresponding $m_L$ and $R_c$ are presented in **Figure 5c**.

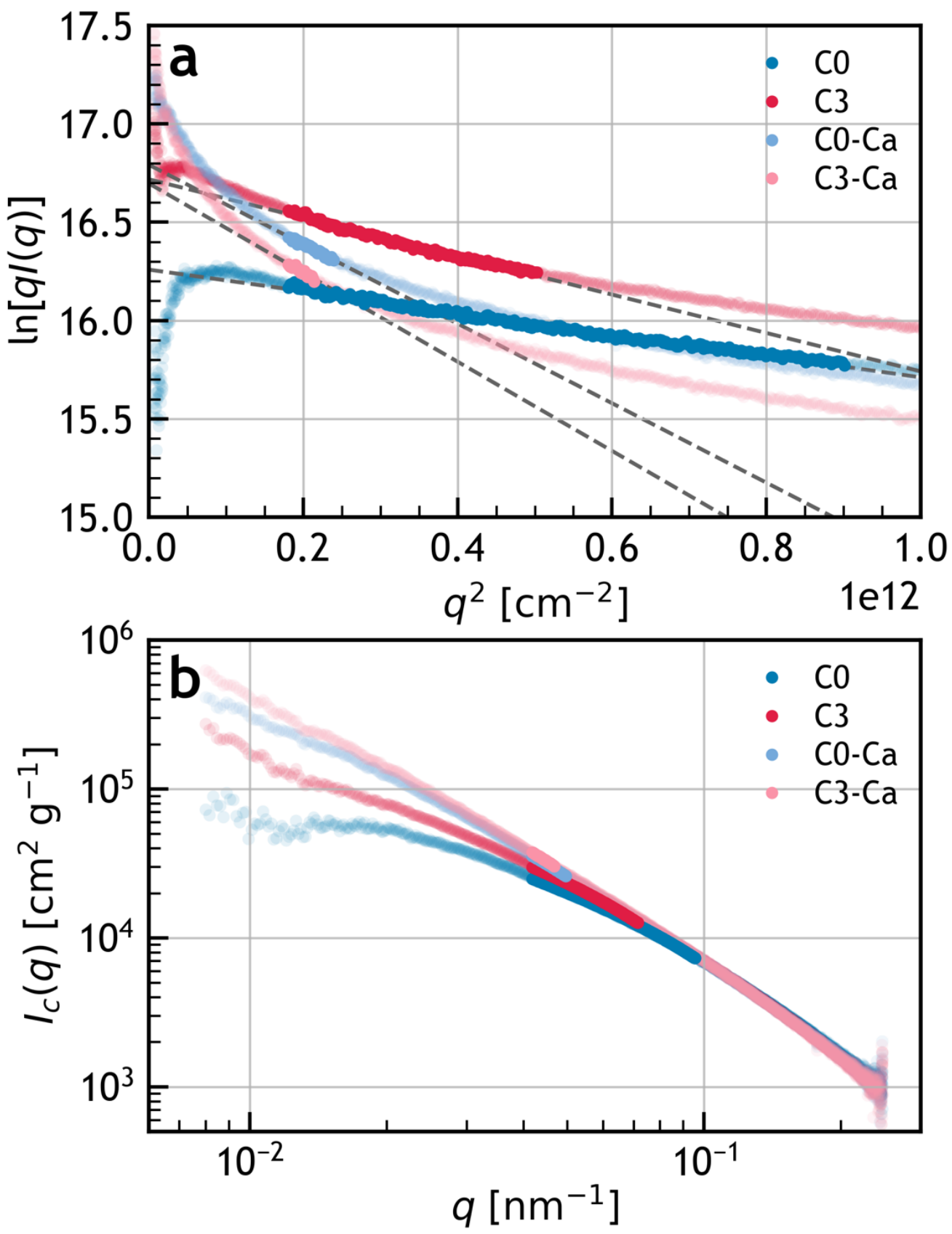

**Figure S11**. Overview of the Porod analysis with **(a)** evolution of ln [$q$ $I$($q$)] against $q^2$, overlayed with the points used for the Porod analysis and corresponding fit from **Equation** S11 and **(b)** concentration normalized intensity as a function of $q$ overlayed with the prediction from the Porod fit.

# S5. Electron Diffraction Analyses

## S5.1 Typical Scanning Nanobeam Electron Diffraction (SNBED) Data

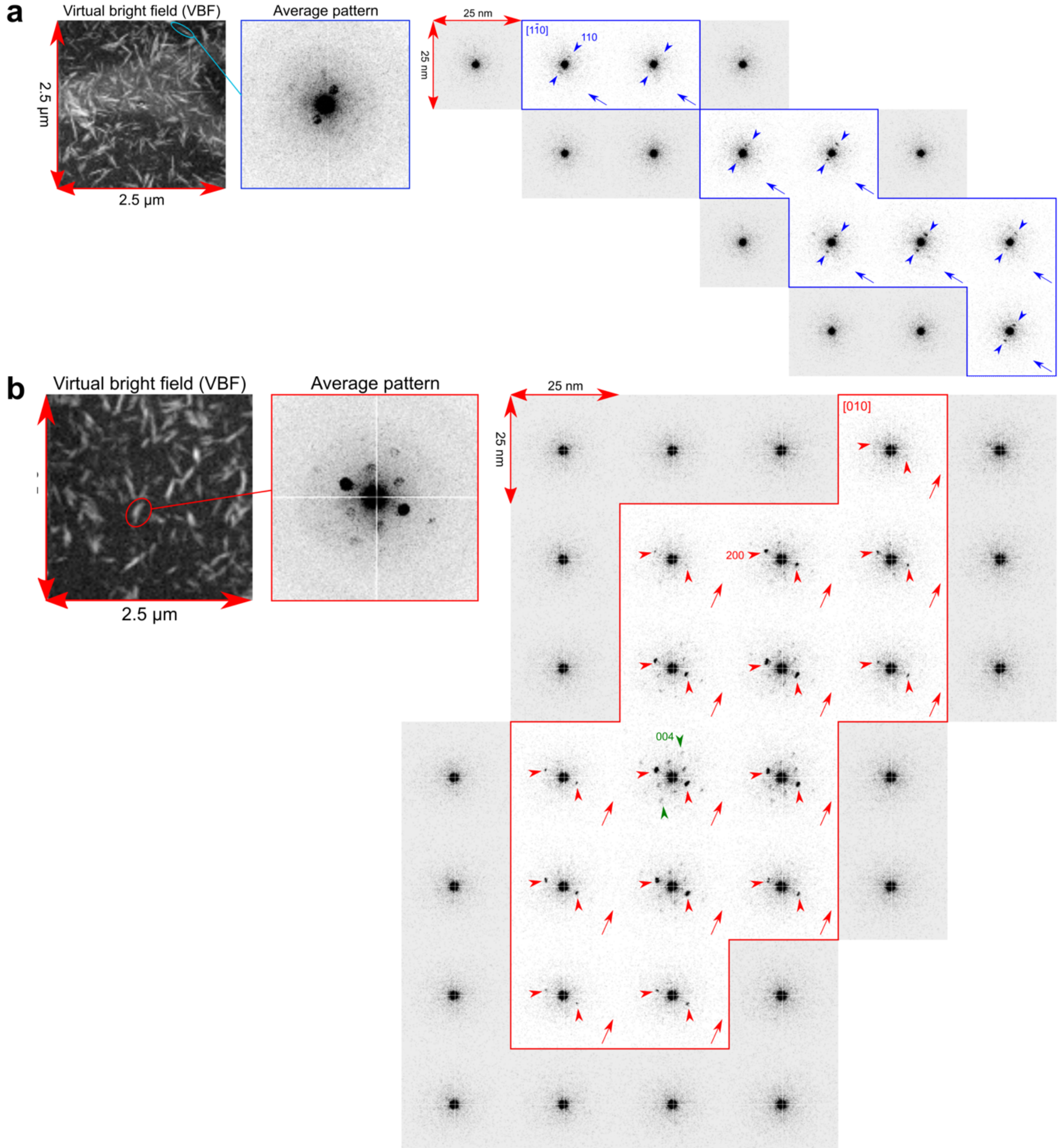

**Figure S12.** SNBED analysis with virtual bright field image (*left*), average diffraction pattern and ED patterns along a CNC (*right*) for (**a**) a never-centrifuged CNC (**C0**) displaying a [1-10] zone axis and (**b**) a centrifuged CNC (**C3**) displaying a [100] zone axis. The arrows on the corner of each ED pattern indicate the fiber axis.

### S5.2 Impact of Favored Interaction Between the Crystallites and the Grid

In a scenario where crystallite association would not be favored along any preferential crystal face, each crystal orientation on the grid would be equiprobable for any population of bundle particles, and as such we would not observe a prevalent orientation. In such a case, even a preferential interaction between the (010) crystal faces and the carbon film on the grid would not account for the observed prevalence of [010] zone-axis orientation. Indeed, upon formation of bundle particles (i.e. from **C0** to **C3**), the (010) faces would be less available (as all the faces would be equally involved in the formation of bundles) resulting in a similar or lower proportion of observed [010] zone axis. Therefore, the sole prevalence of the observation of the [010] zone axis in bundled particles, because of how they lie flat onto the grid, also supports that there is a preferential association of the crystallites through their hydrophobic (100) planes.

## S5.3 Selected Area Electron Diffraction (SAED)

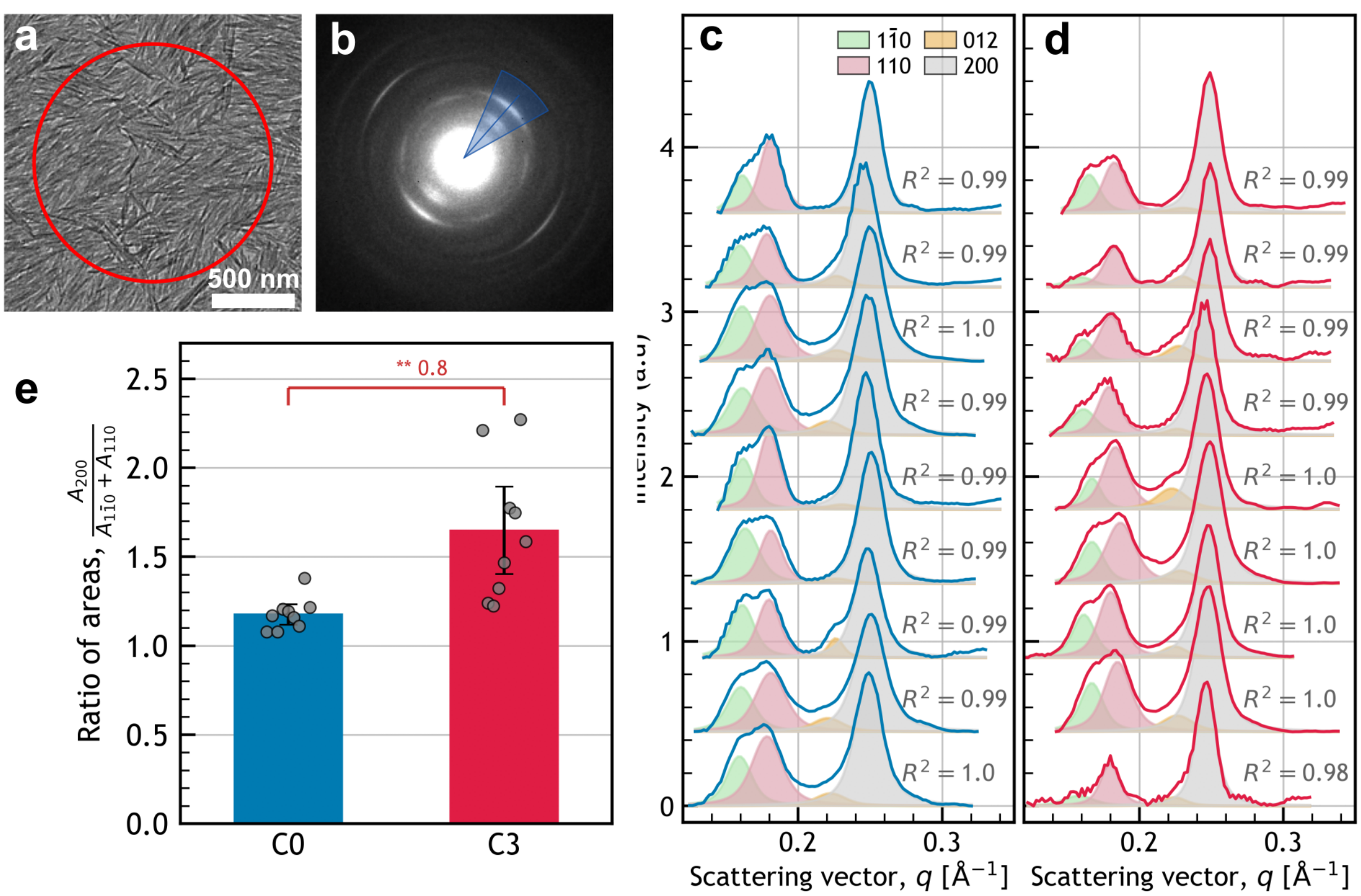

**Figure S13**. Selected area electron diffraction analysis workflow with **(a)** a typical image and its corresponding area of measurement (*ca* 0.8 µm$^2$), **(b)** the typical 2D diffraction pattern from a measurement along with the area used to calculate the equatorial profile. **(c-d)** Evolution of the relative intensity against the scattering vector for all the profiles, along with their peak deconvolution, for **(c) C0** and **(d) C3**, and **(e)** the corresponding ratio of the area of the 200 peak over the sum of the area of the 1-10 and 110 peaks, with significance of the difference between the two series estimated through Mann–Whitney $U$ test ($p$-value < 0.01) and importance of the effect through the absolute point biserial correlation coefficient ($r$ = 0.78).

## S6. Interfacial Tension Measurements

### S6.1 Extraction of Interfacial Tension from Drop Images

The formation of bundles through the preferential association of crystallites along their hydrophobic faces results to a reduction of the proportion of hydrophobic faces, which could have practical implications for their use as emulsion stabilizer. To investigate if the reduction of the proportion of hydrophobic faces had a significant impact on the CNC amphiphilicity, the interfacial tension between drops of CNC suspensions and hexadecane was measured through pendant drop experiments. The surface charges of the CNCs were screened by NaCl through the preparation of 0.9 wt% CNC suspension in 20 mM NaCl.[9]

To extract the interfacial tension (γ) from the picture of pendant drops, one can rewrite the Young-Laplace equation as:[10]

$$\gamma = \frac{\Delta\rho g D_e^2}{H} \quad \text{(S12)}$$

where $\Delta\rho$ is the density difference between the phases (taken as 224 kg $m^{-3}$), $g$ is the acceleration of gravity (9.81 m $s^{-2}$), $D_e$ is the equatorial diameter of the droplet and $H$ is a function of $D_s/D_e$, with $D_s$ the diameter of the droplet at a height $D_e$ (see example **Figure S14**). The evolution of $H$ as a function of $D_s/D_e$ was previously tabulated, allowing the estimation of $\gamma$ from $D_s/D_e$.[11] This approach was implemented in a custom-made Python script to treat the collected images, the software is available on Github under the name DropPyTension.[12] To challenge the accuracy of the script, the surface tension of water in air was calculated from the profile of 20 drops, yielding $\gamma = 71 \pm 4$ mN $m^{-1}$. This value is close to the accepted value of 72 mN $m^{-1}$ validating the use of this workflow to treat the collected pictures.

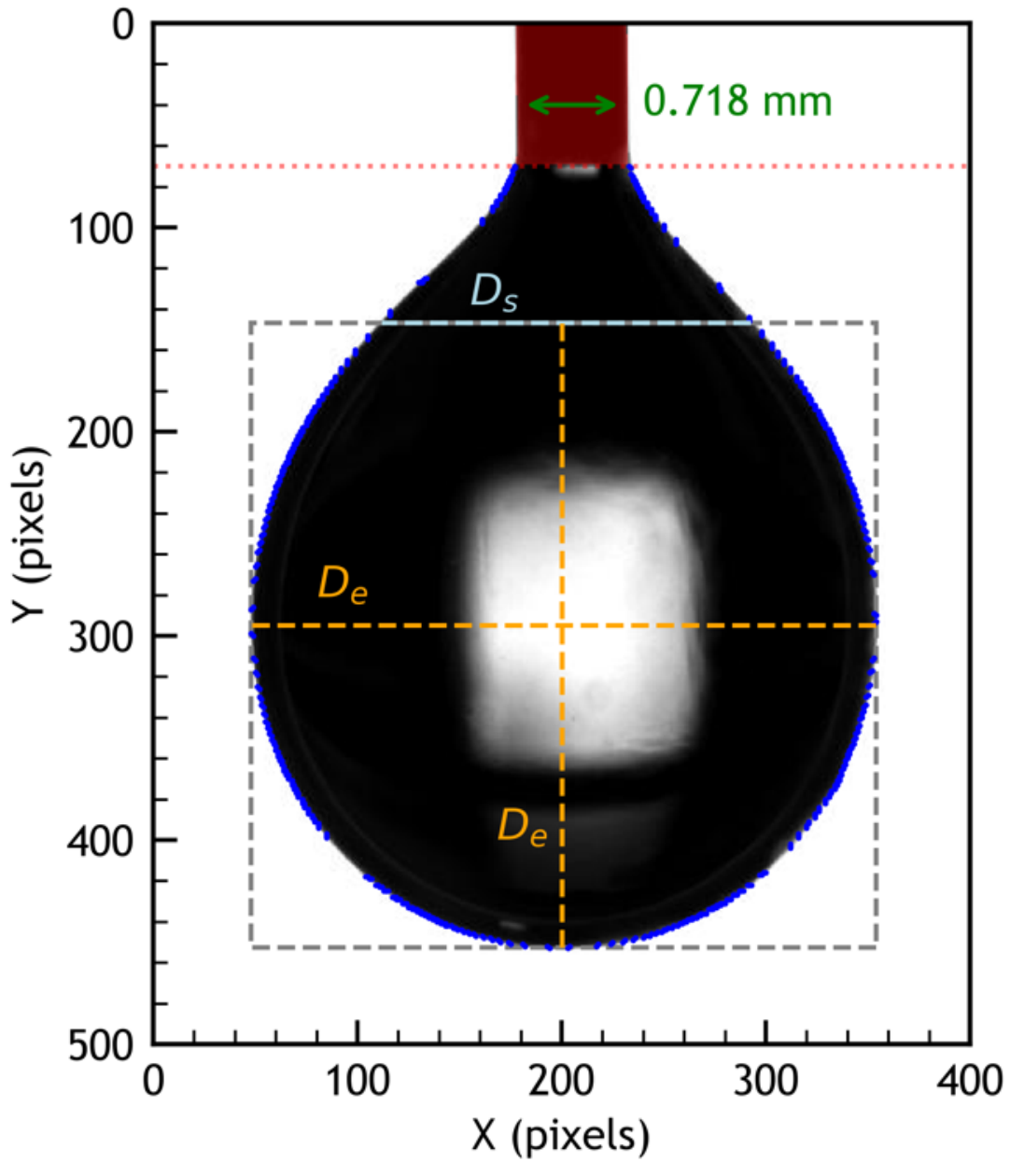

**Figure S14**. Illustration of pendant drop analysis.

## S6.2 Interfacial Tension Measurement of CNC Suspensions

The interfacial tension measurements of the CNC suspensions with hexadecane are presented in **Figure S15**. The γ between water and hexadecane was of 50 ± 3 mN m$^{-1}$. This is slightly lower than the expected value of 53 mM m$^{-1}$, which might be attributed to the presence of residual impurities in the hexadecane.[13] The significance of the differences of γ between the samples was estimated through a Student-t test (see **Figure S15** and **Table S12**), but a Mann-Whitney *U* test led to similar conclusions (see **Table S13**). All CNC suspensions, led to a significant lowering of γ, as previously described.[14] The difference of γ between the CNC samples was too small to present any practical relevance but was significant nonetheless, with γ decreasing in the order: **C0-Na**, **C3-Na**, and **C3-Ca**. This small difference is not trivial to relativize and out of the scope of the present work.

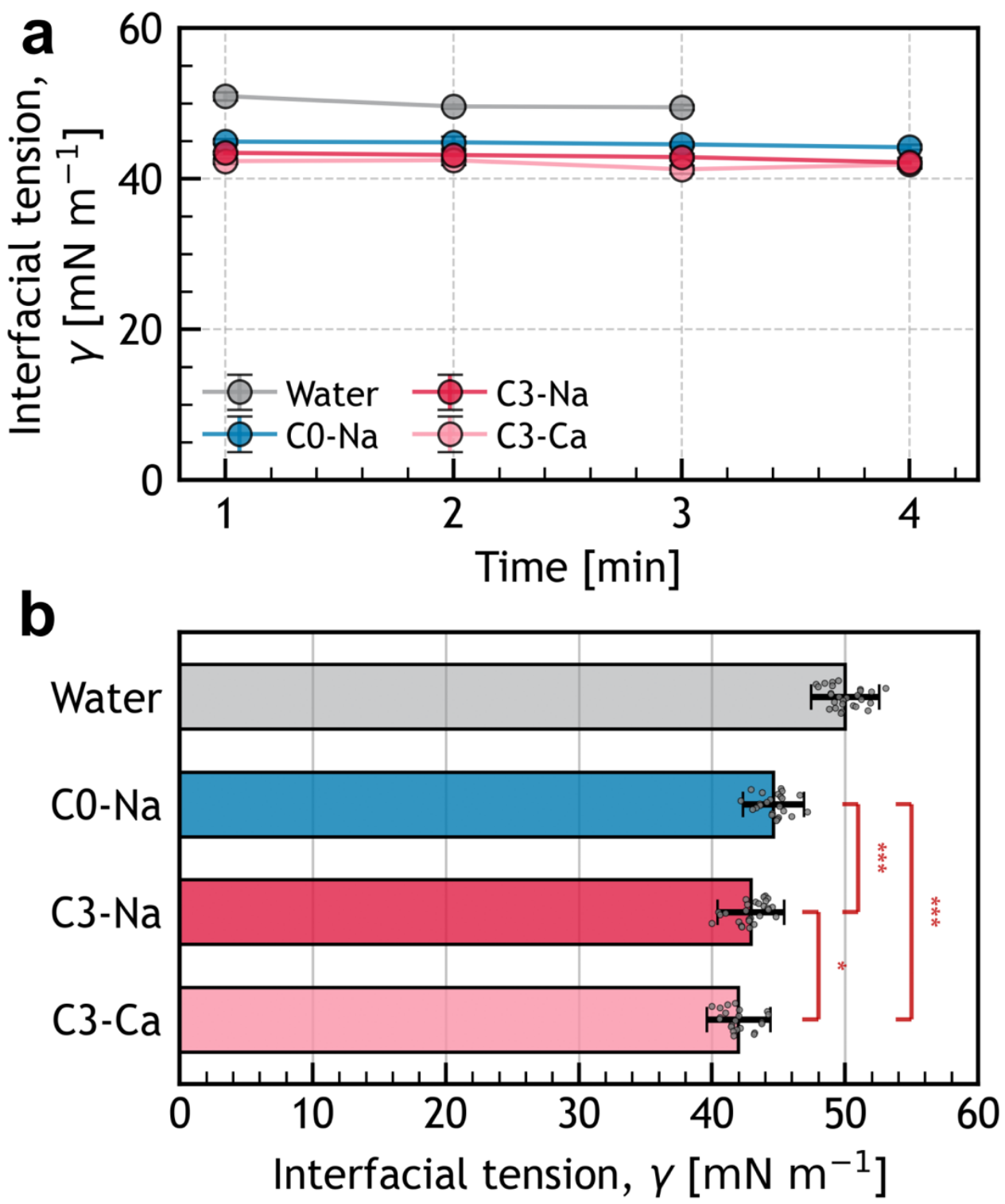

**Figure S15**. Average interfacial tension (γ) between water or CNC suspension (0.9 wt% CNCs, 20 mM NaCl) drops and hexadecane as a function of time **(a)** and independently of time **(b)**. The distributions were compared pairwise with a Student t-test, statistical significance is indicated by the following *p*-value thresholds: * ($0.05 > p > 0.01$), ** ($0.01 > p > 0.001$), and *** ($p < 0.001$).

**Table S12**. Absolute Student-t statistics and corresponding *p*-value resulting from the pairwise comparison of interfacial tension values for all samples.

| | **Water** | **C0-Na** | **C3-Na** | **C3-Ca** |
|---|---|---|---|---|
| **Water** | / | 15.5 (0.000) | 20.2 (0.000) | 22.0 (0.000) |
| **C0-Na** | 15.5 (0.000) | / | 4.9 (0.000) | 7.5 (0.000) |
| **C3-Na** | 20.2 (0.000) | 04.9 (0.000) | / | 2.6 (0.013) |
| **C3-Ca** | 22.0 (0.000) | 07.5 (0.000) | 2.6 (0.013) | / |

**Table S13**. Absolute Mann-Whitney *U* statistics and corresponding *p*-value resulting from the pairwise comparison of interfacial tension values for all samples.

| | **Water** | **C0-Na** | **C3-Na** | **C3-Ca** |
|---|---|---|---|---|
| **Water** | / | 648 (0.000) | 729 (0.000) | 594 (0.000) |
| **C0-Na** | 648 (0.000) | / | 551 (0.000) | 494 (0.000) |
| **C3-Na** | 729 (0.000) | 551 (0.000) | / | 424 (0.011) |
| **C3-Ca** | 594 (0.000) | 494 (0.000) | 424 (0.011) | / |

## S7. Capillary images of C3-Ca

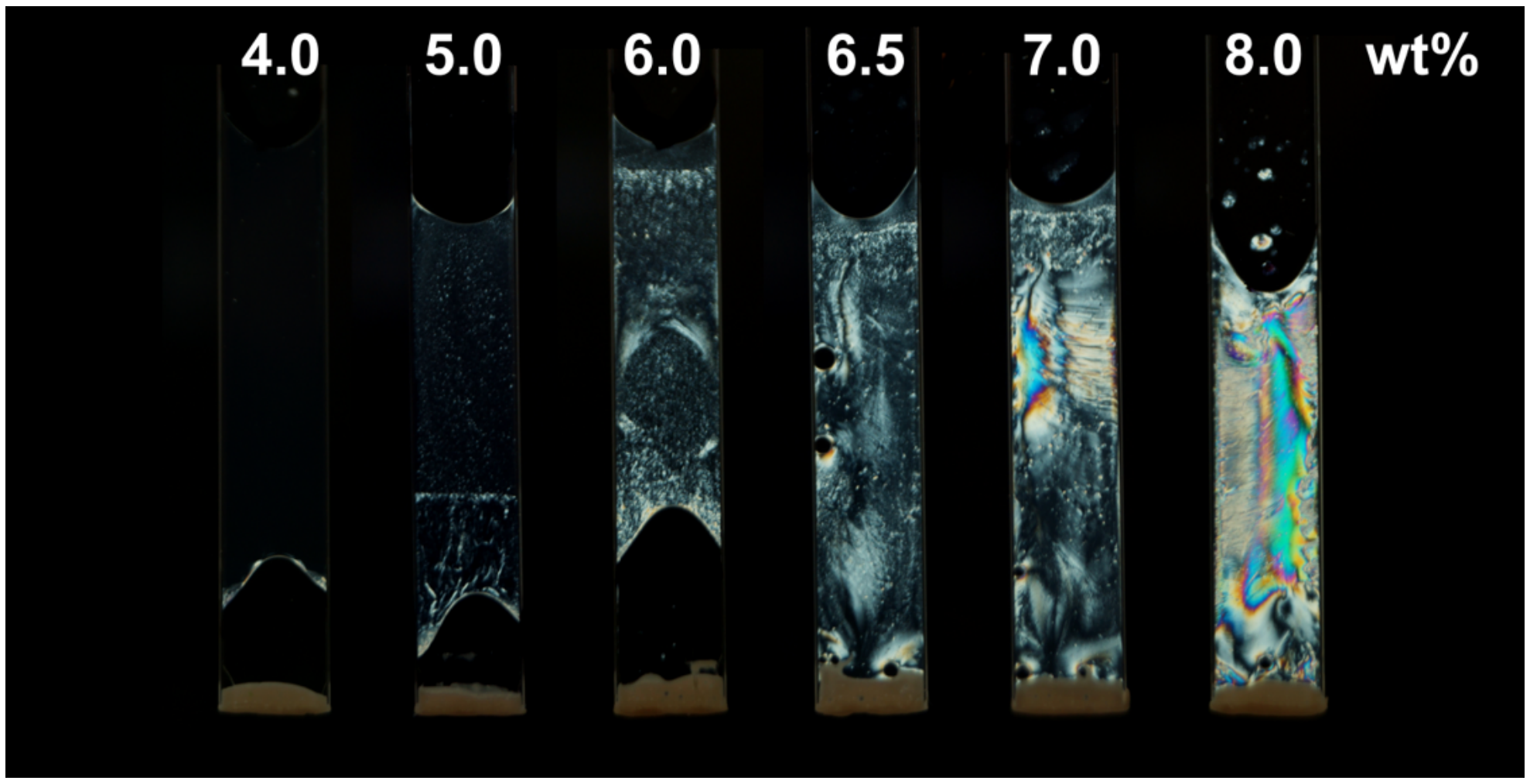

**Figure S16**. Photograph of a series of capillaries containing CNC suspensions that were exposed to calcium and then dialyzed against ultrapure water (**C3-Ca**) with increasing CNC weight fraction from left to right.

## S8 Impact of Salt Condition on CNC Self-Organization

To illustrate the practical significance of our findings for CNC self-organization, we performed a series of experiments using sonicated CNCs (sonication doses 12, 192 and 768 J $mL^{-1}$) from our previous study.[15] For each sample, a mother suspension of concentrated CNCs (> 10 wt%) was prepared in the presence of NaCl (0.12 mmol $g^{-1}$ of CNC) before being diluted with ultrapure water to obtain a range of CNC volume fractions ($\Phi$) at fixed ratio of NaCl per CNC. The self-assembly behavior of these suspensions was investigated and the Z-average diameter of the CNCs after dilution (0.1 wt%, 1 mM NaCl) was measured and compared to their value before concentration in the presence of salt.

The evolution of the anisotropic volume fraction ($\varphi_{ani}$) as a function of $\Phi$ for each sample is presented in **Figure S17a**. The $\varphi_{ani}$ for all samples were superimposed, with an appearance of the anisotropic phase around 2 vol% followed by a sigmoid increase until a fully anisotropic phase is reached around 7 vol%. In **Figure S17b**, the evolution of the corresponding pitches ($p$) with $\Phi$ started to converge whatever the sample sonication history. These results are in stark contrast with the behavior of the same suspensions without salt,[15] for which the self-assembly behavior was dependent of the sonication dose. Moreover, all samples that were concentrated in salt also exhibited an increase of Z-average diameter (**Figure S17c**).

Overall, this experiment shows that a suspension of CNCs reaching 12 mM of NaCl can lead to irreversible morphological changes that impact subsequent self-organization. While adding salt to a CNC suspension is known to shift its biphasic regime to higher $\Phi$ and its pitch to lower values,[16] to our knowledge, such superimposition of the self-organization values for different samples was never reported. These observations are coherent with the irreversible formation of bundles upon concentration of the CNCs in the presence of salt, leading to a shift of the biphasic regime toward lower $\Phi$ and to lower pitches. However, further work is needed to elucidate the exact origin of the superimpositions of the $\Phi$ and pitch values, despite the size differences between the samples.

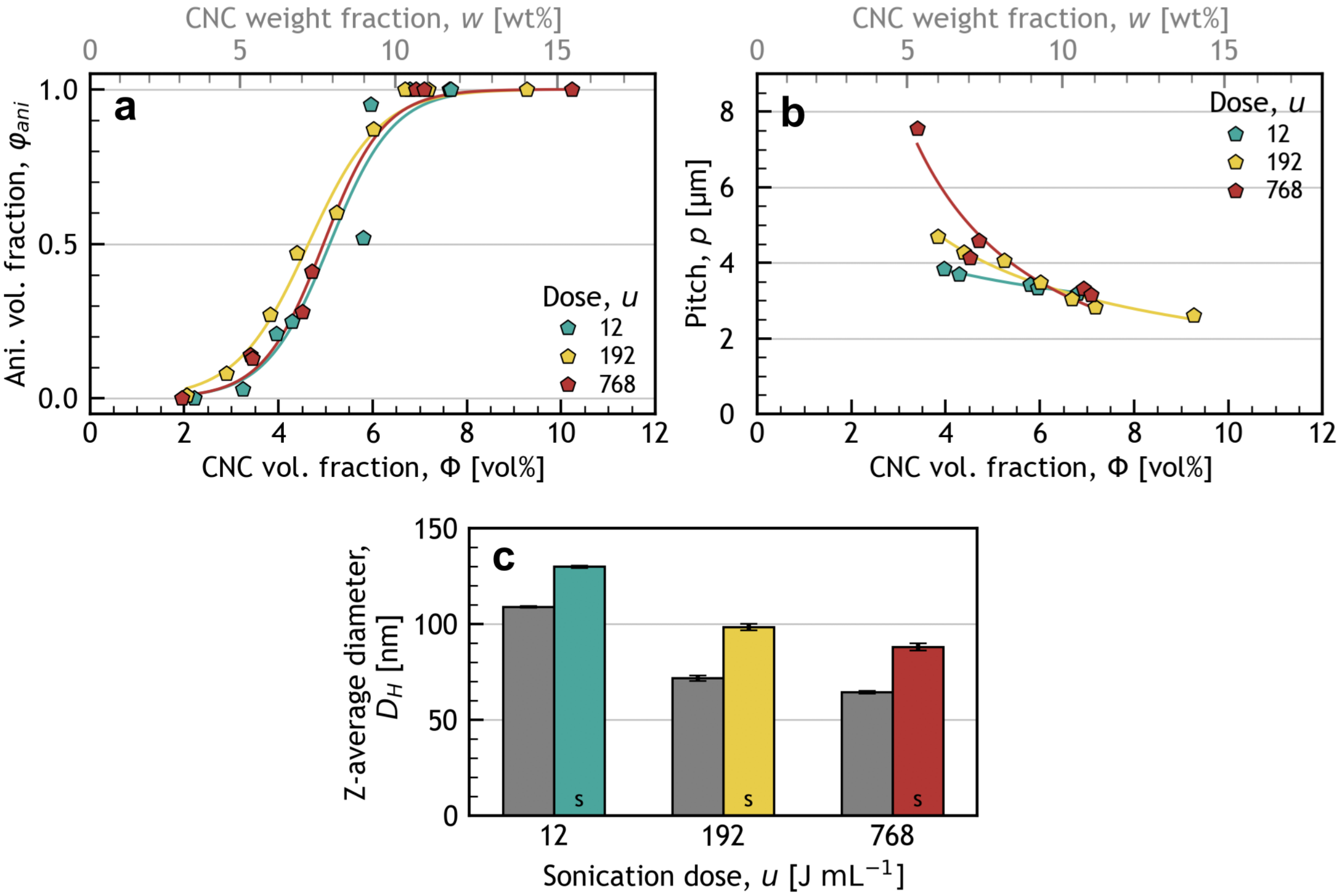

**Figure S17**. Evolution of the **(a)** anisotropic phase volume fraction ($\varphi_{ani}$) and **(b)** pitch ($p$) as a function of the CNC volume fraction ($\Phi$) for sonicated samples (12, 192 and 768 J $mL^{-1}$) at fixed ratio of NaCl per CNC (0.12 mmol $g^{-1}$), after concentration in the presence of NaCl (> 10 wt% CNCs, > 12 mM NaCl). Lines are guides to the eye. **(c)** Z-average diameter ($D_H$) of the CNCs (at 0.1 wt% in 1 mM NaCl) before (grey) and after (colored, marked with letter 's') concentrating CNC in presence of NaCl salt.

## S9. Viscometry

### S9.1 Fitting of the Relative Viscosity

In most previous studies, the intrinsic viscosity of CNC suspensions was obtained by fitting the data using a Fedors plot, according to the following expression:[17–23]

$$\frac{1}{2\left(\sqrt{\eta_r}-1\right)}=\frac{1}{[\eta]c}-\frac{1}{[\eta]c_m} \tag{S13}$$

where $\eta_r$ and $[\eta]$ are respectively the relative and intrinsic viscosities, and $c$ and $c_m$ are respectively the particle concentration and the concentration at maximum packing. However, this approach often yields physically incoherent (i.e. negative or huge) critical concentrations,[24] particularly when applied on low concentrations where small uncertainties lead to huge change of $\sqrt{\eta_r}-1$. Therefore, the authors of the method recommend neglecting data points at low concentrations (*i.e.* below 0.01 g mL$^{-1}$). In the present study, the viscosity was mainly investigated at low concentration to stay in the dilute regime. Consequently, applying this method to our data still requires to consider low concentrations points (*i.e.* from 0.004 g mL$^{-1}$ to have enough data for a meaningful fit, as illustrated in **Figure S18** and **Table S14**. Instead, we used the Huggins approach, presented in **Equation** 4, which does not suffer from this issue. The intrinsic viscosity values extracted from both methods are quite similar (see **Table S14** and **Table S15**).

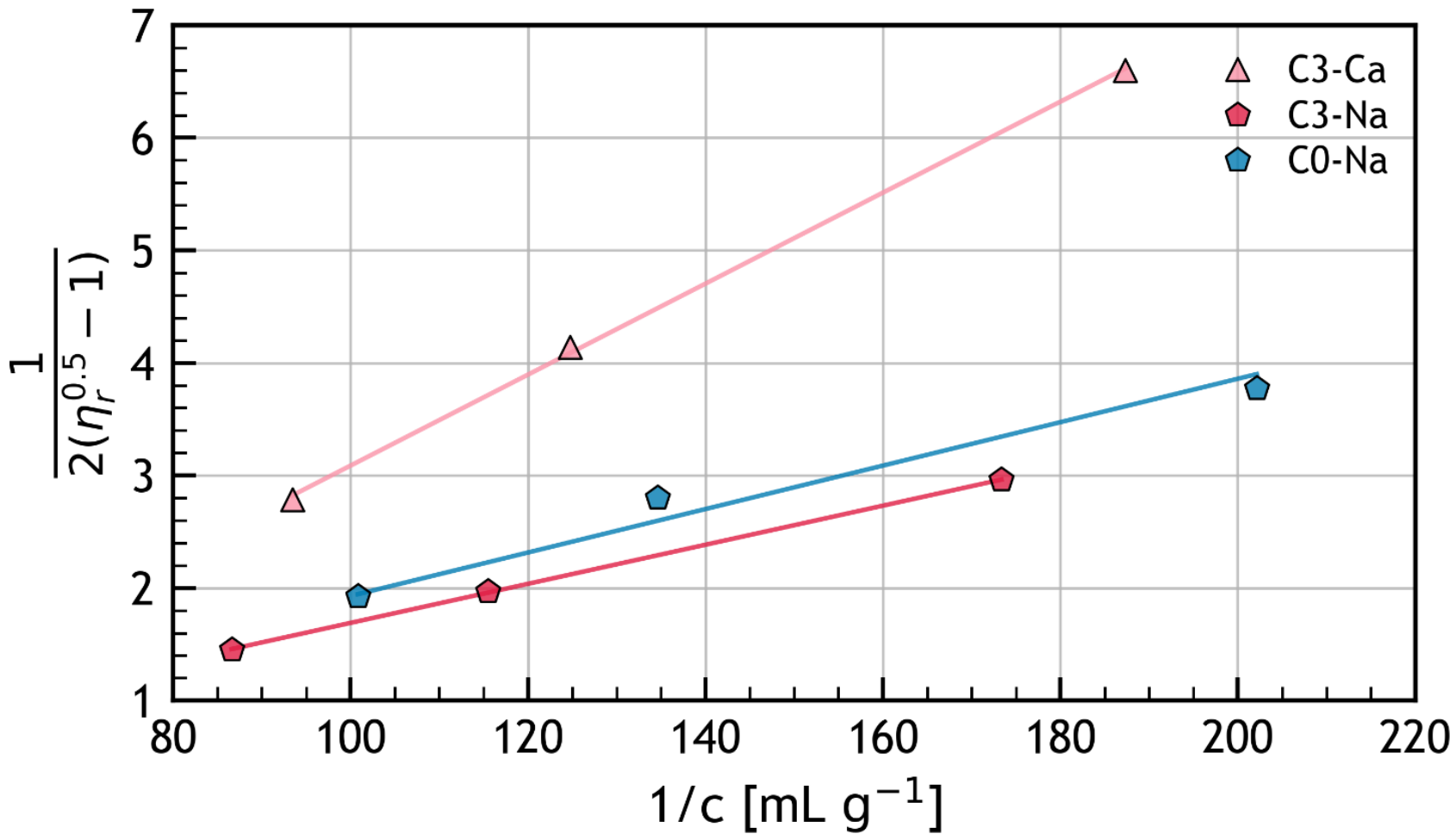

**Figure S18**. Fedors plot of the relative viscosity data. Values below 0.004 g mL$^{-1}$ (i.e. above than 250 mL g$^{-1}$) were excluded.

**Table S14**. Results from the Fedors fit using **Equation** S13, with intrinsic viscosity ([$\eta$]), concentration at maximum packing ($c_m$) and corresponding coefficient of determination ($R^2$).

| Sample | [$\eta$] [mL g$^{-1}$] | $c_m$ [g mL$^{-1}$] | $R^2$ |
|---|---|---|---|
| **C0** | 52 ± 8 | 6.6E5 ± 9E12 | 0.967 |
| **C3** | 58 ± 1 | 0.4 ± 0.2 | 1.000 |
| **C3-Ca** | 25 ± 1 | 4.2E-2 ± 0.5E-2 | 0.999 |

**Table S15**. Results from the Huggins fit using **Equation** 4, with intrinsic viscosity ([$\eta$]), Huggins coefficient ($k_H$) and corresponding coefficient of determination ($R^2$).

| Sample | [$\eta$] [mL g$^{-1}$] | $k_H$ | $R^2$ |
|---|---|---|---|
| **C0** | 56 ± 6 | 0.1 ± 0.2 | 0.992 |
| **C3** | 54 ± 2 | 0.5 ± 0.1 | 0.999 |
| **C3-Ca** | 25 ± 3 | 1.6 ± 0.8 | 0.994 |

### S9.2 Expression for 3D Aspect Ratio

Various expressions for rods and spheroids can be used to relate the 3D aspect ratio ($r$) to the intrinsic viscosity ($[\eta]$), the Simha relation being the most widespread.[25] However, this expression is specific to rods and the CNC bundles in this work are closer in shape to a spheroid than a cylinder. Therefore, it is more appropriate to use an expression derived from Doi and Edwards, and Brenner for prolate spheroids,[26,27] for which the rotational friction constant is exactly resolved.

The intrinsic viscosity is defined as:

$$[\eta] = \lim_{c \to 0} \frac{1}{c} \frac{\eta - \eta_s}{\eta_s} \tag{S14}$$

where $\eta$ and $\eta_s$ are the viscosities of the polymer and of the solvent respectively. In steady shear flow rate for rod-like polymers, the viscosity can be expressed as:[26]

$$\eta = \eta_s + \frac{2}{15} \frac{c}{\rho_p V_p} \zeta_{rot} \tag{S15}$$

where $c$ is the particle concentration, $\rho_p$ is the particle density, $V_p$ is the volume of a particle, and $\zeta_{rot}$ is their rotational friction constant. According to Brenner,[27] the rotational friction constant can be expressed as:

$$\zeta_{rot} = 6\eta_s V_p K \tag{S16}$$

where $K$ is a dimensionless scalar coefficient defined in **Equation** S18. Consequently, from **Equation** S14 to S16, the intrinsic viscosity can be simplified as:

$$[\eta] = \frac{4}{5} \frac{K}{\rho_{CNC}} \tag{S17}$$

For spheroids:[27]

$$K = \frac{2(r^2 + 1)}{3(r^2 \alpha_{\parallel} + \alpha_{\perp})} \tag{S18}$$

with:

$$\alpha_{\perp} = \frac{r^2}{r^2 - 1}(1 - \beta) \tag{S19}$$

$$\alpha_{\parallel} = \frac{2}{r^2 - 1}(r^2 \beta - 1) \tag{S20}$$

and where $r$, the 3D aspect ratio of the spheroid, is expressed as the ratio of its major axis over its minor axis. Finally, for prolate spheroids (i.e. with $r > 1$):

$$\beta = \frac{\cosh^{-1}(\mathrm{r})}{\mathrm{r}\sqrt{\mathrm{r}^2 - 1}} \tag{S21}$$

This yields **Equation** 5, as presented in the Results and Discussion section of the main article.

## S10. Fitting the Z-average Diameter as a Function of Ultrasonication Dose

To model the evolution of the hydrodynamic diameter of the CNCs samples with the sonication dose, we propose to use a modified dissociation expression. Thereby, an infinitesimal change of size ($dD_H$) as a function of an infinitesimal change of ultrasonication dose ($du$) is expressed by equation:

$$\frac{dD_H}{du} = -k'(u)\, D_H \tag{S22}$$

with:

$$k'(u) = k^{\alpha}\, \alpha\, u^{\alpha-1} \tag{S23}$$

where $k'$ (mL J$^{-1}$) is a dose-dependent factor of size decrease, $k$ is a constant and $\alpha$ is a stretching exponent. Integrating this expression leads to:

$$D_H(u) = [D_H^0 - D_H^\infty]\, \exp(-[k\, u]^{\alpha}) + D_H^\infty \tag{S24}$$

where $D_H^0$ is the size before sonication and $D_H^\infty$ is the size at infinite sonication, fixed at 59 nm for all samples.

To independently isolate the rate of dissociation of the composite particles formed during calcium-induced aggregation, the Z-average diameter of Ca-CNCs ($D_{H,Ca}$) included the fit of the parent sample according to:

$$D_{H,Ca}(u) = \left[D_{\mathrm{H,Ca}}^0 - D_H^0\right]\, \exp(-[\mathrm{k_{Ca}}\, u]^{\alpha_{Ca}}) + D_H(u) \tag{S25}$$

where $D_{H,Ca}^0$ is the size before sonication and $D_H(u)$ is the fitted size of the corresponding parent sample using **Equation** S24. **Equation** S24 and **Equation** S25**,** were used to fit the data presented in **Figure 9** and the corresponding fitting parameters are presented in **Table S16**.

For all samples, the model fitted the data very well for all sonication doses (**Table S16**). Moreover, this approach permits to extract independent factors of size decrease for each type of composite particle, providing information on their relative disassociation behavior. The constants $k$ for bundle particles were an order of magnitude smaller than their corresponding $k_{Ca}$ for calcium induced composite particles (**Table S16**). The $\alpha$ coefficients for the bundle particles were also similar and slightly lower than the $\alpha_{Ca}$ for calcium induced composite particles (**Table S16**). Overall, this indicates that sonication energy had a bigger effect on the size of calcium induced composite particles than on that of bundle particles.

**Table S16**. Best fitting parameters (*k* or $k_{Ca}$ and $\alpha$ or $\alpha_{Ca}$) and goodness of fit ($R^2$ and MSE) for the evolution of the Z-average diameter as a function of the sonication dose using a modified dissociation function (**Equation** S24 and **Equation** S25).

| Sample | $k$ or $k_{Ca}$ [mL J$^{-1}$] | $\alpha$ or $\alpha_{Ca}$ | $R^2$ | MSE |
|---|---|---|---|---|
| **C0** | 0.030 ± 0.003 | 0.57 ± 0.04 | 1.00 | 1.0 |
| **C3** | 0.036 ± 0.004 | 0.53 ± 0.04 | 0.99 | 3.5 |
| **C0-Ca** | 0.300 ± 0.030 | 0.66 ± 0.09 | 1.00 | 6.4 |
| **C3-Ca** | 0.340 ± 0.030 | 0.74 ± 0.09 | 1.00 | 5.2 |